\documentclass{article}

\usepackage{graphicx} 
\usepackage{booktabs}
\usepackage{float}
\usepackage{setspace}
\usepackage{adjustbox}
\usepackage{threeparttable}
\usepackage{multirow, makecell}
\usepackage{PRIMEarxiv}
\usepackage{wrapfig,lipsum,booktabs}
\usepackage{titletoc}

\usepackage[utf8]{inputenc} 
\usepackage[T1]{fontenc}    
\usepackage{hyperref}       
\usepackage{url}            
\usepackage{booktabs}       
\usepackage{amsfonts}       
\usepackage{amsmath} 
\usepackage{multirow}
\usepackage{array}
\usepackage{nicefrac}       
\usepackage{microtype}      
\usepackage{lipsum}
\usepackage{tcolorbox}
\usepackage{lmodern}
\usepackage{fancyhdr}       
\usepackage{graphicx}       
\usepackage{listings}       
\usepackage{chngcntr}       
\usepackage[table]{xcolor}
\usepackage{soul}
\usepackage{url}
\usepackage{amsmath, amssymb, amsfonts}
\usepackage{hyperref}
\usepackage{algorithm}
\usepackage{algorithmic}
\usepackage{amsmath}
\usepackage{bm}
\usepackage{caption}

\graphicspath{{figures/}}

\makeatletter
\NewDocumentCommand{\matskraft}{}{M\lowercase{at}{\textsc{skraft}}%
  \@ifnextchar.{}{
    \@ifnextchar,{}{%
      \@ifnextchar;{}{%
        \@ifnextchar:{}{%
          \@ifnextchar?{}{%
            \@ifnextchar!{}{}%
          }%
        }%
      }%
    }%
  }
}
\makeatother

\graphicspath{{figures/}}

\title{\textsc{MatSKRAFT}: A framework for large-scale materials knowledge extraction from scientific tables}

\author{
    Kausik Hira$^{1}$, Mohd Zaki$^{2}$, \textbf{Mausam Mausam}$^{1,3,\#}$, \textbf{N. M. Anoop Krishnan}$^{1,2,\#}$ \\
    $^{1}$Yardi School of Artificial Intelligence, Indian Institute of Technology Delhi \\
    $^{2}$Department of Civil Engineering, Indian Institute of Technology Delhi \\
    $^{3}$Department of Computer Science and Engineering, Indian Institute of Technology Delhi \\
    $^\#$Corresponding authors: \texttt{\{mausam, krishnan\}@iitd.ac.in} \\
}

\begin{document}
\maketitle

\begin{abstract}
Scientific progress increasingly depends on synthesizing knowledge across vast literature, yet most experimental data remains trapped in semi-structured formats that resist systematic extraction and analysis. Here, we present \matskraft, a computational framework that automatically extracts and integrates materials science knowledge from tabular data at unprecedented scale. Our approach transforms tables into graph-based representations processed by constraint-driven GNNs that encode scientific principles directly into model architecture. \matskraft{} significantly outperforms contemporary frontier large language models, achieving F1 scores of 89.33 for property extraction and 71.35 for composition extraction, while processing data $6$-$496\times$ faster---compared to the fastest and the slowest models, respectively---with modest hardware requirements. Applied to 66,267 tables from more than 45,500 research publications, we construct a comprehensive database containing 509,281 entries, including 104,000 compositions that expand coverage beyond major existing databases. This systematic approach reveals previously overlooked materials with distinct property combinations and enables data-driven discovery of composition-property relationships forming the cornerstone of materials and scientific discovery.
\end{abstract}

\section{Introduction}
Scientific progress increasingly depends on synthesizing knowledge across vast scientific literature to discover composition--structure--processing--property relationships and accelerate discovery cycles. However, the exponential growth of scientific publications has created a fundamental computational challenge: an estimated 80\% of experimental data remains locked in semi-structured formats including tables and figures~\cite{olivetti2020data, fortunato2018science, bornmann2015growth, price1963little, zeng2017science}. This creates a bottleneck for knowledge-driven discovery through traditional manual methods, particularly in scientific domains, such as chemistry, physics, or materials, where composition-property relationships documented across decades of research hold the key to designing next-generation technologies~\cite{venugopal2021looking, hira2024reconstructing}.

Tables in material science articles form a particularly rich repository of experimental data, where composition--property relationships are documented across diverse experimental conditions and material classes~\cite{hira2024reconstructing,olivetti2020data}. Several computational approaches have been developed for materials table extraction using regular expressions~\cite{mit_table_parser,swain2016chemdataextractor}, fine-tuned language models~\cite{gupta2022matscibert,zhao2023opticalbert, song2023matsci}, specialized graph neural networks (GNNs)~\cite{gupta-etal-2023-discomat}, and more recently, large language models (LLMs)~\cite{jablonka2024leveraging, jablonka202314, m2024augmenting, dagdelen2024structured,yi2024matablegpt, song-etal-2023-honeybee}. However, these methods face fundamental computational limitations: regular expressions lack generalizability, small language models require expensive manual annotation, and LLMs face critical limitations in handling complex table structures (as further demonstrated in later sections), computational accuracy, scalability, and domain-specific knowledge when processing heterogeneous scientific data at scale~\cite{schilling2024text, miret2024llms, circi2024well, hira2024reconstructing, zaki2024mascqa, circi2024extracting, mirza2024large, alampara2024mattext}.

The computational challenge extends beyond data extraction to knowledge synthesis at scale. Manual literature review cannot systematically analyze hundreds of thousands of experimental results distributed across decades of research, while existing automated approaches either lack accuracy for reliable scientific knowledge bases (KBs) or demand computational resources that render large-scale analysis impractical. This prevents systematic identification of novel relationships, rare property combinations, and underexplored regions of design space that could guide discovery efforts~\cite{olivetti2020data,kim2017materials}.

Here, we present \matskraft{} (acronym for Materials Science Knowledge Repository Accumulated From Tables), a computational framework addressing these limitations through: (i) constraint-driven GNNs that encode scientific principles directly into model architecture, (ii) automated training data generation through distant supervision, annotation algorithms and data augmentation---eliminating dependence on large-scale manual annotation, and (iii) computational efficiency enabling processing of entire research domains with modest hardware requirements (for instance, 1 V100 GPU with 32 GB RAM). By processing over 45,500 research articles, we demonstrate how specialized computational methods can systematically extract and synthesize scientific knowledge, creating comprehensive databases with over 509,000 entries. This work establishes how computational approaches can make the collective experimental record systematically accessible for knowledge synthesis, the first step of accelerated materials development~\cite{krishnan2024machine, venugopal2024matkg, song2023matsci,gupta-etal-2023-discomat,walker2021impact,zhao2023opticalbert,buehler2024mechgpt,song-etal-2023-honeybee,chemllm,zaki2022extracting,zaki2022natural,zaki2024mascqa}.

\section{Results}
\label{sec:results}

\begin{figure}[t!]
    \centering
    \includegraphics[width=0.9\textwidth, keepaspectratio]{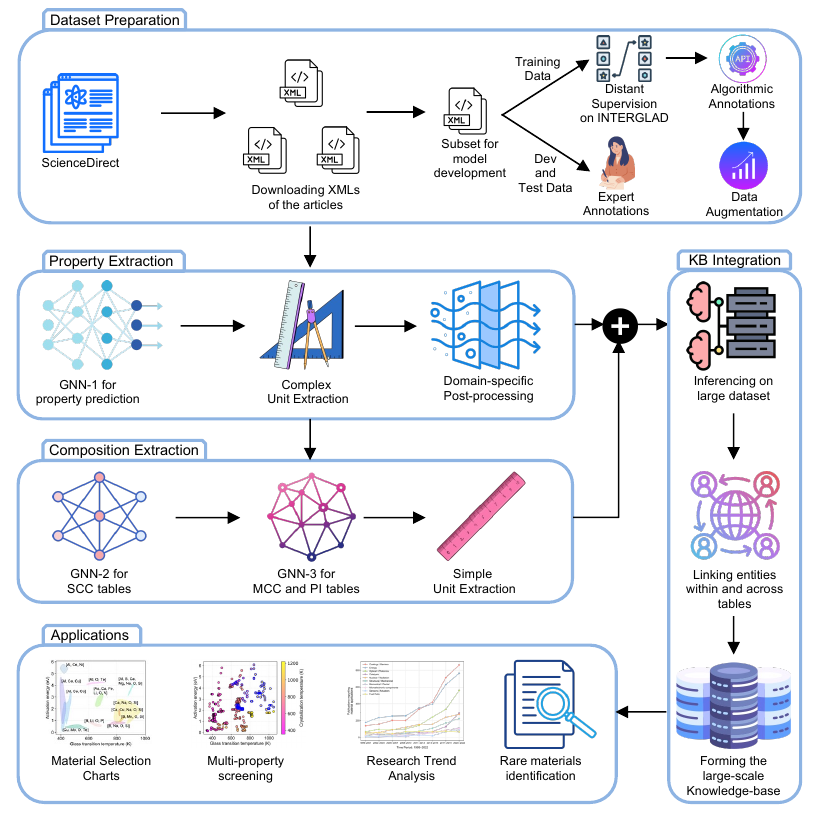}
    \caption{\textbf{\textsc{MatSKRAFT} framework for automated knowledge extraction from scientific tables.}
    The pipeline comprises: (1) Dataset preparation involving systematic processing of scientific articles and automated training data generation, (2) Property extraction using graph neural network (GNN1) with domain-based post-processing, (3) Composition extraction employing specialized GNNs for different table structures, (4) Knowledge-base construction by linking entities within and across tables, and (5) Applications enabled by large-scale analysis.}
    \label{fig:1}
\end{figure}

\subsection{Overview of \textsc{MatSKRAFT}}
\matskraft{} introduces a comprehensive framework for automated scientific knowledge extraction from tabular data in literature. The framework comprises five integrated components designed to systematically process scientific tables and construct large-scale KBs as shown in Figure~\ref{fig:1}. The pipeline begins with table data processing, which involves systematic collection and parsing of scientific articles from XML sources and automated generation of high-quality training data. Our approach combines distant supervision from existing databases, expert-verified annotation algorithms that encode domain expertise, and strategic data augmentation to generate balanced training datasets across diverse material properties (detailed in Section~\ref{subsec: method_dataset_cons} and Supplementary Information~\ref{algo_for_annotation_code_prop} and~\ref{algo_for_table_augmentation}). This component addresses a critical bottleneck in scientific information extraction by eliminating dependence on expensive manual annotation~\cite{roh2019survey, huang2024data, song2022learning}.

\matskraft{} employs two specialized extraction modules tailored to different information types. Property extraction utilizes a GNN (GNN-1) that represents tables as structured graphs, where nodes correspond to cells, headers, and captions, while edges capture structural relationships and semantic dependencies. The architecture incorporates constraint-driven learning that encodes scientific principles directly into the neural network, followed by domain-specific post-processing rules that implement sophisticated disambiguation mechanisms for accurate property identification across heterogeneous reporting formats~\cite{hira2024reconstructing} (see Section~\ref{subsec: methods_prop_extraction}).

Composition extraction employs separate specialized architectures (GNN-2 and GNN-3) designed to handle different table structures commonly found in materials literature. GNN-2 processes single-cell composition (SCC) tables where complete compositions appear within individual cells, while GNN-3 handles multiple-cell composition (MCC) tables where constituents are distributed across separate cells, and Partial-Information (PI) tables requiring inference from broader document context. This structure-specific approach addresses the heterogeneous reporting conventions prevalent in scientific literature (refer to Section~\ref{subsec: methods_comp_extraction}).

The KB integration component links extracted compositions and properties through dual-pathway integration: linking entities within individual tables (intra-table) and connecting entities across different tables (inter-table) within an article. For intra-table linking, the system performs orientation-based connections within individual tables by analyzing structural patterns and positional relationships. For inter-table linking, the system uses identifier-based association (such as material IDs) to connect information across separate tables that reference the same materials. This integration process constructs coherent composition-property relationships from fragmented tabular data (see Section~\ref{subsec: methods_linking_info}).


Finally, the framework enables diverse applications through the resulting comprehensive database, including materials selection charts for systematic exploration of composition-property relationships, multi-property screening for identifying materials meeting complex design criteria, temporal analysis of research trends, and accelerated selection of rare materials with exceptional property combinations. Thus, the complete pipeline transforms heterogeneous tabular data from scientific literature into structured, query-able KBs that support systematic materials discovery and development.

\subsection{Training Dataset for \textsc{MatSKRAFT}}
We constructed the training dataset for composition and property extraction through a hierarchical pipeline that avoids manual annotation. The approach combines multiple techniques including distant supervision~\cite{mintz2009distant}, annotation algorithms, and power-law guided data augmentation (see Section~\ref{subsec: method_dataset_cons}). Distant supervision leverages existing structured datasets to automatically generate labeled training data by aligning text with known facts or relationships. Here, for distant supervision, we employed INTERGLAD~\cite{interglad}, a commercial dataset of inorganic glasses comprising compositions, properties and the references from which they have been extracted. Distant supervision using INTERGLAD and the full text of the respective journals provided an initial 805 training instances (property-bearing rows or columns), with F1 score of 82.77 on the expert-annotated development set. Improving this further with domain-informed annotation algorithms (see Supplementary Information~\ref{subsec: appendix_hierarch_performance_validation} and ~\ref{algo_for_annotation_code_prop}, coverage expanded to 2,162 instances and the F1 score improved to 88.18, with precision increasing in 17 out of 18 properties and a notable overall precision gain of 11.36 points. 

To address frequency imbalance across materials' properties (that is, some properties having low number of entries), we introduced statistically grounded augmentation based on co-occurrence priors and Gaussian sampling (see Section~\ref{method_data_aug} and Supplementary Information~\ref{algo_for_annotation_code_prop}). This approach increased the training set to 3,561 instances and achieving an F1 score of 89.56, with balanced precision (90.59\%) and recall (88.56\%). The final dataset comprises 2,793 tables (2,009 training, 416 validation, 368 manually annotated test) with comparable property and non-property distributions, enabling robust evaluation of extraction performance across diverse table structures. This automated training pipeline resolves the data scarcity challenge while maintaining scientific-grade accuracy for both common and specialized properties (detailed in Supplementary Information~\ref{subsec:hierarchical_data_generation}).

For composition extraction, we enhanced our earlier work on structure-aware table classification, presented in DiSCoMaT~\cite{gupta-etal-2023-discomat}, by applying targeted annotation algorithms to identify compositional content initially overlooked. In this approach, tables are classified based on the nature in which the compositions are represented. Specifically, they are classified as SCC, MCC, PI, and non-composition (NC) tables~\cite{gupta-etal-2023-discomat}. Our structure-specific annotation algorithms targeting MCC and PI formats reclassified compositional information within the non-compositional (NC) tables, increasing total compositional table coverage from 1,647 to 1,801 out of 4,408 training tables compared to the original training split in DiSCoMaT~\cite{gupta-etal-2023-discomat}.


\subsection{Performance Evaluation}

To evaluate \matskraft, we obtained domain-expert (manually) annotated dev and test dataset. The dev and test sets consist of 416 and 738 tables for composition extraction, and 368 and 737 tables for property extraction, respectively. We conducted a comprehensive evaluation of \matskraft's extraction performance across both property and composition identification tasks, benchmarking against leading LLMs at the time of evaluation.

\subsubsection{Property Extraction}
The property extraction module achieved an F1 score of 89.56\% and 88.68\% on the validation and test datasets respectively. On applying confidence thresholding to properties with imbalanced precision-recall on the validation dataset (see Supplementary Information~\ref{app:confidence_thresholding}), these F1 scores improved to 90.38\% and 89.33\%---with test precision of  92.40\% and recall of 86.45\% across 18 distinct material properties (see Supplementary Table~\ref{tab:threshold_impact}). Performance varied systematically across property categories, with the framework demonstrating robust extraction capabilities for frequently reported properties. For instance, density extraction achieved the highest F1 score of 96.57\%, followed by glass transition temperature (93.00\%) and crystallization temperature (92.99\%). For industrially relevant physical properties, the model exhibited F1 scores of 94.01\% for liquidus temperature, 91.51\% for melting temperature, and 82.02\% for activation energy (detailed in Supplementary Table~\ref{tab:detailed_property_metrics}).


The effectiveness of our automated training data generation strategy (see Section~\ref{subsec: method_dataset_cons} and Supplementary Information~\ref {algo_for_annotation_code_prop}) is particularly evident from the detailed ablation studies performed (see Supplementary Information~\ref{subsec: appenix_data_ablation}). Removal of the annotation algorithm leads to severe degradation in precision over 18 points. Properties with minimal representation in the original training dataset---such as Abbe value, fracture toughness, and shear modulus---achieved test F1 scores of 100.0\%, 77.42\%, and 85.45\% respectively. This represents a classic long-tail distribution challenge in scientific information extraction, which has been resolved through strategic power-law guided training data augmentation (see Section~\ref{subsec: method_dataset_cons}, Supplementary Figure~\ref{fig:hierarchical_data_preparation}, and Supplementary Information~\ref{algo_for_table_augmentation}). For hardness measurements (F1 of 87.34\%), the model successfully extracts values across different hardness scales reported in the literature, including Vickers, Knoop, Rockwell, and Shore scales, enabled by our rule- and dictionary-based unit extraction framework (see Section~\ref{subsec: unit_extraction_method} and Supplementary Information~\ref{algo_unit_extraction}).

However, challenges remain with properties characterized by inconsistent reporting conventions. Examples include thermal expansion coefficient achieving an F1 score of 68.60\%, Poisson's ratio (73.14\%), reflecting the diverse notation systems or unit representations prevalent across different research groups and publication venues for these properties.

\subsubsection{Composition Extraction}
The composition extraction system achieved an overall F1 score of 71.35\%, with precision of 82.31\% and recall of 62.97\% on the test dataset. Performance varied significantly across different table structures, reflecting the heterogeneous reporting conventions found in materials science literature. SCC tables, processed by GNN-2, achieved the strongest performance with an F1 score of 78.62, where entire compositions are written within individual cells. MCC tables, handled by GNN-3, also demonstrated comparable performance with F1 score of 75.99, where compositions are distributed across multiple cells by reporting each constituent separately. The most challenging category proved to be PI tables, where GNN-3 achieved an F1 score of 52.82 with precision of 82.15\% and recall of 38.93\%. The lower recall reflects the inherent difficulty of this task, as complete composition details must be inferred from article text beyond the tables themselves, requiring sophisticated contextual understanding (see Section~\ref{subsec: methods_comp_extraction}).

\subsubsection{Evaluation across LLMs and Tabular Baselines}

\begin{figure}[htbp]
    \centering
    \includegraphics[width=0.9\textwidth, keepaspectratio]{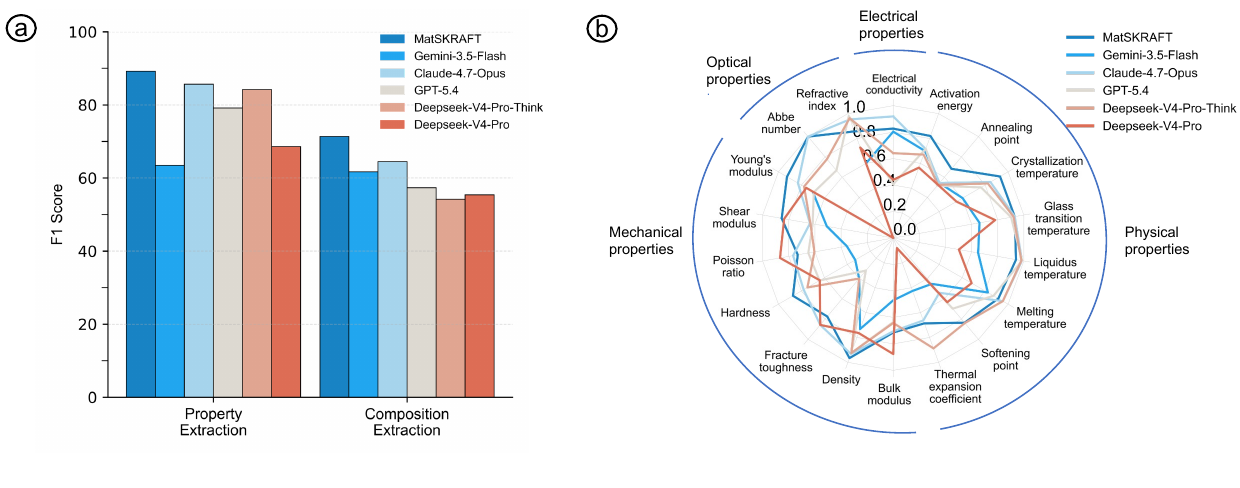}
    \caption{\textbf{Performance comparison of \textsc{MatSKRAFT} with five frontier LLMs released in 2026 for materials science information extraction from tables.} (a) F1 scores for property and composition extraction across MatSKRAFT, Gemini-3.5-Flash, Claude-4.7-Opus, GPT-5.4, DeepSeek-V4-Pro-Thinking, and DeepSeek-V4-Pro. (b) Per-property F1 scores across 18 material properties grouped by category (electrical, physical, mechanical, and optical), enabling direct model comparison at the individual property level.}
    \label{fig:2}
\end{figure}

LLMs have been widely used for extracting compositions and properties from tables in scientific literature.  To contextualize \matskraft{}'s performance, we benchmarked it against five leading LLMs released in 2026---including Gemini-3.5-Flash~\cite{google2026gemini35flash}, GPT-5.4~\cite{openai2026gpt54}, Claude-4.7-Opus~\cite{anthropic2026claudeopus47}, DeepSeek-V4-Pro-Thinking and DeepSeek-V4-Pro (Non-thinking)~\cite{deepseek2026deepseek}. Despite these models having vast parametric scales and generalized capabilities, \matskraft{} outperformed all LLM baselines on both the extraction tasks (Figure~\ref{fig:2}a).

For property extraction, the best-performing LLM (Claude-4.7-Opus) achieved an F1 score of 85.65\% on the test dataset, falling 3.68\% below \matskraft's performance (89.33\%). The margin increased for composition extraction, where the best LLM (again Claude-4.7-Opus, with F1 score of 64.51\%) fell 6.84\% short of \matskraft's performance (71.35\%) (see Supplementary Table~\ref{tab:model_comparison}). However, LLM capabilities have advanced substantially since our previous benchmarks. Previously in 2025, we benchmarked against LLMs released during 2024-2025 --- Gemini-1.5-Pro~\cite{gemini2024}, GPT-4o~\cite{achiam2023gpt}, Claude-3.5-Sonnet~\cite{claude2024}, DeepSeek-V3~\cite{liu2024deepseek}, and DeepSeek-R1~\cite{guo2025deepseek}, and found the best-performing LLM ---DeepSeek-V3 for property extraction with F1 score of 73.28\% and DeepSeek-R1 for composition extraction with F1 score of 54.28\% on the test dataset, trailing \matskraft{} by more than 15 points on both extraction tasks.

The LLM baselines demonstrated stronger performance on a subset of properties with standardized notation, having well-defined value ranges, and no unit extraction burden. For refractive index, where values are physically bounded between approximately 1 and 6 and notation is uniform across the literature, GPT-5.4, Deepseek-V4-Pro-Think, and Claude-4.7-Opus achieved F1 scores of 98.11, 96.30, and 95.38 respectively, performing significantly better than \matskraft{}(F1 score - 85.71\%). However, these models struggled significantly with properties requiring contextual disambiguation or domain knowledge.



The domain-specificity advantage extends to properties that expose fundamental gaps in generalist pretraining. Hardness extraction provides the most striking illustration: \matskraft{} achieves an F1 score of 87.34\%, while the 2026 frontier LLM cohort spans a wide range of 33.12--78.05\% (comprising Claude-Opus-4.7 (78.05\%), DeepSeek-V4-Pro-Think (74.77\%), GPT-5.4 (64.23\%), DeepSeek-V4-Pro (63.96\%), and Gemini-3.5-Flash (33.12\%)), representing a spread of 45 points among models from the same frontier generation. This extreme variance indicates that even the most capable generalist models remain unreliable on this property, reflecting the absence of systematic protocols for disambiguating parallel hardness scales (Vickers, Knoop, Rockwell, Shore) that \matskraft{}'s property-specific framework handles deterministically. Annealing point presents a different challenge: as a viscosity-defined processing temperature specific to glass science only, it is systematically underrepresented in general pretraining corpora. Consequently, all five frontier LLMs converge to a remarkably narrow band of 51.61--54.24\%---a tight clustering that itself constitutes diagnostic evidence of a shared pretraining blind spot rather than model-specific variance, compared to \matskraft{}'s F1 score of 68.24\%. These property-level findings demonstrate that the performance advantages of domain-specialized architecture are most durable precisely where the extraction is hardest (see Figure~\ref{fig:2}b).

In addition to proprietary frontier LLMs, we further extend this benchmarking to two additional baseline categories--- open-source LLMs (Supplementary Information~\ref{sec:appendix_qwen}) and open-source tabular-based models (Supplementary Information~\ref{sec:appendix_table_baselines}). First, we benchmark five compact open-source LLMs from the Qwen3~\cite{bai2025qwen3,yang2025qwen3} and Qwen3.5~\cite{team2026qwen3} variants (4B--9B parameters, including vision--language variants). All the five variants F1 score collapse to 0.76--9.36 for property extraction and 3.57--11.67 for composition extraction (see Supplementary Table~\ref{tab:qwen_strict}). On error analysis, we realize the models often extract the correct composition or property values; but misplace its source coordinates, yielding tuples that cannot be merged  to assemble the knowledge base accurately. Relaxing the strict cell-position requirement to table-level matching restores F1 scores to 13.36--64.16 and 21.84--33.42 for property and composition extraction respectively (see Supplementary Table~\ref{tab:qwen_lenient}). Even under this lenient criterion, the strongest Qwen variant on each task trails \matskraft{} by more than 25 F1 points.

Second, we benchmark three table-specialized architectures---TAPAS~\cite{herzig2020tapas}, TaBERT~\cite{yin2020tabert}, and TAPEX~\cite{liu2021tapex}---in base and MatSciBERT-adapted (MSB)~\cite{gupta2022matscibert} configurations (TaBERT additionally evaluated with generic bert-base-uncased initialisation~\cite{devlin-etal-2019-bert}; detailed in Supplementary Information~\ref{sec:appendix_table_baselines}). Domain adaptation improves performance consistently across all three architectures for both the extraction tasks, raising property-extraction F1 score by 5.5--34.6 points and composition extraction by 7.1--35.7 points. 
Composition evaluation is restricted to MCC tables, the only composition table-type that contains complete compositional information without requiring domain-specific regex parsing, together with NC (non-compositional) tables~\cite{hira2024reconstructing}. The strongest variant (TAPEX-MSB) trails \matskraft{} on both tasks---F1 score of 81.24 vs.\ 89.33 on property and 61.67 vs.\ 68.77 on MCC-NC composition extraction (see Supplementary Table~\ref{tab:table_model_comparison}). A closer look at per-property scores reveals gaps that aggregate metrics obscure (see Supplementary Figure~\ref{fig:d2}). Notably, these adapted encoders surpass every open-source LLM considered (see Supplementary Table~\ref{tab:qwen_strict} and~\ref{tab:qwen_lenient}), and several frontier proprietary LLMs across both extraction tasks (see Supplementary Table~\ref{tab:model_comparison}). Processing each table in 0.27--0.32 seconds on local hardware, these domain-adapted encoders constitute an efficient and underexplored baseline for large-scale literature mining.

\subsubsection{Linking the extracted entities}
After successfully developing high-accuracy extraction models, we focused on evaluating the extracted KB obtained by linking compositions and properties. 
For inter-table linking, we 
achieve F1 scores of 86.25 and 82.90 for composition and property tables respectively (that is, connecting the respective entities with the respective glass ID), enabling connections between entities in separate tables referring to identical materials. Given \matskraft's substantial superiority over baseline models in extraction accuracy, we focused KB construction exclusively on our framework rather than implementing linking capabilities for the baselines such as LLMs(detailed in Section~\ref{subsec: methods_linking_info}). This integration successfully formed over hundred thousands of composition-property pairs, with 73\% derived from intra-table linking and 27\% from inter-table associations. After linking and evaluating at the complete composition-property, we achieved an overall precision 79.04\%, recall 60.90\%, and F1 score of 68.97\%, which is particularly remarkable given the heterogeneity of reporting formats across the literature.

\subsubsection{Dissecting the Performance of \textsc{MatSKRAFT}}
To understand the relative importance of each of the components in \matskraft, rigorous ablation studies were performed by removing each component one-by-one systematically(detailed in Supplementary Information~\ref{sec: abl std}). Removing post-processing resulted in the largest performance degradation (9.38\% F1), confirming its critical role in achieving scientific-grade accuracy. Elimination of annotation codes also caused a substantial decline (9.02\% F1), while removal of data augmentation resulted in moderate degradation (1.18\% F1).

Notably, our fully automated pipeline without commercial database supervision---relying solely on annotation codes and data augmentation---achieved superior precision (93.04\%) compared to the full framework incorporating distant supervision from INTERGLAD (90.35\%), with controlled recall reduction (81.02\% vs. 87.07\%) (see Supplementary Information~\ref{subsec: appenix_data_ablation}). This demonstrates the high quality of our automated training data generation approach and its potential for eliminating dependence on expensive commercial databases, and enabling generalization to domains where such databases do not exist for distant supervision.

\subsubsection{Computational Efficiency}
Beyond accuracy advantages, \matskraft{} demonstrated substantial computational efficiency benefits that transform the practical feasibility of large-scale scientific knowledge extraction. Our specialized graph neural modules process extraction tasks 6–496$\times$ faster than LLM alternatives, with processing times of 0.22 seconds per table for property extraction and 0.39 seconds for composition extraction, compared to 1.38-187.59 seconds per table for LLM approaches (detailed in Supplementary Information~\ref{sec: appendix_llms_comp}).

This efficiency advantage, combined with modest hardware requirements (single consumer-grade GPU versus extensive cloud-based infrastructure required for LLMs), enabled us to scale the extraction pipeline to 66,267 tables across 11 journals in 11.23 hours on a single 32GB V100 GPU. The same scale of processing would require approximately 3 days using the fastest LLM baseline (GPT-5.4 at 3.81~s/table) and up to 8 months using the slowest (DeepSeek-R1 at 302.53~s/table), with associated API costs ranging over tens of thousands of dollars --- compared to \matskraft{}'s minimal computational expenses on standard hardware (see Supplementary Information~\ref{sec: appendix_llms_comp}).


\begin{figure}[htbp]
    \centering
    \includegraphics[width=0.9\textwidth, keepaspectratio]{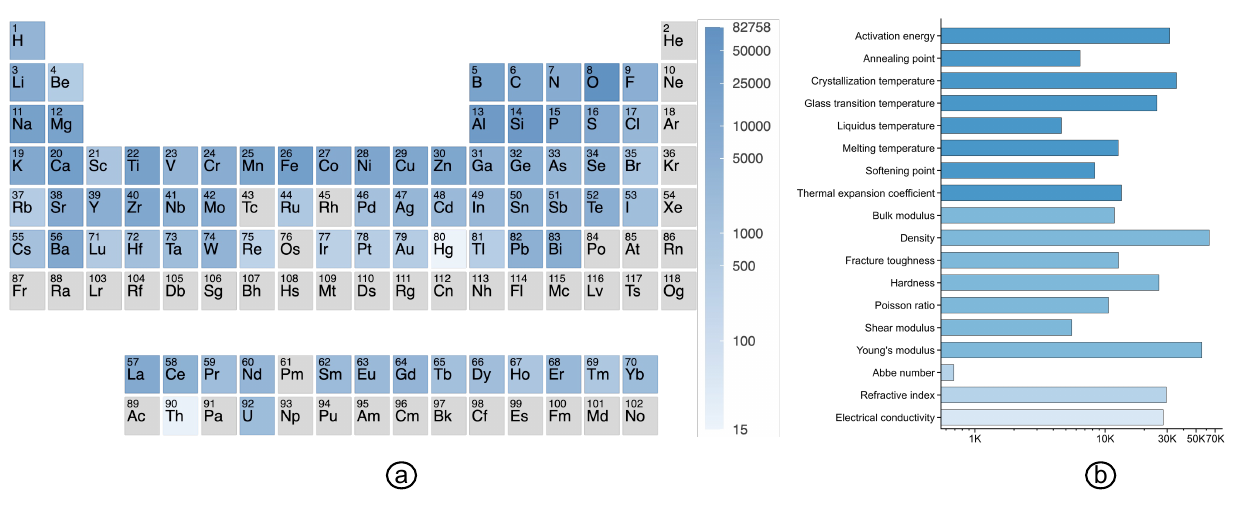}
    \caption{\textbf{Statistical analysis of extracted scientific knowledge.} (A) Periodic table visualization of elemental frequency in extracted compositions, with color intensity representing occurrence frequency. (B) Distribution of extracted properties showing coverage across 18 distinct material properties on logarithmic scale.}
    \label{fig:3}
\end{figure}

We applied \matskraft{} at scale to extract information from over 66,000 tables reported in research journals (detailed in Supplementary Information~\ref{subsec: appendix_database_coverage}). Statistical analysis of the resulting KB reveals extensive compositional diversity and scale. Figure~\ref{fig:3}a demonstrates comprehensive elemental coverage across the periodic table, with silicon, aluminum, and oxygen showing highest prevalence reflecting their fundamental role in silicate-based systems. However, our extraction also captures significant representation of technologically critical elements often underrepresented in conventional databases—including rare-earth elements (La, Ce, Nd), transition metals crucial for electronic applications (Nb, Ta, Mo), and semiconductor dopants (Ga, In, Sb). The substantial presence of lithium, sodium, and potassium indicates robust extraction of ionic conductors critical for energy storage technologies. Comparative analysis reveals that \matskraft{} contains more than 104,000 material compositions entirely absent from both INTERGLAD~\cite{interglad} and SciGlass~\cite{epam_sciglass}, representing a critical expansion of the known inorganic materials design space.

\begin{figure}[htbp]
  \centering
  \includegraphics[page=1, width=0.9\textwidth]{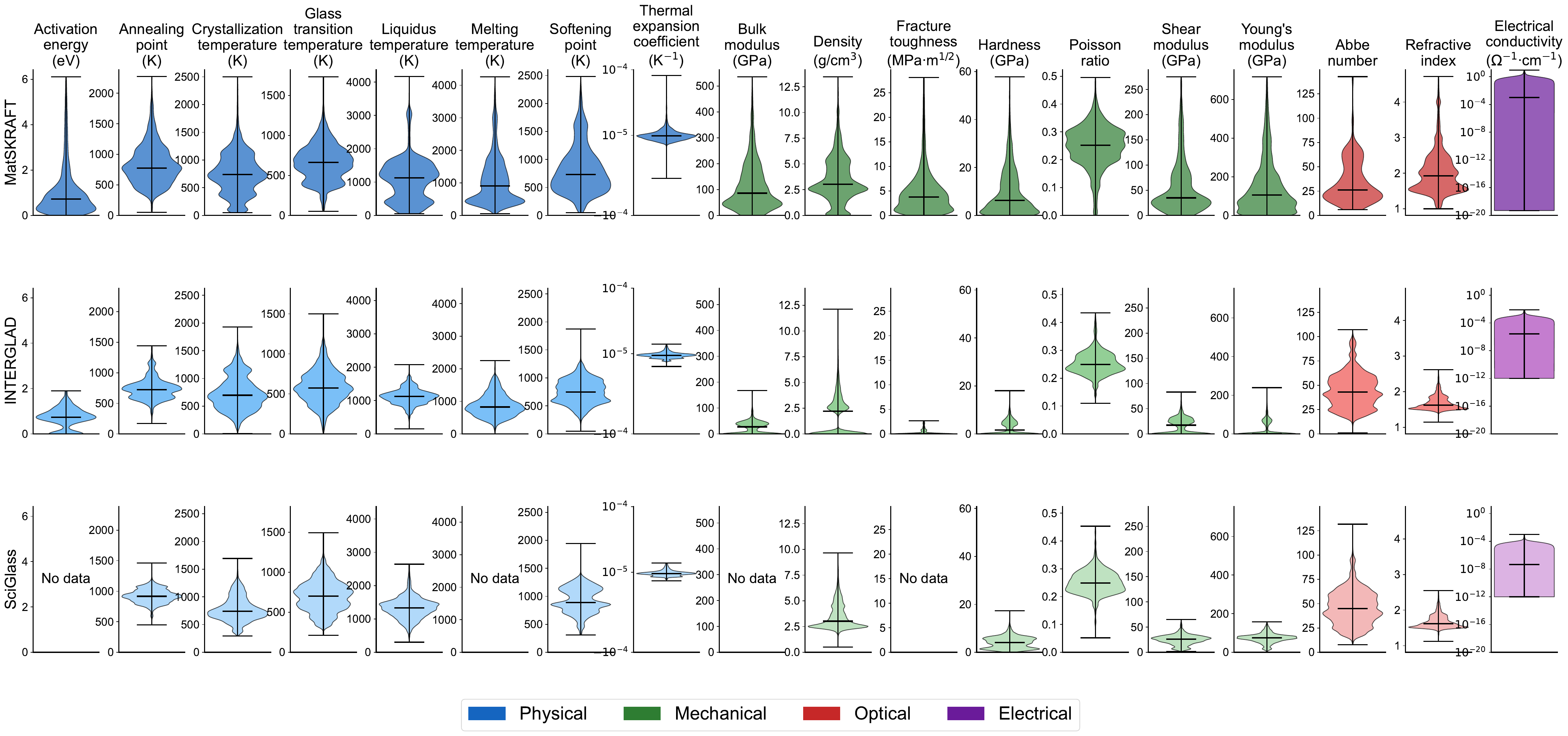}
  \caption{\textbf{Comparison of property value distributions across materials databases.} Violin plots comparing range and distribution of 18 key properties in \matskraft{} (top row), Interglad (middle row), and SciGlass (bottom row). \matskraft{} demonstrates substantially broader coverage across nearly all properties, indicating successful extraction of both conventional materials and those with specialized properties from scientific tables.}
  \label{fig:4}
\end{figure}

Property distribution analysis demonstrates \matskraft{'s} broad coverage across 18 distinct material properties spanning physical, mechanical, optical, and electrical domains (Figure~\ref{fig:3}b). The database contains extensive records for industrially important properties, including density (63,129 entries), Young's modulus (55,352 entries), crystallization temperature (35,414 entries), and activation energy (31,335 entries). When comparing property value distributions with established databases (Figure~\ref{fig:4}), \matskraft{} demonstrates substantially broader coverage—with electrical conductivity ranging from insulating to semi-conducting regimes, physical properties spanning low-temperature transitions through high-temperature melting and liquidus regimes, and mechanical properties covering ultra-soft to super-hard materials across an unprecedented range of compositions. Several properties exhibit distinctive multi-modal distributions absent in established databases, revealing our ability to capture diverse material families with distinctive property regimes. This comprehensive property coverage creates a resource for materials informatics towards accelerated materials discovery~\cite{olivetti2020data,kim2017materials, national2011materials}.

\subsection{Discovery-enabling Applications}

\begin{figure}[htbp]
    \centering
    \includegraphics[width=0.9\textwidth, keepaspectratio]{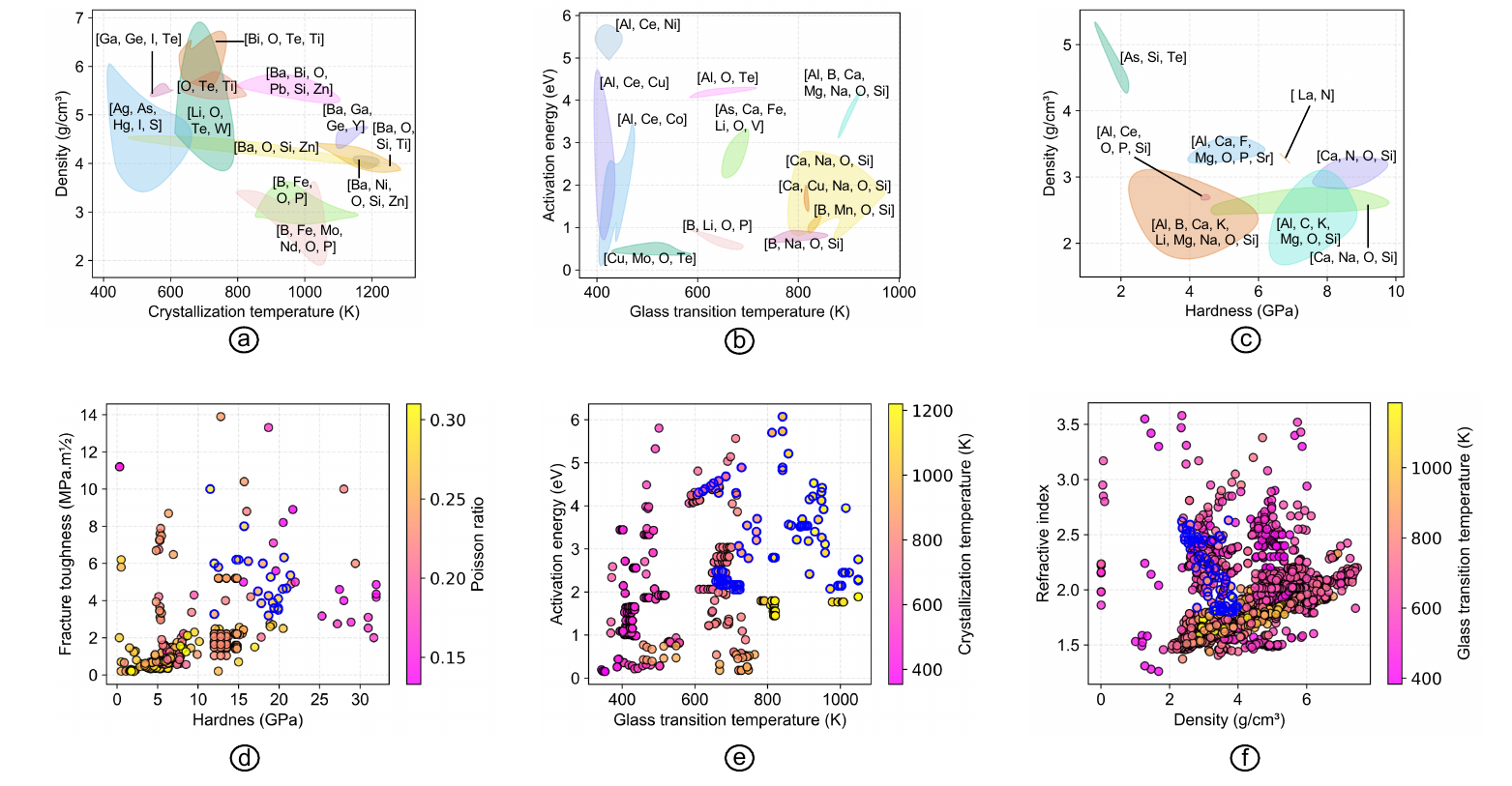}
    \caption{\textbf{Applications for materials discovery and screening.} (a--c): Systematic analysis reveals distinct material families achieving remarkable property combinations across thermal and mechanical property spaces through compositional interdependencies. (d--f): Multi-property screening identifies materials satisfying complex design criteria (represented by blue borderline) for technological applications.}
    \label{fig:5}
\end{figure}

The comprehensive KB enables exploration of composition-property relationships that can potentially accelerate materials selection. To this end, Ashby plots, also known as Materials selection charts (Figure~\ref{fig:5}), provide visual frameworks for identifying optimal compositions across multi-dimensional property spaces, while cross-compositional screening facilitates the discovery of rare materials with exceptional property combinations that traditional databases often overlook. The KB's extensive coverage of underrepresented compositions---including over 104,000 materials absent from existing databases---creates opportunities for identifying novel structure-property relationships and filling critical gaps in materials knowledge. Furthermore, the temporal analysis capabilities enable tracking of research trends and emerging focus areas, providing strategic intelligence for future research directions. These analyses reveal distinct property clusters corresponding to different compositional families, demonstrating how exceptional property combinations emerge through synergistic elemental interactions~\cite{varshneya2013fundamentals,bhattoo2023understanding, wang2018new}. Here, we discuss several case studies to demonstrate how the KB accelerates discovery for targeted applications across diverse technological domains.

Analysis of crystallization temperature-density relationships (Figure~\ref{fig:5}a) identifies [B, Fe, Mo, Nd, O, P] systems achieving low density (<3.5 g/cm$^3$) with controlled crystallization (800–1000 K) for applications such as nuclear waste immobilization as well as specialized radiation shielding, catalysis, and bioactive materials~\cite{wang2022effects}. The unique combination of low density and controlled crystallization allows these materials to be both lightweight and highly durable under specific heat conditions. This unique combination emerges through Mo incorporation as [MoO$_4$]$^{2-}$ species that enhance thermal stability while Nd promotes selective crystallization through NdPO$_4$ nucleation. Analysis of activation energy-glass transition temperature space (Figure~\ref{fig:5}b) identifies [Al, B, Ca, Mg, Na, O, Si] systems achieving high activation energies (>4 eV) and elevated glass transitions (>800 K) having exceptional thermal and chemical stability and are suitable for high-temperature processing applications~\cite{jindal2011synthesis}. Analysis of density-hardness relationships (Figure~\ref{fig:5}c) reveals material families achieving low density (<3.5 g/cm$^{3}$) with high hardness (>6 GPa), including enstatite-leucite glass-ceramics [Al, C, K, Mg, O, Si]~\cite{khater2021preparation}, alkali aluminosilicate systems [Ca, Na, O, Si]~\cite{limbach2014strain, de2018lateral}, and oxynitride glasses [Ca, N, O, Si]~\cite{sharafat2009hardness}.

Beyond pairwise relationships, our database enables identification of materials satisfying complex multi-property criteria addressing documented challenges in materials science.
\begin{itemize}
\item \textbf{Advanced Safety Materials:} 
Screening identified 29 materials (from 251 candidates) exhibiting high hardness ($\geq$10 GPa), high fracture toughness ($\geq$3.0 MPa$\cdot$m$^{1/2}$), and moderate Poisson's ratio ($\geq$0.25), addressing structural material challenges~\cite{mauro2014two}(Figure~\ref{fig:5}d). Examples include submicron $\beta$Si$_3$N$_4$~\cite{hou2020high} and Ba$^{2+}$-modified Mg–$\alpha\beta$–Sialon glass-ceramics~\cite{qin2019investigation}.


\item \textbf{High-Temperature Stable Materials:} Our analysis identified 84 materials with high glass transition temperature ($\geq$600 K), elevated crystallization onset ($\geq$ 800 K), and substantial activation energy ($\geq$ 2 eV) for thermal cycling applications (Figure~\ref{fig:5}e), including diopside-jadeite glasses~\cite{jindal2011synthesis} and Fe-Y-B metallic glasses~\cite{ma2015role}.

\item \textbf{Optical Materials for AR/VR:} We obtained 69 materials (from 1,310 candidates) with high refractive index ($\geq$ 1.8), low density ($\leq$ 4 g/cm$^3$), and thermal stability (Tg $\geq$ 750K) for lightweight optical systems (Figure~\ref{fig:5}f), including niobium-rich borophosphate glasses~\cite{el2022fiber} and NaNbGeO$_5$-based systems~\cite{chayapiwut2005synthesis}.
\end{itemize}

\begin{figure} [!t]
    \centering
    \includegraphics[width=0.9\textwidth, keepaspectratio]{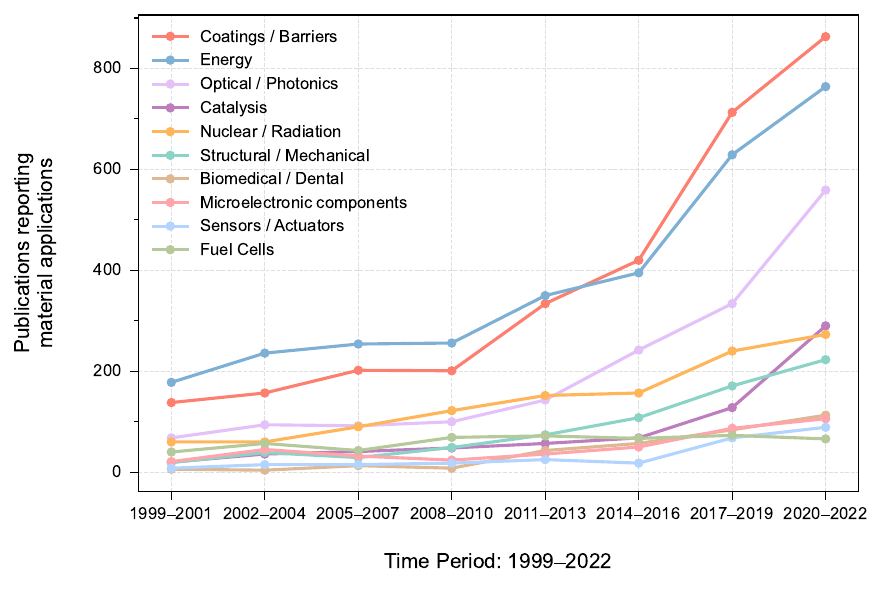}
    \caption{\textbf{Temporal analysis of materials research evolution (1999–2022).} Publication trends across application domains derived from analysis of abstracts, revealing strategic shifts in research focus with explosive growth in Energy, Coatings/Barriers, and Optical/Photonics after 2010.}
    \label{fig:6}
\end{figure}

Another interesting insight that the extracted KB provides is the evolution of research priority across materials science domains through temporal analysis. Figure~\ref{fig:6} visualizes publication trends across ten application domains from 1999-2022, derived through MatSciBERT~\cite{gupta2022matscibert} named entity recognition analysis of paper abstracts. This analysis reveals four distinct growth patterns: explosive growth in Coatings/Barriers, Energy, and Optical/Photonics domains coinciding with sustainability initiatives; significant growth in Catalysis, Nuclear/Radiation materials, and Structural/Mechanical systems after 2014; consistent gradual growth in Biomedical/Dental and Microelectronic components; and plateauing in Fuel Cells research. This temporal mapping provides strategic intelligence for research planning.

The KB enables identification of rare materials with transformative potential. Table~\ref{tab:rare_materials} presents materials exhibiting property combinations that defy conventional design constraints, such as high glass transition temperature with low thermal expansion (0.19\% occurrence rate)~\cite{rocherulle1998heat}. Our framework provides metadata linking each entry to its original research publication, enabling investigation of synthesis methodologies and composition-structure-property relationships, transforming traditionally serendipitous materials exploration into systematic, data-driven processes.

\begin{table}[htbp]
\small
\begin{tabular}{p{3.5cm}p{2cm}cccp{5.5cm}}
\toprule
\textbf{Description} & \makecell{\textbf{Property} \\ \textbf{Criteria}} & \textbf{Total} & \textbf{Rare} & \textbf{Rarity(\%)} & \textbf{Application Significance} \\
\midrule
High Glass Transition Temperature with Low Thermal Expansion & 
$T_g > 1073$ K and CTE $< 3 \times 10^{-6}$ K$^{-1}$ & 1352 & 3 & 0.22 & 
Three-dimensional photonic devices for integrated optics via femtosecond-laser-writing in silica glass~\cite{ehrt2004femtosecond} \\
\\
Low Electric Conductivity with High Melting Temperature & 
$\sigma < 10^{-10}$ $\Omega^{-1}$cm$^{-1}$ and $T_m > 2273$ K & 158 & 1 & 0.63 & 
Electrical insulation with extreme thermal resilience for substrates in power electronic modules~\cite{akhtar2017development}\\
\\
Low Softening Point with High Glass Transition Temperature & 
$T_{soft} < 1073$ K and $T_g > 973$ K & 1663 & 31 & 1.86 & 
Glass sealing for thermally compatible, crystallization-resistant SOFC applications~\cite{lara2004sintering}\\
\\
High Young's Modulus with Low Poisson's Ratio & 
$E > 150$ GPa and $\nu < 0.15$ & 6696 & 187 & 2.79 & 
Superhard materials exhibiting high stiffness with low compressibility~\cite{kou2019stability}\\
\\
High Activation Energy with High Glass Transition Temperature & 
$E_a > 2.0$ eV and $T_g > 973$ K & 871 & 29 & 3.33 & 
Thermally robust glasses for biomedical implants and bone tissue regeneration~\cite{ma2011synthesis} \\
\bottomrule
\end{tabular}
\caption{Identification of rare materials combinations with transformative potential, automatically identified by \matskraft{} from over 509,000 materials entries.}
\label{tab:rare_materials}
\end{table}

\matskraft{} addresses the open problem of information extraction from scientific tables by introducing a graph-based framework that transforms tables into structured knowledge representations, enabling automated database construction at scales previously unattainable. Processing over 45,500 research papers to extract 509,000+ entries, we demonstrate that KB development and database curation from literature could be automated avoiding prohibitively expensive manual efforts to tractable automated processes. The breadth and throughput capabilities demonstrated here establish new possibilities for knowledge synthesis across scientific domains. Analysis of the KB developed using \matskraft{} reveals patterns and relationships challenging to obtained through traditional literature review methods. Demonstrations of this fact include the identification of 104,000+ novel material compositions absent from existing databases and systematic discovery of rare property combinations with transformative technological potential. The proposed approach achieves computational speedups of several orders of magnitude over existing methods, making large-scale knowledge extraction significantly more accessible. This capability has the potential to democratize information extraction by enabling institutions with limited computational resources to perform analyses that were previously impractical.

\section{Discussion}
\label{sec: discussion}

\matskraft{} addresses the challenge of extracting structured information from scientific tables through a graph-based framework that converts tabular data into machine-readable knowledge representations. By processing over 45,500 research papers to extract 509,000+ entries, this work demonstrates how KB construction can transition from manual curation---traditionally requiring extensive human expertise and resources---to scalable automated processes. The resulting database reveals compositional and property relationships that would be difficult to identify through conventional literature analysis, including 104,000+ material compositions not present in existing databases and systematic identification of rare property combinations. The computational efficiency of the approach---processing times of 0.22--0.39 seconds per table compared to 1.38--187.59 seconds for large language models---makes comprehensive literature analysis accessible to researchers with modest computational resources.

Several technical limitations constrain the current framework. Properties with inconsistent reporting conventions across the literature, particularly thermal expansion coefficient and Poisson's ratio, present ongoing extraction challenges due to variational notation systems and unit representations. Complex table structures with non-standard layouts or multiple materials in unconventional orientations remain difficult to process automatically. A significant portion of the KB contains property measurements without corresponding compositional data, reflecting standard reporting practices where compositional information appears in article text rather than tables. The 62.97\% recall for composition extraction varies substantially across table types, with PI tables requiring contextual inference beyond tabular content proving particularly challenging. Additionally, specialized formats such as chemical reaction tables and incremental doping studies often require synthesis of information distributed across multiple document sections.

These limitations do not diminish the utility of the extracted data. Property measurements with complete source traceability support applications including multi-property materials screening for specific design requirements and identification of materials with unusual property combinations relevant to emerging technologies. Future development should prioritize text-based composition extraction to complement the table-focused approach, extension to synthesis and characterization methods, integration with predictive modeling frameworks, and adoption of standardized unit vocabularies such as QUDT~\cite{qudt_ontology} or UCUM~\cite{ucum_v2_2}. While the current implementation of \matskraft{} uses ScienceDirect~\cite{elsevier} XML files as input, the framework can potentially be extended to other publishers through PDF-to-XML conversion pipelines. However, the reliability of such approaches requires careful benchmarking because errors introduced during document parsing and conversion may propagate to downstream information extraction tasks. In addition, several important material properties, such as stress-strain curves and phase diagrams, are often reported as figures rather than text. Extracting such information would require the development of multimodal extraction models capable of jointly processing textual and visual content. Also, a hybrid approach could selectively employ large language models (LLMs) to annotate training data for properties where \matskraft{} exhibits lower extraction accuracy. However, contemporary frontier LLMs do not offer a significant extraction gain on these properties, and no single model consistently outperforms \matskraft{} by a significant margin across any subset of properties. We have therefore not pursued this strategy here; nevertheless, it remains a promising avenue for future exploration as next-generation LLMs continue to evolve. The availability of both the \matskraft{} framework and the resulting knowledge base provides a foundation for accelerated materials discovery across research communities, with potential applications in energy storage, electronics, and sustainable material development.


\section{Methods}
\label{sec: methods}
\subsection{Data acquisition and pre-processing}
To develop the training data for \matskraft{}, we extracted information from publications across several materials journals within the ScienceDirect database (see Supplementary Information~\ref{subsec: appendix_database_coverage}). The pipeline first retrieved full-text XMLs from research articles using Elsevier API\cite{elsevierapi}. From these XMLs, we extracted two components: the complete text of each paper, and all the respective tables with their captions. We used the MIT Table parser ~\cite{mit_table_parser} to process the raw XML tables and convert them to a two-dimensional list, with each list corresponding to an individual row. The extracted tables received a unique identifier in the form of PII\_table\_index from the Publisher Item Identifier (PII) and table number, ensuring traceability to the original publication and facilitating cross-reference during entity linking.

\subsection{Dataset construction}
\label{subsec: method_dataset_cons}
Deep learning models require high-quality training data that captures the full complexity of domain-specific knowledge representation ~\cite{olivetti2020data, song2023matsci, roh2019survey, huang2024data, song2022learning}. Tables in materials literature exhibit high structural heterogeneity, often reporting information without any standardized formatting conventions, thereby challenging automated information extraction~\cite{hira2024reconstructing}. Instead of resorting to resource intensive manual annotation, we constructed the training dataset through an automated hierarchical pipeline that integrates: (i) distant supervision from INTERGLAD to generate initial labels, (ii) deterministic knowledge encoding via property-specific annotation algorithm to expand and refine training coverage, and (iii) power-law guided data augmentation to correct long-tail imbalances(see Supplementary Figure~\ref{fig:hierarchical_data_preparation}). Material identifiers were manually annotated to enable precise entity linking during knowledge base construction, while development and test datasets were annotated manually to establish scientifically rigorous evaluation benchmarks (detailed in Supplementary Information~\ref{subsec:hierarchical_data_generation}).

\subsubsection{Distant Supervision}
First, we discuss the methodology employed for distant supervision. We started by leveraging commercial databases, in this case, International Glass Database (INTERGLAD)~\cite{interglad} as a structured source to initiate training data annotation through distant supervision. By aligning numeric entries in research tables with property and composition values from INTERGLAD, we developed a labeled data to build our training corpus.

The alignment procedure parses all cell-wise numeric entries and compares them against INTERGLAD entries via exact match followed by composition and separate property-specific tolerance thresholds. For dimensional properties such as density and glass transition temperature, we reconcile units through value transforms (e.g., $\times 1000$ and $\div 1000$ for density, $\pm$273 for temperature values), selecting a single consistent transformation per table. Matched values are tracked as candidate tuples (database index, observed value, inferred unit), and their distribution across rows and columns is used to identify valid property tables and infer orientation. Only tables with sufficient match density ($\geq 30\%$ along one axis) are retained. 

Initial validation revealed several limitations in this approach. Models trained solely on distantly supervised data achieved moderate validation scores, with particularly nominal precision on property extraction tasks. Detailed error analysis showed that numerous properties present in research tables remained undetected because they were absent from INTERGLAD. 
Thus, while distant supervision offers a viable initial approach, it fundamentally constrains the model's learning capacity to properties tracked in the reference database.




\subsubsection{Property-Specific Annotation Algorithms}
\label{method_anno_codes}


To overcome the limitations of distant supervision, we developed property-specific annotation algorithms that encode materials science expertise through multi-dimensional pattern recognition. These rule-based systems transcend conventional string matching by simultaneously analyzing unit patterns, value ranges, structural relationships, and semantic contexts—effectively replicating expert domain knowledge in automated form.

Our annotation framework includes multi-criteria verification protocols. For instance, consider the symbol ``$n$''---potentially representing either refractive index, Poisson's ratio, Avrami exponent, or various crystallographic parameters. Our solution deploys multi-criteria verification protocols: methodically examining the value ranges (refractive indices exceed 1.0, whereas Poisson ratios fall between -1.0 and 0.5), analyzing unit patterns (dimensionless vs. wavelength-dependent reporting), detecting wavelength references (632.8 nm, 546.1 nm) used for reporting refractive index measured at a particular wavelength, and cross-referencing caption semantics for discriminating terms, or direct reference to one of the properties in the caption, explained more in Supplementary Information~\ref{algo_for_annotation_code_prop}. Beyond simple disambiguation, the annotation algorithms accommodate diverse property categories, each presenting distinct complexity challenges---from thermophysical properties like glass transition temperature requiring recognition of 15+ notation variants ($T_g$, Tg, T-glass) to electrical conductivity extraction demanding sophisticated unit analysis patterns ($S/cm$, $\Omega^{-1}\text{m}^{-1}$) and caption context filtering to distinguish from dielectric properties. Most critically, this disambiguation resolves the challenge where identical symbols represent entirely different properties, achieving automation of domain expertise previously requiring manual expert validation. The performance of domain-specific algorithms are discussed in Supplementary Information~\ref{subsec: appendix_hierarch_performance_validation} and~\ref{subsec: appenix_data_ablation}.

\subsubsection{Data Augmentation for Representational Balance}
\label{method_data_aug}
Despite improvements using annotation codes, significant frequency imbalances persisted. 
To address this limitation, we developed a statistically guided augmentation strategy leveraging co-occurrence patterns in materials science tables. By identifying natural ``property neighborhoods''---for instance, Abbe value with refractive index, and fracture toughness with Young's modulus, we systematically augmented tables by integrating columns of underrepresented target properties into tables containing their frequently co-occurring ``neighbor'' properties. Crucially, we employed a power-law scaling function to determine the augmentation factor for each property, defined as $n_{\text{new}} = \lceil a \cdot n_{\text{original}}^{\alpha} \rceil$, where $a=10$ and $\alpha=0.65$. This amplifies rare properties while moderately increasing the common ones, creating a balanced distribution which imitates the real data (see Supplementary Information~\ref{algo_for_table_augmentation} and Supplementary Figure~\ref{fig:hierarchical_data_preparation}).

For generating synthetic values, we used Gaussian-based augmentation which preserves the natural statistical distribution of each property: $X_{\text{augmented}} \sim \mathcal{N}(\mu, \sigma)$, where $\mu$ and $\sigma$ represent the mean and standard deviation of observed values in the source column, constrained within $\pm 3\sigma$ to prevent implausible outliers. For column-oriented tables, we implemented a precise reconciliation process by adjusting the column length of the source table ($CL_{s}$) to match that of the augmented target table ($CL_{t}$). When $CL_{s} < CL_{t}$, we sampled additional values from the Gaussian distribution; conversely, when $CL_{s} > CL_{t}$, we clipped the source column accordingly. We implemented automatic failure detection with three safeguards: (1) abortion if no valid numerical values detected; (2) controlled noise ($\sigma = 0.05$) when $\sigma = 0$; and (3) discarding values exceeding $\pm 3\sigma$ ranges.

The power-law scaling approach substantially rebalanced representation across the property spectrum while maintaining original frequency hierarchy. 
For validation and test datasets, we maintained strict separation from the training data and manually annotated them, ensuring accurate evaluation. This augmentation strategy proved essential for developing extraction models that generalize effectively across both common and specialized materials properties without introducing artificial biases toward rare property classes (see Supplementary Figure~\ref{fig:hierarchical_data_preparation} for more details).

\subsection{Developing the models for extraction}
\label{subsec: model dev}
The extraction of materials knowledge at scale requires models that balance computational efficiency with high accuracy across diverse table structures. For \matskraft, we developed GNN-based architectures to process the structural complexities of materials science tables while maintaining production-level inference speeds and low computational costs. These models recognize compositional patterns, property relationships, and contextual cues with a level of sophistication approaching domain experts while remaining computationally feasible for processing thousands of research papers efficiently.

\subsubsection{Property Extraction Model}
\label{subsec: methods_prop_extraction}

Our GNN-based property extraction model represents each table as a structured graph graph $G = (V, E)$, where:
\begin{itemize}
    \item $V = V_\text{cell} \cup V_\text{row} \cup V_\text{col} \cup V_\text{caption}$, representing nodes for table cells, row headers, column headers, and captions

    \item Initial node embeddings $\mathbf{h}_v^{(0)} \in \mathbb{R}^d$ for $v \in V_\text{cell} \cup V_\text{caption}$ are obtained using MatSciBERT~\cite{gupta2022matscibert}:
    \begin{align}
        \mathbf{h}_v^{(0)} &= \text{MatSciBERT}(\text{content}(v))
    \end{align}
    
    \item Every row and column is represented by a dedicated header node $\mathbf{h}_v^{(0)}$ for $v \in V_\text{row} \cup V_\text{col}$, which are initialized randomly

    \item The edge set $E$ consists of three key connection types:
    \begin{align}
      E &= E_\text{cell-cell} \cup E_\text{cell-header} \cup E_\text{caption-header}\\
      E_\text{cell-cell} &= \{(u, v), (v, u) \mid u, v \in V_\text{cell}, \text{sameRow}(u, v) \lor \text{sameCol}(u, v)\} \quad \text{(bidirectional cell connections)}\\
      E_\text{cell-header} &= \{(v, h) \mid v \in V_\text{cell}, h \in V_\text{row} \cup V_\text{col}, \text{belongsTo}(v, h)\} \quad \text{(cell-to-header connections)}\\
      E_\text{caption-header} &= \{(c, h) \mid c \in V_\text{caption}, h \in V_\text{row} \cup V_\text{col}\} \quad \text{(caption-to-header information flow)}
    \end{align}
    
    These connections establish bidirectional information flow between cells, directional propagation from cells to headers, and contextual enrichment from captions to headers, enabling the model to interpret tables as integrated knowledge structures.
    
    \item Node representations are updated through a two-layer graph attention network:
    \begin{align}
    \mathbf{h}_v^{(l+1)} = \|_{k=1}^{K} \sigma\left(\sum_{u \in \mathcal{N}(v)} \alpha_{vu}^{k} \mathbf{W}^{(l,k)} \mathbf{h}_u^{(l)}\right)
    \end{align}
    where $K=4$ is the number of attention heads, $\|$ denotes concatenation, $\mathcal{N}(v)$ is the neighborhood of node $v$, and $\alpha_{vu}^{k}$ represents the attention coefficient computed as:
    \begin{align}
    \alpha_{vu}^{k} = \frac{\exp\left(\text{LeakyReLU}\left(\mathbf{a}^{(l,k)T}[\mathbf{W}^{(l,k)}\mathbf{h}_v^{(l)} \| \mathbf{W}^{(l,k)}\mathbf{h}_u^{(l)}]\right)\right)}{\sum_{w \in \mathcal{N}(v)} \exp\left(\text{LeakyReLU}\left(\mathbf{a}^{(l,k)T}[\mathbf{W}^{(l,k)}\mathbf{h}_v^{(l)} \| \mathbf{W}^{(l,k)}\mathbf{h}_w^{(l)}]\right)\right)}
    \end{align}
    with hidden dimensions of 2048 and 1024 for the first and second layers respectively
\end{itemize}

This graph architecture captures both the local relationships between cells and the global context provided by headers and captions, which is crucial for accurate property identification in complex tabular layouts. Essentially, our framework remains flexible and orientation independent, enabling our model to seamlessly process both row-oriented and column-oriented tables through the same computational framework — a significant advantage while extracting information from materials science literature where table structures vary substantially across different journals and research groups. We also incorporate positional embeddings to distinguish between different functional zones of the table, improving its ability to correctly interpret property-related information based on structural position. The node representations are further refined through dropout (0.2) between GAT layers to prevent overfitting, ensuring robust generalization across diverse table formats.




To encode domain knowledge and structural patterns directly into our model, we implemented a constraint-driven learning framework that guides the neural network to follow the inherent logic of materials science data representation. This approach enforces four key constraints that reflect the organizational principles of scientific tables:

\begin{enumerate}
    \item \textbf{Material-Property Association}: Material identifiers and property labels should occupy the same structural position (both as row headers or both as column headers) to ensure consistent table organization. This constraint penalizes mixed configurations where materials and properties are in complementary positions:
    \begin{equation}
    \mathcal{L}_{\text{gid-prop}} = \mathbb{E}_{(i,j) \in \text{CrossPairs}} \left[ p_{\text{gid}}(i) + p_{\text{prop}}(j) - 1 + p_{\text{prop}}(i) + p_{\text{gid}}(j) - 1 \right]
    \end{equation}
    where CrossPairs represents row-column pairs, discouraging scenarios where material identifiers are in rows while properties are in columns, or vice versa.
    
    \item \textbf{Material Identifier Exclusivity}: Only one header should serve as the material identifier within a table to avoid ambiguous data interpretation. This constraint discourages multiple headers from being simultaneously classified as material identifiers:
    \begin{equation}
    \mathcal{L}_{\text{gid-gid}} = \mathbb{E}_{(i,j) \in \text{AllPairs}} \left[ p_{\text{gid}}(i) + p_{\text{gid}}(j) - 1 \right]
    \end{equation}
    
    \item \textbf{Property Exclusivity}: Properties should maintain structural exclusivity between row and column headers to preserve logical table organization. This prevents both row and column headers from being simultaneously classified as properties:
    \begin{equation}
    \mathcal{L}_{\text{prop-prop}} = \mathbb{E}_{(i,j) \in \text{CrossPairs}} \left[ p_{\text{prop}}(i) + p_{\text{prop}}(j) - 1 \right]
    \end{equation}
    
    \item \textbf{Material Identifier Uniqueness}: Material identifiers typically contain unique values for each material (e.g., ``Sample-1'', ``Sample-2'' rather than repeated generic labels). This constraint favors headers with higher uniqueness ratios:
    \begin{equation}
    \mathcal{L}_{\text{gid-id}} = \mathbb{E}_{i: p_{\text{gid}}(i) > 0.25} \left[ 0.5 - \frac{|\mathcal{U}_i|}{|\mathcal{C}_i|} \right]
    \end{equation}
    where $|\mathcal{U}_i|/|\mathcal{C}_i|$ represents the ratio of unique values to total values in header $i$.
\end{enumerate}

The total constraint loss combines all components through ReLU activation:
\begin{equation}
\mathcal{L}_{\text{constraint}} = \frac{1}{N}\sum_{i}\text{ReLU}\left(\mathcal{L}_{\text{gid-prop}} \cup \mathcal{L}_{\text{gid-gid}} \cup \mathcal{L}_{\text{prop-prop}} \cup \mathcal{L}_{\text{gid-id}}\right)_i
\end{equation}
\vspace{-0.1in}
\begin{equation}
\mathcal{L}_{\text{total}} = \mathcal{L}_{\text{CE}} + \lambda \cdot \mathcal{L}_{\text{constraint}}
\end{equation}

where $\mathcal{L}_{\text{CE}}$ is the standard cross-entropy loss, and $\lambda=50$ is a hyperparameter controlling the contribution of constraint-based regularization. The ReLU function ensures that only positive violations contribute to the constraint loss, effectively creating soft rules that guide the optimization process toward scientifically coherent table interpretations. This constraint-driven approach represents a fundamental reimagining of how structural knowledge can be encoded within neural architectures, pushing our model to respect the inherent tabular structures.

While training constraints focus on tabular organization, our post-processing pipeline encodes materials science knowledge through validation mechanisms that eliminate false positives across all 18 target properties. The system implements multi-layered validation: physicochemical constraint checking (rejecting negative densities, sub-zero temperatures, impossible Poisson ratios), unit analysis distinguishing legitimate measurements from misclassified parameters, and contextual semantic validation preventing compositional percentages being extracted as properties. For complex cases like thermal expansion coefficients, the framework automatically reconstructs fragmented scientific notation where base values (2.5) appear in cells while exponential modifiers ($\times 10^{-x}$) reside in headers. Property-specific disambiguation protocols resolve ambiguous symbols through multi-criteria analysis of value ranges, units, and caption semantics. This validation architecture achieves high precision by eliminating false positive contamination that typically compromises automated extraction, delivering scientifically reliable databases suitable for materials discovery (Algorithm in Supplementary Information~\ref{algo_post_processing}).

We assess the calibration of GNN-1 confidence scores (Supplementary Information~\ref{sec:appendix_calibration}). The raw confidence scores carry sufficient discriminative 
signal, thereby motivating confidence threshold based filtering (see Supplementary Figure~\ref{fig:c2}). We leverage this through per-property confidence thresholding (detailed in Supplementary Information~\ref{app:confidence_thresholding}).

\subsubsection{Composition Extraction Model}
\label{subsec: methods_comp_extraction}
For extracting material compositions from tables, we employed the DiSCoMaT framework~\cite{gupta-etal-2023-discomat} with several enhancements to improve both accuracy and computational efficiency. Our modifications include re-annotation of training data using refined rule-based constituent detectors (see Supplementary Information~\ref{algo_composition_annotation}), expanded unit detection to atomic percentage (at\%) alongside weight percentage (wt\%) and molar percentage (mol\%), and extending unit detection to ``Results'' and ``Discussion'' sections of the article, addressing cases where unit information is provided in explanatory text rather than in tables. Also, the chemical compound lexicons was expanded for specialized materials. 
Our optimizations increased the processing speed by up to 5× while maintaining same computational efficiency, enabling the system to scale with modest hardware requirements.

Our property extraction model employs a GNN with two hidden layers of dimensions 2048 and 1024 respectively, providing sufficient representation capacity to distinguish between 18 diverse material properties spanning physical, mechanical, optical, and electronic domains — each with distinct units, scales, and physical significance. In contrast, the composition extraction architecture utilizes a hierarchical approach with two sequential GNNs, each comprising three hidden layers with progressively narrower dimensions (256→128→64 and 128→128→64). This architectural difference reflects the distinct nature of the two tasks: property extraction requires broader representations to capture diverse property characteristics and deep domain understanding to correctly classify properties with similar notational representations, while composition extraction benefits from a hierarchical decision-making approach to first classify table types (Single Cell vs. Multi-Cell composition) and then engage specialized extractors for each category separately.

\subsubsection{Unit extraction for composition and properties}
\label{subsec: unit_extraction_method}
Accurate unit extraction is essential in materials science for converting extracted values into physically meaningful records. 
To address this, we developed a modular, rule-based framework for unit extraction that favors interpretability and scientific fidelity over predictive modeling. 

Specifically, our framework implements specialized pattern recognition combining dictionary-based matching with property-specific regular expressions and validation protocols. For electrical conductivity, the system recognizes and canonicalizes over 15 unit variants (e.g., S/cm, S cm$^{-1}$, $\Omega^{-1}$ cm$^{-1}$, (O-cm)$^{-1}$) while distinguishing ionic from electronic conductivity through caption analysis and value range validation. The system implements progressive fallback strategies when initial unit detection fails—searching neighboring cells, then table captions, and finally methodology sections for property-specific unit keywords.

The framework further supports intelligent corrections based on physical reasoning. For example, if resistivity units like $\Omega\cdot$cm are detected in a column predicted as electrical conductivity, the system applies reciprocal transformation and verifies whether the resulting values lie within valid conductivity ranges. Likewise, when temperature-like units are found in mechanical property columns, the pipeline reassigns the predicted property as false positive to improve precision. This context-sensitive validation enables creates a self-correcting mechanism where unit analysis enhances the property classification accuracy.

Property-specific disambiguation is another strength of the system. For hardness values, the pipeline resolves a wide variety of scale notations---e.g., Vickers (HV), Knoop (HK), Rockwell (HR), and Shore (ShA, ShD). Unlike predictive models limited to predefined unit categories, our rule-based framework can identify previously unseen but valid unit conventions, enabling extensibility through minimal dictionary updates rather than costly retraining. This rule-based design enables ongoing extension of scientific coverage through expert-interpretable logic, allowing researchers to trace unit assignments to specific patterns and verify each processing stage---from pattern matching to validation---ensuring transparent knowledge extraction rather than relying on black-box neural network predictions. The framework assigns blank strings when extraction fails, maintaining source traceability without hallucination (see Supplementary Information~\ref{algo_unit_extraction} for algorithms and Supplementary Table~\ref{tab:unit_accuracy_grouped} for unit extraction accuracy).

\subsection{Knowledge Base Integration Framework}
\label{subsec: methods_linking_info}

The development of comprehensive materials knowledge repositories requires not only accurate extraction of individual compositions and properties but also their meaningful integration into a unified knowledge structure.  This integration operates through two complementary processes: intra-table integration based on structural orientation and inter-table linking through material identifiers, each addressing different patterns of knowledge distribution within scientific publications.

\subsubsection{Structural Orientation-Based Integration}
For compositions and properties within the same table, \matskraft{} employs a structural orientation approach leveraging inherent organizational patterns of scientific tables. Each extracted entity receives a unique identifier encoding source information—PII\_TID\_R\_C\_ID for properties and PII\_TID\_R\_C\_0\_ID for compositions (where PII is the article identifier, TID the table number, and R,C the row/column indices). By analyzing positional indices, the system determines the table orientation (row- or column-oriented), then links entities accordingly: in row-oriented tables, compositions and properties sharing column indices are linked; while in column-oriented tables, entities with matching row indices are associated. This approach successfully integrated composition-property pairs from within tables, representing 73\% of all paired entries.

\subsubsection{Cross-Table Entity Linking}
Materials science publications frequently distribute related information across multiple tables, necessitating sophisticated cross-reference resolution mechanisms. \matskraft{} addresses this challenge through material identifier prediction components integrated into both the composition and property extraction models, achieving F1 scores of 86.25 for composition tables and 82.90 for property tables. These components recognize various identifier formats (numeric designations, compositional abbreviations, standard glass names) and incorporate them into the unique identifier structure (PII\_TID\_R\_C\_ID), establishing connections between compositions and properties in separate tables referring to the same material. This cross-table linking contributed 27\% of entries, transforming isolated extraction results into a unified knowledge structure. The substantial volume of inter-table associations underscores the fragmented nature of materials knowledge in scientific literature and highlights the essential role of identifier-based integration. Together, these integration mechanisms created a unified knowledge structure with linked compositions and properties, establishing an unprecedented resource for materials informatics research.

\subsection{Baselines}
To contextualize \matskraft's performance, we conducted evaluation against leading LLMs: Gemini 1.5 Pro, Claude 3.5 Sonnet, DeepSeek-R1, and GPT-4o. Our evaluation methodology provided these baseline systems with substantial advantages to ensure fair comparison through meticulously engineered prompts developed via iterative optimization across multiple refinement cycles. These prompts incorporated detailed instructions, normalization guidelines, few-shot demonstrations with diverse table structures, comprehensive property lists with standardized nomenclature, explicit unit normalization instructions for all properties, and detailed identifier construction guidelines. We further provided substantial contextual information including table captions, article abstracts, and reference examples to simulate domain knowledge, while allocating generous token limits (8,192 tokens) to process large tables without truncation—representing current best practices for LLM-based information extraction(see Supplementary Information~\ref{sec: appendix_llm_fairness} for more details).

All models were evaluated on identical datasets comprising 737 tables for composition extraction and 368 tables for property extraction, with standardized scoring protocols to ensure comparability(see Supplementary Information~\ref{subsec: appendix_dataset_split}). To minimize potential variability, we set temperature parameters to zero for deterministic outputs and implemented robust error handling with automatic retries for rate-limit exceptions (see Supplementary Information~\ref{sec: appendix_llm_fairness}). This rigorous evaluation framework allowed us to assess \matskraft's specialized GNN architecture against generalist LLMs across both extraction accuracy and computational efficiency metrics, providing insights into the comparative advantages of architecture specialization in scientific knowledge extraction tasks (see Supplementary Information~\ref{sec: appendix_llms_comp}).

\subsection{Evaluation Metrics}
\begin{itemize}
    \item \textbf{Training Metrics:} Models were optimized using combined loss:
    \begin{equation}
        \mathcal{L}_{\text{total}} = \mathcal{L}_{\text{CE}} + \lambda \cdot \mathcal{L}_{\text{constraint}}
    \end{equation}
    where $\mathcal{L}_{\text{CE}}$ is cross-entropy loss and $\mathcal{L}_{\text{constraint}}$ encodes structural constraints, and $\lambda$ is a tunable hyper-parameter (see Section~\ref{subsec: methods_prop_extraction} for details).
    
    \item \textbf{Inference Metrics:} We evaluated models using standard precision, recall, and F1 scores:
    \begin{align}
        \text{Precision} &= \frac{|\text{Correct Extractions}|}{|\text{Total Extractions}|}, \quad \text{Recall} = \frac{|\text{Correct Extractions}|}{|\text{Ground Truth Entities}|} \\
        \text{F1} &= 2 \cdot \frac{\text{Precision} \cdot \text{Recall}}{\text{Precision} + \text{Recall}}
    \end{align}
    Metrics were calculated at entity level, requiring correct identification of materials, properties, values, and units against expert-annotated ground truth. We employed strict evaluation criteria where extractions were deemed correct only upon complete accuracy of all components (entity name, value, unit). This conservative methodology ensures high-confidence data suitable for scientific databases, though it may underestimate practical utility.
    
    \item \textbf{Unit Extraction Assessment:} Unit extraction accuracy was calculated conditionally on successful entity extraction:
    \begin{equation}
        \text{Unit Accuracy} = \frac{|\text{Correctly Extracted Units}|}{|\text{Total Entities with Extracted Units}|}
    \end{equation}
    This metric evaluates unit extraction performance conditionally on successful entity extraction, rather than on the total number of ground truth entities. For example, in a dataset containing 100 property entities requiring units, if \matskraft{} extracts 80 property entities and correctly identifies units for 65 of them, the Unit Accuracy would be $\frac{65}{80} = 81.25\%$. By conditioning on successful entity extraction, we isolate unit extraction capability from broader entity detection, providing clearer assessment of this specialized subtask's performance. This conditional evaluation is particularly appropriate for materials science, where unit identification accuracy represents a distinct technical challenge from detecting the associated property entities themselves.
\end{itemize}


\section{Declaration Statements}

\subsection{Data Availability}
Three versions of the extracted dataset are released on Zenodo \url{https://zenodo.org/records/20684754}: (i) the raw MatSKRAFT extraction output (509,281 entries); (ii) a property-normalised variant with consistent target units applied across all entries (502,317 entries); and (iii) a fully harmonised variant in which properties are unit-normalised and compositions are converted to mol\% and scaled to a 0--100 basis (501,777 entries).

\subsection{Code Availability}
The codes used in this work are shared on the \matskraft{} GitHub repository \url{https://github.com/M3RG-IITD/MatSKRAFT} and Zenodo \url{https://doi.org/10.5281/zenodo.20684916}.

\subsection{Acknowledgement}

N.M.A.K. acknowledges the funding support received from BRNS YSRA (53/20/01/2021-BRNS), ISRO RESPOND as part of the STC at IIT Delhi, Google Research Scholar Award, Intel Labs, and Alexander von Humboldt Foundation. M.M. acknowledges grants by Google, IBM, Microsoft, Wipro, and a Jai Gupta Chair Fellowship. M.Z. acknowledges the funding received from the PMRF award by the Ministry of Education, Government of India. 

The authors thank the High-Performance Computing (HPC) facility at IIT Delhi for computational and storage resources. We thank Tanishq Gupta (Tower Research Capital) and Devanshi Khatsuriya (Google) for their crucial contribution to the project during their research at IIT Delhi.

\subsection{Author Contributions}
M.M. and N.M.A.K. conceived and supervised the work. K.H. wrote the code and performed all the experiments. M.Z. assisted K.H. in verification of the results, manual annotations, and figures. All the authors contributed to the analysis of the results. K.H. wrote the main manuscript with input from all the authors. All the authors read, edited, and approved the manuscript.

\subsection{Competing Interests}
The authors declare no competing financial or non-financial interests.

\bibliographystyle{unsrt}
\bibliography{sample}












\clearpage

\appendix
\begin{center}
    {\textbf{\huge Supplementary Materials}}
\end{center}
\setcounter{page}{1}

\startcontents[appendices]
\section*{Contents}
\printcontents[appendices]{}{1}{\setcounter{tocdepth}{2}}

\newpage


\section{Hierarchical Training Data Generation and Performance Analysis}
\label{subsec:hierarchical_data_generation}

\subsection{Dataset Preparation and Annotation Strategy}
\label{subsec: appendix_dataset_split}

The development of robust extraction models for scientific domains require high-quality labeled training data that captures the full complexity of domain-specific knowledge representation. In materials science literature, this challenge is compounded by extreme heterogeneity in notation systems, inconsistent reporting conventions, and characteristic long-tail distributions where technologically critical but specialized properties have minimal representation compared to commonly studied properties. Our hierarchical data preparation strategy directly addresses these constraints through a three-component approach designed to systematically overcome the limitations inherent in each individual methodology while preserving scientific validity essential for materials informatics applications.

For property extraction, we constructed datasets with balanced desired property-containing (Prop) and non-property-containing (Non-Prop) table distributions across training, validation, and test splits (Supplementary Table~\ref{tab:prop_data_split}). For composition extraction, we applied targeted annotation algorithms to reclassify misidentified tables, improving the training distribution from the original structure-aware dataset categorization (SCC, MCC, PI, NC) as shown in Supplementary Table~\ref{tab:comp_data_split}. While training datasets were generated through automated pipelines, validation and test sets were manually annotated by experts to ensure rigorous evaluation benchmarks.

\renewcommand{\tablename}{Supplementary Table} 
\renewcommand{\thetable}{\arabic{table}}       
\setcounter{table}{0}                          

\begin{table}[h!]
\centering
\captionsetup{justification=raggedright,singlelinecheck=false}
\small
\begin{tabular}{lccc}
\toprule
Split & Prop & Non-Prop & Total \\
\midrule
Train & 1021 & 988 & 2009 \\
Validation & 223 & 193 & 416 \\
Test & 194 & 174 & 368 \\
\bottomrule
\end{tabular}
\vspace{0.1in}
\caption{Data distribution of tables for property extraction across training, validation, and test splits.}
\label{tab:prop_data_split}
\end{table}

\begin{table}[h!]
\centering
\captionsetup{justification=raggedright,singlelinecheck=false}
\small
\begin{tabular}{lccccc}
\toprule
Split & SCC & MCC & PI & NC & Total \\
\midrule
Train & 704 & 658 & 439 & 2607 & 4408 \\
Validation & 110 & 132 & 109 & 387 & 738 \\
Test & 114 & 132 & 111 & 380 & 737 \\
\bottomrule
\end{tabular}
\vspace{0.1in}
\caption{Data distribution of tables for composition extraction across training, validation, and test splits.}
\label{tab:comp_data_split}
\end{table}

\subsection{Performance Validation and Component Analysis}
\label{subsec: appendix_hierarch_performance_validation}

Our hierarchical approach achieves systematic performance gains with F1 scores progressing from 82.77\% (distant supervision) to 88.18\% (with annotation algorithms) to 88.88\% (complete framework) on the development (dev/val) dataset, demonstrating that each component addresses different extraction challenges. Distant supervision from INTERGLAD provides broad coverage but exhibits systematic bias toward thermal properties, achieving exceptional performance for glass transition temperature (F1=96.68\%) while showing characteristic limitations for specialized properties like electrical conductivity (F1=52.36\%) and thermal expansion coefficient (F1=47.67\%). This foundation captures established materials knowledge but fails to represent the full diversity of property-structure relationships in contemporary research.

\renewcommand{\figurename}{Supplementary Figure} 
\renewcommand{\thefigure}{\arabic{figure}}       
\setcounter{figure}{0}                          


Annotation algorithms address these limitations through sophisticated rule-based pattern recognition that encodes materials science expertise — distinguishing ambiguous notations, reconstructing fragmented scientific representations, and validating extracted properties against physically plausible ranges. These domain-specific algorithms expand the training data coverage while concurrently maintaining high precision. Specifically, our methodology identified 185 additional density instances (rows or columns), 245 glass transition temperatures, and achieved substantial enhancement of previously underrepresented properties, such as melting temperature and activation energy, having none to 118 and 119 instances respectively. The synergistic integration of distant supervision with property-specific annotation codes amplified total identified properties from 805 to 2162 instances---a 168.57\% increase in the training data, illustrated in Supplementary Figure~\ref{fig:hierarchical_data_preparation}(a).The annotation modules demonstrate outstanding precision enhancement capabilities, with their removal causing the most severe precision collapse from 90.35\% to 72.34\% (18-point degradation)—the largest single-component impact on precision, as shown in the ablation studies performed on the test dataset (see Supplementary Table~\ref{tab:ablation}). Furthermore, training the model on data generated using annotation codes only achieves exceptional performance(F1 = 86.20, precision = 91.17\%), confirming their ability to generate exceptionally accurate training examples independently for the 18 properties with minimal false positive contamination (see Supplementary Information~\ref{subsec: appenix_data_ablation}). Diving into more details during development, we observe the impact of this strategy in properties requiring specialized domain knowledge: annealing point detection advances from F1=73.91\% to 86.67\%, and mechanical properties like hardness achieve near-perfect performance (F1=96.90\%), demonstrating how expert knowledge encoding enables accurate interpretation of complex materials science semantics. The annotation algorithms provide precision enhancements across 17 out of 18 properties, with notable improvements in many properties such as liquids temperature (+21.80 precision points), while maintaining high precision for well-represented properties also like glass transition temperature and density. Therefore, our methodology demonstrates how specialized scientific domains approach the high-quality training data bottleneck that constrains deep learning model development across countless applications, establishing annotation algorithms as a novel automated method for achieving expert-level training data generation without human supervision or commercial database dependencies.

Data augmentation completes the hierarchical strategy by addressing long-tail distribution problems through scientifically-grounded enhancement that preserves natural co-occurrence patterns observed in materials characterization studies. The power-law scaling function strategically amplifies underrepresented properties while maintaining natural frequency hierarchies. Optical properties exhibit pronounced improvements, with refractive index advancing from F1=73.81\% to 85.28\%, while Abbe number recovers from annotation code degradation (F1=38.71\%) to achieve perfect performance (F1=100.0\%). Thermophysical properties demonstrate consistent enhancement: liquidus temperature improves from F1=76.77\% to 95.73\%, softening point advances from F1=86.06\% to 91.03\%, and annealing point reaches F1=92.06\%.

Systematic validation reveals that optimal performance for different properties emerges from distinct components, scientifically justifying our complete hierarchical framework. Mechanical properties demonstrate this principle: while shear modulus and bulk modulus achieve perfect performance (F1=100.0\%) through annotation codes alone, Young's modulus requires the complete framework to reach F1=94.85\%, and fracture toughness benefits incrementally across all stages (F1=66.67\% → 70.00\% → 72.41\%). This heterogeneous optimization profile reflects diverse challenges in materials property extraction—some properties benefit primarily from expert knowledge encoding, others require coverage enhancement, while  thermophysical properties like glass transition temperature achieve peak performance through distant supervision alone. The complete framework achieves optimal balance with precision of 89.97\%, recall of 88.44\%, and F1 score of 88.88\% on the dev set, demonstrating that the synergistic combination of all three components is essential for reliable scientific knowledge extraction. This represents a significant improvement in the model's capability to recognize and interpret a broad range of materials property semantics, establishing a systematic methodology for addressing training data limitations in specialized scientific domains.

\subsection{Algorithms for Property-Specific Annotation}
\label{algo_for_annotation_code_prop}

In order to annotate material science tables, we developed property-specific expert-verified handcrafted extraction rules. This hybrid annotation framework combines symbolic matching, semantic priors, and contextual cues derived from numerical values, physical units, and surrounding text. The main annotation loop (algorithm described in Supplementary Information~\ref{algo: 1}) operates over each unannotated header $h$ present in the table and checks whether it belongs to one of the target properties $p \in \mathcal{P}$. It first checks for exact phrase matches with canonical property names using a high-precision rule-based matcher. If a match is not found, it invokes a more flexible signal-fusion strategy based on domain knowledge, wherein $h$ is compared against a list of known symbolic aliases for each $p$, and passed through the property-validation function \textsc{ValidateProperty}. 

The \textsc{ValidateProperty} function (algorithm described in Supplementary Information~\ref{algo: 2nd}) serves as the decision module within the fallback stage of the annotation process. Rather than relying on a single strong cue, it fuses multiple domain-informed checks—such as unit conformity, numeric plausibility, and contextual signals from the caption and header text.  To accommodate variation across properties, the function dynamically invokes specialized logic for each property, dispatching control to tailored exception checks that encode domain-specific rules. For instance, the criteria for validating activation energy vary considerably from those for thermal expansion or electric conductivity, reflecting variations in expected units, value ranges, and caption phrases. These signals are aggregated into a cumulative score, and a property-specific threshold $t_p$ is used to make the final decision: the header is accepted if its score exceeds $t_p$. This threshold is tunable and calibrated based on signal reliability across properties, enabling automated filtering of noisy candidates without the need for property-specific expert intervention during annotation.  Such tightly scoped symbolic disambiguation is implemented across all 18 target properties to ensure domain-consistent annotations with very low chances of false positives.

This targeted annotation strategy yielded a dramatic improvement in training data volume and quality. By combining distant supervision with these property-specific rules, the framework significantly expanded the number of usable training instances while maintaining high precision. Overall, the number of annotated properties in the training dataset more than doubled (over 168\%), detailed in Section~\ref{method_anno_codes}. Ablation studies (see Supplementary Table~\ref{tab:ablation}) further confirmed the impact of annotation codes, showing that their removal led to the most severe drop in precision among all model components. Even when used in isolation, data generated by this annotation framework yielded near-SOTA performance, underscoring its ability to produce high-quality training data autonomously—without manual effort or reliance on external databases.

\begin{figure}[htbp]
    \centering
    \includegraphics[width=0.9\textwidth, keepaspectratio]{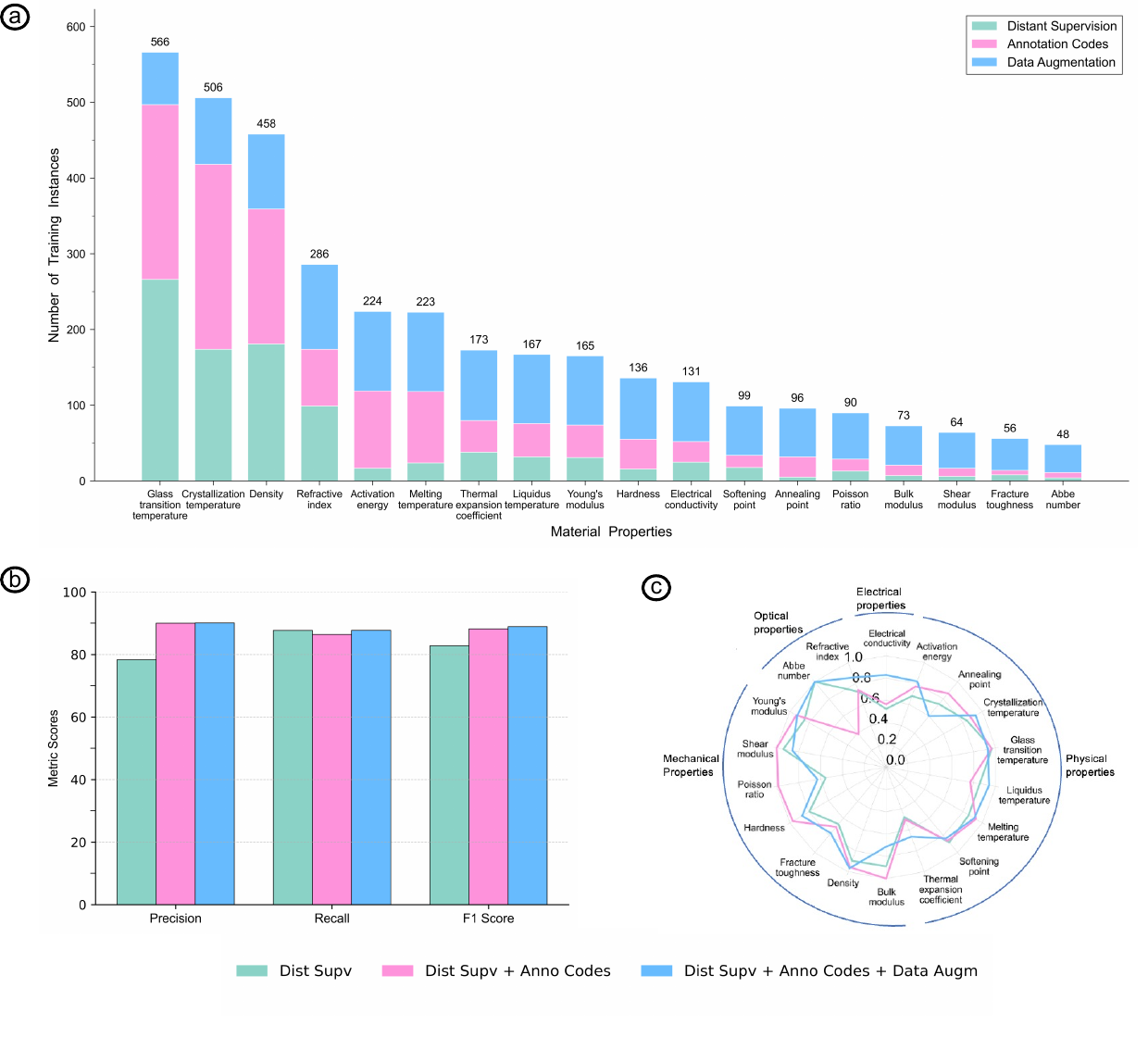}
    \caption{\textbf{Hierarchical data preparation strategy and validation performance.} (a) Training data scaling from 805 to 2,162 to 3,561 instances across three stages (distant supervision, annotation algorithms, data augmentation), showing transformation from sparse to balanced property coverage. (b) Overall performance metrics showing systematic improvement from F1 score of 82.77\% to 88.18\% to 88.88\% on the validation dataset, while maintaining balanced precision-recall characteristics. (c) Property-specific F1 progression demonstrating that different properties achieve optimal performance through distinct framework components, validating the hierarchical approach.}
    \label{fig:hierarchical_data_preparation}
\end{figure}


\begin{algorithm}[H]
\begin{algorithmic}[1]
\REQUIRE Table $T$ with headers $H = \{h_1, h_2, \ldots, h_n\}$, caption $C$, unit list $U$, value matrix $D$, candidate property list $\mathcal{P}$
\ENSURE Annotated headers with assigned properties

\FORALL{$h \in H$}
    \STATE Clean and split $h$ into content and unit components
    \STATE \textbf{Skip} if $h$ already annotated via distant supervision

    \STATE $assigned \gets \textbf{False}$
    \FORALL{$p \in \mathcal{P}$}
        \STATE \COMMENT{Priority 1: Direct semantic match with canonical phrase}
        \IF{\textsc{DirectMatch}($h$, $p$) $\neq -1$}
            \STATE Assign $h$ to property $\textsc{DirectMatch}(h, p)$
            \STATE $assigned \gets \textbf{True}$; \textbf{break}
        \ENDIF

        \COMMENT{Priority 2: Scoring-based validation using symbolic and contextual signals}
        \IF{$h \in$ symbol aliases($p$) \textbf{and} \textsc{ValidateProperty}($p$, $D_h$, $h$, $U_h$, $C$)}
            \STATE Assign $h$ to property $p$
            \STATE $assigned \gets \textbf{True}$; \textbf{break}
        \ENDIF
    \ENDFOR

    \IF{\textbf{not} $assigned$}
        \STATE Assign label $0$ to $h$
    \ENDIF
\ENDFOR

\STATE \textbf{Determine table orientation and suppress misaligned headers:}
\IF{both row and column headers contain property labels}
    \STATE Check if first row contains at least 3 numeric values: $check\_row$
    \STATE Check if first column contains at least 3 numeric values: $check\_col$
    \IF{$check\_row$ \textbf{and not} $check\_col$}
        \STATE Reset column property labels $\geq 4$ to 0
    \ELSIF{not $check\_row$ \textbf{and} $check\_col$}
        \STATE Reset row property labels $\geq 4$ to 0
    \ELSE
        \STATE Count number of assigned properties in row vs column
        \STATE Suppress axis with fewer high-confidence assignments
    \ENDIF
\ENDIF

\RETURN $\mathcal{H}$ \COMMENT{Annotated header list}
\end{algorithmic}
\caption{Property-Specific Annotation Algorithm}
\label{algo: 1}
\end{algorithm}

\begin{algorithm}[H]
\caption{\textsc{ValidateProperty}($p$, $D_h$, $h$, $U_h$, $C$) -- Signal-Fusion Validation with Property-Specific Thresholding}
\label{algo: 2nd}
\begin{algorithmic}[1]
\STATE \textbf{Input:} Property $p$, header $h$, associated unit $U_h$, data values $D_h$, and caption $C$
\STATE \textbf{Output:} \texttt{True} if $h$ validly corresponds to $p$; otherwise \texttt{False}

\STATE Initialize $score \gets 0$

\vspace{4pt}
\STATE \textbf{Fuse multiple signals based on heuristics:}
\begin{itemize}
    \item Unit match: Add 1 if unit $U_h$ matches unit patterns of $p$
    \item Range check: Add 1 if at least one value in $D_h$ lies within the known numerical range for $p$
    \item Caption context: Add 1 if caption $C$ contains keywords associated with $p$
    \item Symbol match: Add 1 if $h$ appears in symbolic aliases of $p$
    \item Name match: Add 1 if $h$ exactly matches the canonical phrase for $p$
\end{itemize}

\vspace{4pt}
\STATE \textbf{Threshold decision:}
\STATE Use a tunable property-specific threshold $t_p$
\STATE \textbf{return} \texttt{True} if $score \geq t_p$; else \texttt{False}

\end{algorithmic}
\end{algorithm}

\newpage

\subsection{Algorithms for Property-Centric Table Augmentation}
\label{algo_for_table_augmentation}

To mitigate the data scarcity for underrepresented properties in the training corpus, we propose a strategic augmentation framework that combines power-law resampling with domain-informed co-occurrence priors. Rarely occurring properties are upsampled more aggressively than frequently seen ones, following a power-law scaling strategy. We also introduced augmentation based on `friend co-occurence', whereby a property is augmented only into tables that contain other properties it is frequently studied alongside—while ensuring the target property is not already present. This design ensures that the augmented tables faithfully mimic the structure and semantics of real-world scientific tables, extending the utility of our framework beyond this work to broader materials informatics applications.
\begin{algorithm}[H]
\caption{\texttt{\textsc{GenerateAugmentationPlan}($\mathcal{T}$)} -- Frequency Scaling and Friend Mapping Function}
\label{algo:augment_plan}
\begin{algorithmic}[1]
\STATE \textbf{Input:} Property-labeled table list $\mathcal{T}$
\STATE \textbf{Output:} Dictionary \texttt{aug\_dict} with keys: \texttt{prop\_id}, \texttt{prop\_names}, \texttt{orig\_freq}, \texttt{friend\_ids}, \texttt{new\_freq}

\STATE Initialize dictionary $\mathcal{F}_{\text{orig}}[p] \gets 0$ for all 18 property labels $p \in \{4, 5, \dots, 21\}$

\STATE For each $p \in \mathcal{P}$, compute $\mathcal{F}_{\text{orig}}[p]$ as total occurrences in row/column labels across all property tables in $\mathcal{T}$

\STATE Use the original frequency distribution $\mathcal{F}_{\text{orig}}$ to visualize and compare augmentation targets
\STATE Tune power-law parameters $a$ and $\alpha$ based on plot alignment between $\mathcal{F}_{\text{orig}}$ and $\mathcal{F}_{\text{target}}$ (desired scaling)
\STATE Assign final scaling parameters: $a \gets 10$, $\alpha \gets 0.65$

\STATE Build list \texttt{prop\_names} and \texttt{prop\_id} using known property mappings
\STATE Assign \texttt{friend\_ids} for each property $p$ with the help of domain expert (co-occurrence priors)

\STATE Compute scaled target frequency for each property:
\[
\texttt{new\_freq}(p) \gets \left\lceil a \cdot \mathcal{F}_{\text{orig}}[p]^\alpha \right\rceil
\]

\STATE Store all lists as keys in dictionary \texttt{aug\_dict}

\RETURN \texttt{aug\_dict}
\end{algorithmic}
\end{algorithm}
The augmentation process begins with \textsc{GenerateAugmentationPlan}($\mathcal{T}$), which estimates how many synthetic entries are required for each property (algorithm described in Supplementary Information~\ref{algo:augment_plan}). We first compute the original frequency $\mathcal{F}{\text{orig}}(p)$ of each property $p$ by scanning the row and column labels across all input tables in the training dataset $\mathcal{T}$. A power-law scaling function, parameterized by $a = 10$ and $\alpha = 0.65$, is then applied to determine a smoothed target frequency $\mathcal{F}{\text{target}}(p) = \left\lceil a \cdot \mathcal{F}_{\text{orig}}(p)^\alpha \right\rceil$. These parameters were empirically tuned to align the source and target property frequency distributions, ensuring that the augmentation introduces balanced coverage without distorting the natural occurrence patterns present in the source training corpus. To promote realistic property combinations, friend property sets $\mathcal{F}_p$ were manually defined for each property $p$ by domain experts. These sets restrict augmentation to tables containing commonly co-reported properties, ensuring that synthetic entries resemble patterns observed in real materials literature.

The core augmentation procedure (algorithm described in Supplementary Information~\ref{algo:augmentation_main}) targets each underrepresented property $p$ where $\mathcal{F}_{\text{target}}(p) > \mathcal{F}_{\text{orig}}(p)$. For every such property, we sample $n_{\text{needed}} = \mathcal{F}_{\text{target}}(p) - \mathcal{F}_{\text{orig}}(p)$ destination tables from the pool of eligible candidates—those that do not already contain $p$ but include at least one friend property $f \in \mathcal{F}_p$. This friend-based filtering ensures that augmentation occurs only in scientifically plausible table contexts. Source vectors $v_p$ are obtained from existing tables where $p$ is already reported.

Since the dimensionality of the source vector may not match the row or column structure of the destination table, we employ the \textsc{RowOrColAugmentor} function (algorithm described in Supplementary Information~\ref{algo:roworcol_augment}) to align the vector length using Gaussian synthesis. If the vector $v_p$ is shorter than required, we append synthetic values sampled from a normal distribution $\mathcal{N}(\mu, \sigma)$, where $\mu$ and $\sigma$ are the empirical mean and standard deviation of $v_p$. If $v_p$ is longer, it is truncated to the desired length. We reject any augmentation where the standard deviation is zero or NaNs are encountered, returning a sentinel value $-50$ to skip the current insertion.

\newpage

\begin{algorithm}[H]
\caption{Strategic Property Data Augmentation with Gaussian Synthesis and Co-occurrence Filtering}
\label{algo:augmentation_main}
\begin{algorithmic}[1]
\REQUIRE Set of source input tables $\mathcal{T}$ with structural metadata and property annotations
\REQUIRE Desired property frequency function $\mathcal{F}_{\text{target}}: \mathcal{P} \rightarrow \mathbb{N}$
\REQUIRE Observed property frequency function $\mathcal{F}_{\text{orig}}: \mathcal{P} \rightarrow \mathbb{N}$
\REQUIRE Friend mapping $\mathcal{M}: \mathcal{P} \rightarrow 2^{\mathcal{P}}$
\STATE Compute augmentation plan: $\texttt{aug\_dict} \gets \textsc{GenerateAugmentationPlan}(\mathcal{T})$
\STATE Extract $\mathcal{F}_{\text{target}}, \mathcal{F}_{\text{orig}}, \mathcal{M}$ from $\texttt{aug\_dict}$
\STATE \textbf{Initialize:} \texttt{augment} flag $\gets [\vec{0}_{\text{rows}}, \vec{0}_{\text{cols}}]$ for all $T \in \mathcal{T}$

\FORALL{$p \in \mathcal{P}$ such that $\mathcal{F}_{\text{target}}(p) > \mathcal{F}_{\text{orig}}(p)$}
    \STATE Let $n \gets \mathcal{F}_{\text{target}}(p) - \mathcal{F}_{\text{orig}}(p)$
    \STATE Let $\mathcal{F}_p \gets \mathcal{M}(p)$ be the set of co-occurring (friend) properties

    \STATE \textbf{Collect source entries:}
    \STATE $\mathcal{S}_p \gets$ all rows or columns from $\mathcal{T}$ that contain property $p$

    \STATE \textbf{Collect candidate destination tables:}
    \STATE $\mathcal{D}_p \gets \{ T \in \mathcal{T} \mid T$ is a property table, $p \notin T$, and $\exists f \in \mathcal{F}_p$ such that $f \in T \}$
    
    \STATE Select $n$ destinations: sample indices $\{i_1, \ldots, i_n\} \sim \textsc{UniformSample}([0, |\mathcal{D}_p|), n)$
    
    \STATE Let $\{T_1, \ldots, T_n\} \gets \{ \mathcal{D}_p[i_j] \text{ for } j = 1, \ldots, n \}$

    \STATE $s \gets 0$
    \WHILE{$s < n$}
        \STATE Select a source entry $v_s \in \mathcal{S}_p$ containing property $p$ \\
        \hspace{1.2em}(cycled using $v_s \gets \mathcal{S}_p[s \bmod |\mathcal{S}_p|]$ if $n > |\mathcal{S}_p|$)
        
        \STATE Let $T_i$ be the $s$-th sampled destination: $T_i \gets \mathcal{D}_p[\texttt{dst\_idx}[s]]$

        \STATE Let $L$ be the number of columns in target table $T_i$ if row-oriented; number of rows if column-oriented
        \STATE Generate aligned property entry $v_p \gets \textsc{RowOrColAugmentor}(v_s, L)$

        \STATE \textbf{if} $v_p = -50$: \textbf{continue} \COMMENT{Indicates augmentation failure}


        \STATE Insert $v_p$ into $T_i[\texttt{act\_table}]$ along the property axis (row or column, \textbf{based on table orientation})
        \STATE Append $p$ to the corresponding label list and $1$ to the augmentation flag
        \STATE Increment the appropriate count: $T_i[\texttt{num\_rows}]$ or $T_i[\texttt{num\_cols}]$, based on table orientation

        \STATE Update cell count to keep record: $T_i[\texttt{num\_cells}] \gets T_i[\texttt{num\_rows}] \cdot T_i[\texttt{num\_cols}]$
        \STATE Raise augmentation error if $T_i[\texttt{num\_rows}] \ne |\texttt{act\_table}|$ or $T_i[\texttt{num\_cols}] \ne |\texttt{act\_table}[0]|$

    \STATE Increment counter $s$ to move to the next augmentation insertion
    \ENDWHILE
\ENDFOR

\RETURN Augmented table collection $\mathcal{T}$
\end{algorithmic}
\end{algorithm}

\begin{algorithm}[H]
\caption{\textsc{RowOrColAugmentor}($v_s$, $L$) -- Gaussian-Based Property List Augmentation}
\label{algo:roworcol_augment}
\begin{algorithmic}[1]
\STATE \textbf{Input:} Source list of property values $v_s$; target length $L$ for alignment
\STATE \textbf{Output:} Augmented list $v_p$ of length $L$, or $-50$ if augmentation fails

\vspace{3pt}
\IF{$|v_s| = L$}
    \STATE Return $v_s$ unchanged
\ELSIF{$|v_s| > L$}
    \STATE Truncate $v_s$ to the first $L$ elements
    \STATE Return truncated list as $v_p$
\ELSE
    \STATE Extract numeric values $nvs$ from $v_s$ using \texttt{find\_num}
    \IF{$nvs$ is empty}
        \STATE \textbf{return} $-50$ \COMMENT{No valid numeric values for sampling}
    \ENDIF
    \STATE Compute $\mu \gets \texttt{mean}(nvs)$, $\sigma \gets \texttt{std}(nvs)$
    \STATE Sample $(L - |v_s|)$ values from $\mathcal{N}(\mu, \sigma)$
    \STATE Convert sampled values to strings and append to $v_s$
    \STATE Return the augmented list $v_p$
\ENDIF
\end{algorithmic}
\end{algorithm}

\section{Ablation Studies}
\label{sec: abl std}

To further evaluate the contribution of each component within the \matskraft{} framework, we conducted comprehensive ablation studies by selectively removing individual components and measuring the resulting performance degradation. These studies provide crucial insights into the relative importance of different architectural choices and data preparation strategies, validating our methodological choices and identifying the most critical elements for achieving such high extraction accuracy.

Supplementary Table~\ref{tab:ablation} presents the ablation results for both property and composition extraction tasks, systematically evaluating architectural components and data preparation strategies. The complete \matskraft{} framework achieves the highest F1 scores of 88.68 for property extraction and 71.35 for composition extraction, establishing our baseline performance. This systematic removal of individual components reveals distinct patterns of contribution across the two extraction tasks, with performance degradations providing clear evidence of component-specific importance in descending order of F1 scores.

For constructing large-scale scientific knowledge repositories, achieving high F1 scores with elevated precision is essential, as false positive extractions can propagate errors throughout downstream analyses and compromise the reliability of materials research studies. Scientific databases should contain accurate data to support critical applications such as materials design optimization and safety-critical engineering decisions, where incorrect property-composition associations could lead to material failures with significant economic and safety consequences. This precision-centric requirement distinguishes scientific knowledge extraction from general information retrieval tasks, where moderate false positive rates may be acceptable. All the log files or required codes for ablation studies have been uploaded in our repository on \href{https://github.com/M3RG-IITD/MatSKRAFT}{GitHub}.

\begin{table}[htbp]
\centering
\begin{tabular}{lccc}
\toprule
\textbf{Model Configuration} & \textbf{F1 Score} & \textbf{Precision} & \textbf{Recall} \\
\midrule
\multicolumn{4}{l}{\textit{Property Extraction (\matskraft{}\_P)}} \\
\midrule
\cellcolor{pink!30}\matskraft{}\_P  & \cellcolor{pink!30}\textbf{88.68} & 90.35 & 87.07 \\
\quad w/o Constrained learning* & 88.38 & 88.04 & 88.72 \\
\quad w/o Caption information & 86.94 & 85.94 & 87.95 \\
\quad w/o Post-processing & 79.30 & 76.63 & 82.16 \\
\quad w/o Data augmentation & 87.50 & 88.37 & 86.65 \\
\quad w/o Annotation algorithms & 79.66 & 72.34 & 88.64 \\
\quad w/o Distant supervision & 86.62 & \cellcolor{pink!30}\textbf{93.04} & 81.02 \\
\quad w/o (Distant supervision + Data augmentation) & 86.20 & 91.17 & 81.11 \\
\quad w/o (Annotation algorithms + Data augmentation) & 84.31 & 80.09 & \cellcolor{pink!30}\textbf{89.01} \\
\midrule
\multicolumn{4}{l}{\textit{Composition Extraction (\matskraft{}\_C)}} \\
\midrule
\cellcolor{pink!30}\matskraft{}\_C  & \cellcolor{pink!30}\textbf{71.35} & \cellcolor{pink!30}\textbf{82.31} & \cellcolor{pink!30}\textbf{62.97} \\
\quad w/o Thresholding & 68.73 & 78.19 & 61.32 \\
\quad w/o Constrained learning & 64.83 & 76.86 & 56.05 \\
\quad w/o Annotation algorithm & 62.42 & 73.57 & 54.21 \\
\quad w/o Caption information & 61.64 & 73.12 & 53.27 \\
\bottomrule
\multicolumn{4}{l}{*ID constraint improved the F1 score of ID classification in property tables from 81.4\% to 82.9\%.}
\end{tabular}
\vspace{0.1in}
\caption{Ablation study results for \matskraft{} framework components. Performance metrics show the contribution of each component to overall extraction accuracy for both property extraction (\matskraft{}\_P) and composition extraction (\matskraft{}\_C).}
\label{tab:ablation}
\end{table}

\subsection{Property Extraction Ablation Analysis}
\label{subsec : appendix_prop_extr_ablation}

Property extraction represents a complex task within \matskraft{}, requiring sophisticated disambiguation of ambiguous scientific notation and handling of diverse property reporting conventions across materials science literature~\cite{hira2024reconstructing}. The ablation analysis reveals distinct importance hierarchies between architectural components that provide structural guidance and data preparation strategies that address training data quality and coverage. Performance degradations range from minimal (0.30 F1 points) to substantial (9.38 F1 points), quantifying the relative importance of each framework component and validating our design philosophy that prioritizes domain-specific expertise over generic machine learning approaches.

\subsubsection{Architectural Components}
\label{subsec: appendix_property_arch_abl_std}

We systematically evaluate the core architectural elements by measuring performance degradation upon their removal, revealing dependencies in our graph neural network design and the relative importance of structural constraints, contextual information integration, and domain-specific post-processing within the extraction pipeline:

\textbf{1. MatSKRAFT\_P - F1 = 88.68 :} The complete framework represents our optimal configuration, incorporating all architectural and data preparation components in a synergistic manner. This baseline demonstrates excellent precision (90.35) and a strong recall (87.07), indicating effective extraction across both common and rare properties while maintaining the stringent quality standards essential for scientific knowledge repositories.

\textbf{2. MatSKRAFT\_P w/o Constrained Learning - F1 = 88.38 :} Removing the constraint-driven learning component results in the smallest performance degradation of only 0.30 F1 points. This configuration does achieve the second-highest recall in the entire ablation studies (88.72, +1.65 points), while experiencing a moderate precision reduction (88.04, -2.31 points). The constrained learning framework enforces four key structural principles through differentiable constraint loss terms: material-property association, material identifier exclusivity, property exclusivity, and material identifier uniqueness, described in Section~\ref{sec: methods}. The minimal F1 impact suggests these structural constraints provide valuable but not critical architectural guidance for overall property extraction, while the precision-recall trade-off indicates that constraint removal leads to more liberal property identification—capturing additional true positives at the cost of introducing false positives. Notably, the ID constraint component specifically improved the F1 score of material identifier classification in property tables from 81.4\% to 82.9\%, demonstrating its targeted effectiveness in distinguishing material identifiers from other table elements—a crucial subtask for enabling proper composition-property linking during knowledge base construction.

\textbf{3. MatSKRAFT\_P w/o Caption Information - F1 = 86.94 :} The removal of caption information leads to a moderate F1 decrease of 1.74 points, driven primarily by precision reduction (85.94, -4.41 points) while maintaining comparable recall (87.95, +0.88 points). Caption information serves multiple essential architectural functions: providing semantic context for property identification, disambiguating identical symbols across different material systems, and supplying unit information when table cells contain only numerical values. The precision degradation demonstrates that without caption context, the graph neural network loses critical disambiguation capability, leading to increased false positive identification where semantically similar but distinct properties are conflated. The maintained recall indicates that the cell-header relationship modeling within the graph architecture can still identify most true properties, but lacks the contextual sophistication necessary for precise classification in ambiguous scenarios prevalent in materials science literature.

\textbf{4. MatSKRAFT\_P w / o post-processing - F1 = 79.30 :} The elimination of post-processing produces highest F1 degradation of 9.38 points, making it the most vital architectural component for property extraction. This performance loss manifests through severe precision reduction (76.63, -13.72 points) and moderate recall degradation (82.16, -4.91 points), revealing the importance of domain-specific knowledge encoding. The post-processing component represents a methodological breakthrough that elevates raw neural network predictions into scientifically rigorous, fully interpretable knowledge entries through sophisticated disambiguation mechanisms encoding deep materials science expertise.

The system inherently handles challenging distinctions in materials science literature through sophisticated error detection and correction mechanisms that operate beyond simple pattern recognition. Post-processing implements intelligent property boundary validation that rejects extractions violating physical principles---automatically filtering negative density measurements, unrealistic temperature ranges, and elastic constants exceeding theoretical material limits. A critical validation challenge involves disambiguating properties with identical symbolic representations, exemplified by Poisson ratio extraction where symbols ``s'' or ``n'' frequently overlap with electrical conductivity and refractive index notation. The post-processing pipeline resolves these conflicts through dual-criteria validation—semantic analysis detecting ``Poisson'' terminology while excluding "conductivity" or "refractive" references, combined with strict value-range enforcement ensuring extracted measurements fall within physically plausible bounds (-1 to 0.5 for Poisson ratio versus >1 for refractive index). A particularly sophisticated capability involves hardness measurement validation, where the system processes over 50 notation variants across multiple scales (Vickers, Rockwell, Shore, Knoop, Brinell) through hierarchical recognition---first attempting direct standardization, then employing contextual assignment based on paper content and value ranges, and finally validating against theoretical limits for each scale. Additionally, the system implements intelligent composition-property separation protocols that prevent extraction of compositional percentages as material properties, while simultaneously detecting and correcting systematic misclassifications where temporal entities with units (hour, minutes, or seconds) are erroneously identified as temperatures due to similar notations. 

The severe precision degradation without post-processing demonstrates that these domain-specific heuristics are not superficial corrections but deeply embedded scientific reasoning that mirrors expert materials scientist decision-making. The moderate recall reduction indicates that post-processing occasionally filters out true positives through conservative validation, but this approach is essential for ensuring that the final extracted database is scientifically trustworthy, capable of supporting high-impact materials discovery and rigorous downstream analyses.

\subsubsection{Data Preparation Strategy Analysis}
\label{subsec: appenix_data_ablation}

Our automated data preparation strategy comprises three sequential components: distant supervision, annotation codes, and data augmentation. The systematic ablation of these components provides insights into the challenge of obtaining high-quality labeled data in specialized scientific domains---one of the major bottlenecks in deep learning applications.

\textbf{1. MatSKRAFT\_P w/o Data Augmentation - F1 = 87.50 :} Removing the data augmentation component results in a moderate F1 decrease of 1.18 points, characterized by precision reduction (88.37, -1.98 points) and slight recall degradation (86.65, -0.42 points). Our power-law guided augmentation strategy specifically targets the long-tail distribution problem endemic to scientific property extraction, where rare but technologically important properties like Abbe number and fracture toughness have minimal representation in training data in comparison to commonly studied properties like density and glass transition temperature. The recall reduction confirms that without strategic augmentation, the model exhibits reduced sensitivity to underrepresented property classes, failing to generalize from limited training examples — demonstrated by the degraded performance on rare properties such as Abbe value (recall drops from 100\% to 0\%) and shear modulus (recall drops from 74.6\% to 57.9\%). Similarly, precision degradation is evident across multiple properties, with notable impacts on Young's modulus (precision drops from 86.1\% to 72.5\%) and liquidus temperature (precision drops from 91.9\% to 77.3\%). The overall precision reduction (88.37 vs. 90.35) indicates that data augmentation not only improves coverage of rare properties but also enhances overall extraction accuracy, validating that our power-law guided augmentation strategy successfully addresses the long-tail distribution problem without introducing classification noise.

\textbf{2. MatSKRAFT\_P w/o Annotation Algorithm - F1 = 79.66 :} Removing the domain-specific annotation algorithms results in the most notable F1 decrease of 9.02 points among data preparation components, characterized by severe precision degradation (72.34, -18.01 points) while maintaining high recall (88.64, +1.57 points). This performance shift reveals the critical role of annotation codes in maintaining extraction precision across diverse property types. Physical and mechanical properties exhibit severe degradation, with density precision plummeting from 95.7\% to 57.0\% and Poisson ratio declining from 63.3\% to 43.9\%. Both optical and electrical properties demonstrate substantial deterioration, as refractive index precision falls from 84.3\% to 61.4\%, while electric conductivity drops from 90.9\% to 67.1\%. For some properties, the impact extends beyond precision to even recall degradation, as observed for annealing point where the recall plummets from 67.4\% to 23.3\%. This systematic performance decline demonstrates widespread over-classification where the model incorrectly identifies numerous non-property cells due to superficial pattern similarities, compromising the reliability required for scientific knowledge repositories.

Annotation codes represent a critical training data enhancement component that supplements distant supervision from INTERGLAD through sophisticated rule-based property identification. These codes systematically improve data quality by identifying properties absent from reference databases through multi-dimensional pattern analysis—examining unit patterns, value ranges, structural relationships, and semantic contexts to generate high-quality labeled examples beyond the scope of commercial databases. The substantial precision reduction without annotation codes demonstrates their essential role in enabling neural networks to learn precise classification boundaries, revealing that distant supervision alone, even when followed by data augmentation, is insufficient for achieving the stringent accuracy standards required for scientific knowledge extraction. Annotation codes bridge the critical gap between the limited scope of existing commercial databases and the full diversity of property reporting semantics in materials science literature, systematically encoding domain expertise into the training data generation process to develop models capable of maintaining both high recall and precision specificity required for reliable scientific knowledge repositories.

\textbf{3. MatSKRAFT\_P w/o Distant Supervision - F1 = 86.62 :} The removal of distant supervision reveals a distinctive and highly significant performance profile — achieving the highest precision (93.04, +2.69 points) among all the configurations while experiencing considerable recall reduction (81.02, -6.05 points), resulting in a moderate 2.06 point F1 decrease. This configuration represents a fully automated training data generation approach that eliminates dependency on existing commercial databases, addressing one of the key challenges in deep learning — obtaining high-quality labeled data without expensive manual annotation or external database dependencies~\cite{roh2019survey, huang2024data, song2022learning, olivetti2020data}. The exceptional precision demonstrates that our annotation codes and data augmentation components can independently generate training data of remarkable quality, producing extraction models that rarely commit false positive errors, while the recall limitation indicates that approximately 19\% of true properties remain undetected — a trade-off where data fidelity supersedes coverage for knowledge-base construction applications.

The removal of distant supervision effectively demonstrates that our framework can achieve rigorous scientific knowledge extraction without reliance on existing commercial databases, resolving the data dependency bottleneck that has historically constrained the scope and accessibility of automated scientific information extraction systems. This capability enables researchers with limited access to commercial databases to develop extraction systems of comparable performance using our fully automated data preparation pipeline. This, in turn, democratizes access to advanced materials knowledge extraction technologies and accelerating deep learning model development by eliminating the cost and time barriers associated with high-quality training data generation.

\textbf{4. MatSKRAFT\_P w/o (Distant Supervision + Data Augmentation) - F1 = 86.20 :} This combined ablation retains only annotation codes for training data generation — removing both distant supervision and data augmentation components, resulting in a moderate F1 decrease of 2.48 points through precision increase (91.17, +0.82 points) and recall reduction (81.11, -5.96 points). The performance profile shows higher precision compared to the full model baseline, confirming that annotation codes alone can maintain strong precision performance while the removal of distant supervision and data augmentation primarily impacts the property coverage through reduced recall.

\textbf{5. MatSKRAFT\_P w/o (Annotation Codes + Data Augmentation) - F1 = 84.31 :} This configuration shows F1 degradation of 4.37 points while achieving the highest recall (89.01, +1.94 points) but significant reduction in precision (80.09, -10.26 points). The removal of both components leaves only distant supervision for training data generation, resulting in a system that exhibits the characteristic limitation of distant supervision: high sensitivity to property identification but poor precision due to insufficient training data diversity and quality. The considerable precision drop demonstrates that distant supervision alone generates training data with limited scope and disambiguation capability, leading to models that liberally classify cells as properties but frequently misidentify non-property content, thereby failing to meet the stringent accuracy requirements essential for reliable scientific knowledge extraction.

The systematic ablation analysis reveals that \matskraft{}`s exceptional performance emerges from the strategic integration of components with complementary strengths, where each addresses distinct limitations inherent in scientific information extraction. Comparative analysis across ablation configurations demonstrates three critical performance profiles that validate our multi-component design philosophy. Distant supervision exhibits a characteristic high recall, moderate-to-low precision profile, as evidenced by comparing the w/o distant supervision configuration (93.04 precision, 81.02 recall) with the w/o (annotation codes + data augmentation) configuration, while retaining only distant supervision gives us a 80.09 precision of 80.09 and recall of 89.01. This performance inversion reveals distant supervision's limitation---it successfully identifies diverse property representations across various semantic expressions but generates training data of insufficient quantity for precise classification, resulting in models that achieve high recall through liberal property identification but suffer from poor precision due to inadequate disambiguation capability. Conversely, annotation codes demonstrate high precision, with their removal causing the most severe precision collapse from 90.35 to 72.34 (18-point degradation)---the largest single-component impact across all data preparation strategies. The w/o (distant supervision + data augmentation) configuration, retaining only annotation codes, achieving precision of 91.17 (+0.82 points above baseline) confirms that annotation codes independently excel at generating exceptionally accurate training examples, providing minimal false positive contamination with controlled recall reduction of 81.11.

Data augmentation serves as the critical balancing mechanism, addressing coverage limitations of both distant supervision and annotation codes through strategic enhancement of underrepresented properties without compromising accuracy. This is demonstrated through moderate performance improvements across all metrics (F1 increase of 1.18 points, precision enhancement of 1.98 points, and recall boost of 0.42 points), expanding rare property coverage while simultaneously improving common property extraction — converting complete failures like Abbe value (100\% to 0\% recall without augmentation) into successful extractions while enhancing precision across properties like Young's modulus and liquidus temperature. This comprehensive ablation analysis validates that no single component can independently achieve production-level performance for scientific information extraction, establishing that \matskraft{}'s superior capabilities emerge from the deliberate orchestration of complementary data preparation strategies. The strategic sequencing—distant supervision providing broad coverage, annotation codes contributing high-precision labeling independent of reference databases, and data augmentation ensuring balanced representation—creates a synergistic training data generation system that achieves the accuracy standards required for automated scientific knowledge extraction in domains where precision is critical.

\subsection{Composition Extraction Ablation Analysis}
\label{subsec: appendix composition abl std}

Composition extraction, adapted and enhanced on DiSCoMaT~\cite{gupta-etal-2023-discomat}, exhibits different architectural dependencies compared to property extraction, with component removals revealing the unique challenges inherent in interpreting complex compositional data across three distinct table structures and maintaining chemical consistency across diverse reporting formats. The task must successfully handle Single-Cell Composition (SCC) tables where entire compositions are written in single cells, Multiple-Cell Composition (MCC) tables where compositions are distributed across multiple cells with separate constituent values, and Partial-Information (PI) tables where complete compositional details must be inferred from contextual information beyond the table itself. The ablation analysis demonstrates heightened sensitivity to contextual information and structural guidance essential for accurate chemical formula interpretation and stoichiometric relationship modeling across these heterogeneous representation formats.

\subsubsection{Architectural Components Impact Analysis}

We systematically evaluated the core architectural elements for composition extraction, revealing critical dependencies that differ markedly from property extraction patterns and highlighting the unique challenges in chemical formula recognition, constituent identification, and stoichiometric validation across diverse compositional-tables reporting conventions:

\textbf{1. MatSKRAFT\_C - F1 = 71.35 :} The complete composition extraction framework achieves optimal performance across all metrics — F1 score (71.35), precision (82.31), and recall (62.97). This baseline represents our enhanced DiSCoMaT architecture with comprehensive improvements including re-annotation of training data using refined rule-based constituent detectors, expanded detection mechanisms for atomic percentage (at\%) compositions alongside weight percentage (wt\%) and molar percentage (mol\%) units, and optimized post-processing with improved partial-composition extraction rules and enhanced compound recognition patterns. This framework demonstrates robust extraction capabilities with high precision across all types of table structures, resulting in accurate extraction of materials reported in the composition tables.

\textbf{2. MatSKRAFT\_CC w/o Thresholding - F1 = 68.73 :} Removing the thresholding component decreases F1 score by 2.62 points, characterized by precision reduction (78.19, -4.12 points) and slight recall degradation (61.32, -1.65 points). The thresholding mechanism implements a confidence-based filtering strategy with $\alpha = 0.7$, where compositional predictions with softmax probabilities below this threshold are automatically reclassified as non-compositional elements, effectively preventing low-confidence false positive extractions from contaminating the final knowledge base. This conservative approach prioritizes precision over recall by rejecting ambiguous predictions that could introduce chemically invalid formulations.

The precision-focused degradation pattern demonstrates that thresholding serves as a critical quality control mechanism, systematically filtering out uncertain predictions where the model exhibits insufficient confidence in compositional classification. Without this confidence barrier, the system accepts marginal predictions that often represent misclassified non-compositional content, leading to database contamination with invalid chemical formulations. This filtering proves particularly essential in composition extraction where false positive errors can propagate incorrect stoichiometric relationships throughout materials databases, potentially misleading subsequent computational studies and experimental investigations that rely on accurate compositional data for materials design and property prediction applications.

\textbf{3. MatSKRAFT\_C w/o Constrained Learning - F1 = 64.83 :} The elimination of constrained learning triggers a notable 6.52-point F1 decline, with precision dropping to 76.86 (-5.45 points) while recall simultaneously deteriorates to 56.05 (-6.92 points). This balanced performance loss indicates that constrained learning provides structural guidance essential for composition extraction accuracy across diverse table formats. This notable impact reflects the complex organizational challenges inherent in composition extraction — determining whether compositions span single cells or multiple cells, identifying constituent elements and their respective proportions, and maintaining stoichiometric consistency across diverse table formats and chemical notation systems.

The substantial precision-recall degradation demonstrates that without structural constraint enforcement, the model struggles with both accurate identification and comprehensive coverage of compositional information. Composition extraction requires sophisticated understanding of chemical relationships that extend beyond simple pattern recognition—including constituent-proportion associations, elemental balance requirements, and structural consistency across different representation formats. The constraints enforce chemical logic during training, guiding the optimization process toward chemically coherent interpretations followed by the reported compositional tables.

\textbf{4. MatSKRAFT\_C w/o Annotation Codes - F1 = 62.42:} Annotation codes emerge as one of the most critical components for composition extraction, with their removal resulting in 8.93-point F1 collapse through severe precision (73.57, -8.74 points) and significant recall deterioration (54.21, -8.76 points). This balanced precision-recall collapse reveals annotation codes as the key component for chemical interpretation accuracy, distinguishing \matskraft{} from conventional pattern-matching approaches through nuanced chemical intelligence encoding. For composition extraction, annotation codes implement a multi-layered chemical reasoning system that transforms raw neural network predictions into chemically coherent interpretations through three critical validation mechanisms: chemical formula recognition across diverse notation systems (from simple oxides like SiO$_2$ to complex quaternary compositions), stoichiometric validation ensuring charge balance and chemical plausibility, and compositional unit detection that automatically identifies and normalizes percentage reporting conventions (mol\%, wt\%, at\%) scattered throughout table headers and captions. The system demonstrates remarkable chemical sophistication by implementing regex-based constituent detection that recognizes an extensive library of chemical elements and compounds, processes complex compositional expressions like ``Na$_2$O-CaO-SiO$_2$'' by decomposing them into constituent oxides, and reconstructs complete compositional information from fragmented table representations where individual constituents appear in separate cells with their respective percentages.

The performance degradation without annotation codes demonstrates widespread over-classification where the neural network incorrectly identifies numerous non-compositional cells as valid chemical formulations due to superficial pattern similarities—properties like density values (2.65 g/cm$^3$) are misclassified as compositional percentages, measurement uncertainties ($\pm$0.1) are interpreted as stoichiometric coefficients, and temperature values (80$^\circ$C) are confused with percentage compositions. Advanced chemical validation mechanisms systematically filter out compositionally invalid entries through multi-dimensional verification: examining whether extracted values sum to physically reasonable totals (typically 95-105\% for complete compositions, $\pm$5\% as for doping concentration or error control), validating that constituent ratios fall within established chemical bounds, and distinguishing between mass percentages, molar percentages, and atomic percentages through caption semantic analysis. Annotation codes systematically expand the training data coverage by identifying compositional rows and columns absent in addition to those covered by the distant supervision, enabling detection of diverse chemical notation patterns and percentage reporting conventions that significantly increase both the quantity and quality of labeled training examples---transforming \matskraft{} from a database-dependent extraction system into a comprehensive compositional detection framework capable of recognizing the full spectrum of materials science reporting formats essential for reliable scientific knowledge repositories.

\textbf{5. MatSKRAFT\_C w/o Caption Information - F1 = 61.64 :} Caption information removal produces the most severe impact on composition extraction with a 9.71-point F1 decrease, driven by both precision (73.12, -9.19 points) and recall reduction (53.27, -9.70 points). This severe loss highlights the critical dependence of composition extraction on contextual information. Table captions provide essential semantic context that proves indispensable for composition interpretation: specifying compositional units (wt\%, mol\%, at\%), clarifying nomenclature conventions across different research groups, defining chemical system boundaries and compositional frameworks, and providing stoichiometric context that enables proper constituent identification.

\section{Calibration of the Property-Extraction GNN}
\label{sec:appendix_calibration}


\begin{figure}[htbp]
    \centering
    \includegraphics[width=0.9\textwidth, keepaspectratio]{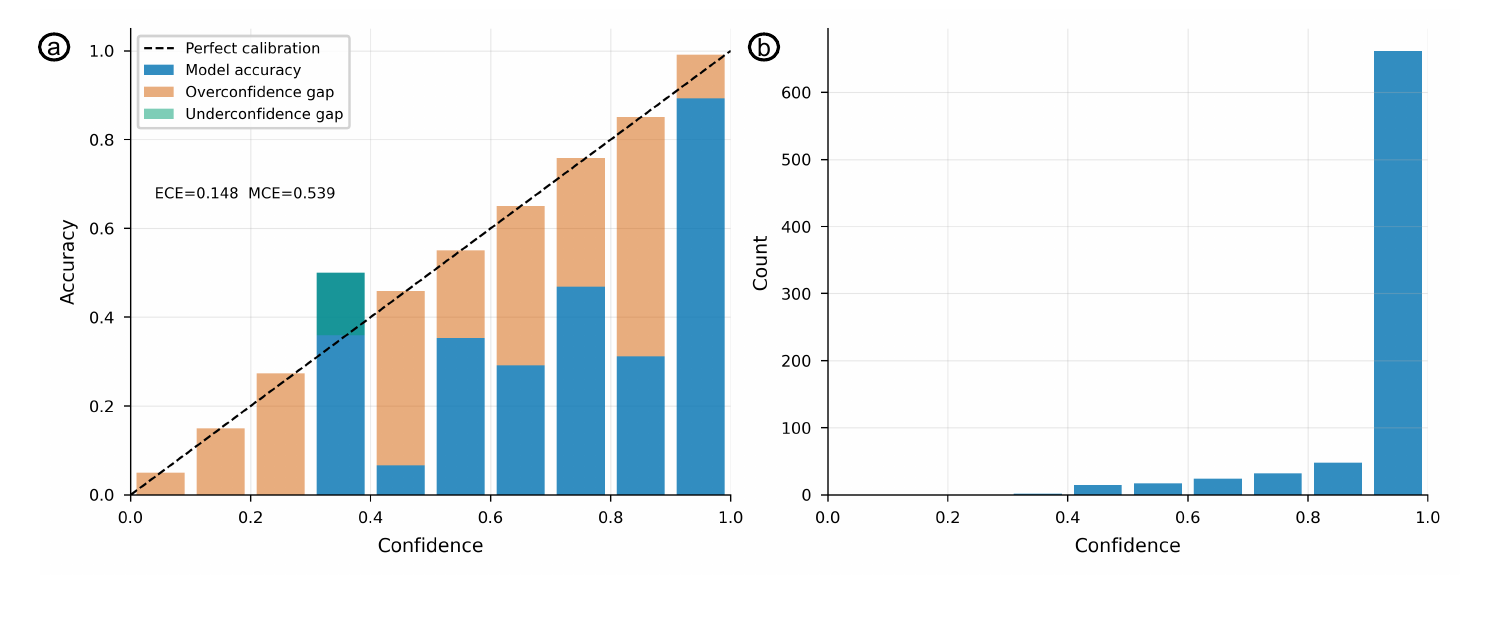}
    \caption{\textbf{Calibration of GNN-1 (\textsc{MatSKRAFT\_P}) softmax confidence.} (a) Reliability diagram for non-zero property predictions. Blue bars show observed accuracy per confidence bin; orange bars indicate the overconfidence gap (mean confidence $-$ accuracy); the green bar at [0.30--0.40] is the sole underconfident bin. Dashed diagonal: perfect calibration. (b) Distribution of softmax confidence values for the same predictions, showing strong concentration near [0.90--1.00].}
    \label{fig:c1}
\end{figure}

We analyzed the GNN classifier's maximum softmax score as a proxy for prediction confidence across the 18 target properties. The analysis below characterizes how faithfully these scores reflect actual prediction correctness.

\begin{figure}[htbp]
    \centering
    \includegraphics[width=0.9\textwidth, keepaspectratio]{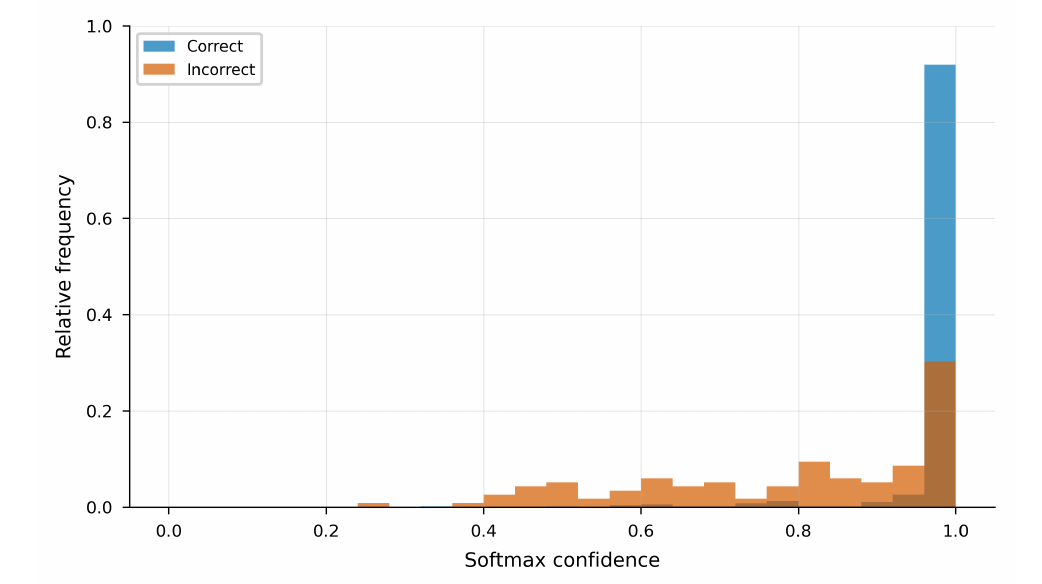}
    \caption{\textbf{Confidence distribution: correct vs.\ incorrect predictions.} Softmax confidence distributions for correct (true-positive) and incorrect (false-positive) property-class predictions on the test split.}
    \label{fig:c2}
\end{figure}

We note that the model is overconfident. The highest-confidence bin [0.90–1.00] contains around 82\% of predictions (see Supplementary Figure~\ref{fig:c1}b), with an observed accuracy of 0.893, indicating that these high-confidence predictions are largely trustworthy, making our model reliable. The critical failure zone is the [0.80–0.90] bin, producing the Maximum Calibration Error (MCE) of 0.539, which is significantly higher than the aggregate Expected Calibration Error (ECE = 0.148). These are the ambiguous predictions where the graph structure provides conflicting evidence and the model neither commits fully nor abstains cleanly.

\begin{figure}[htbp]
    \centering
    \includegraphics[width=0.9\textwidth, keepaspectratio]{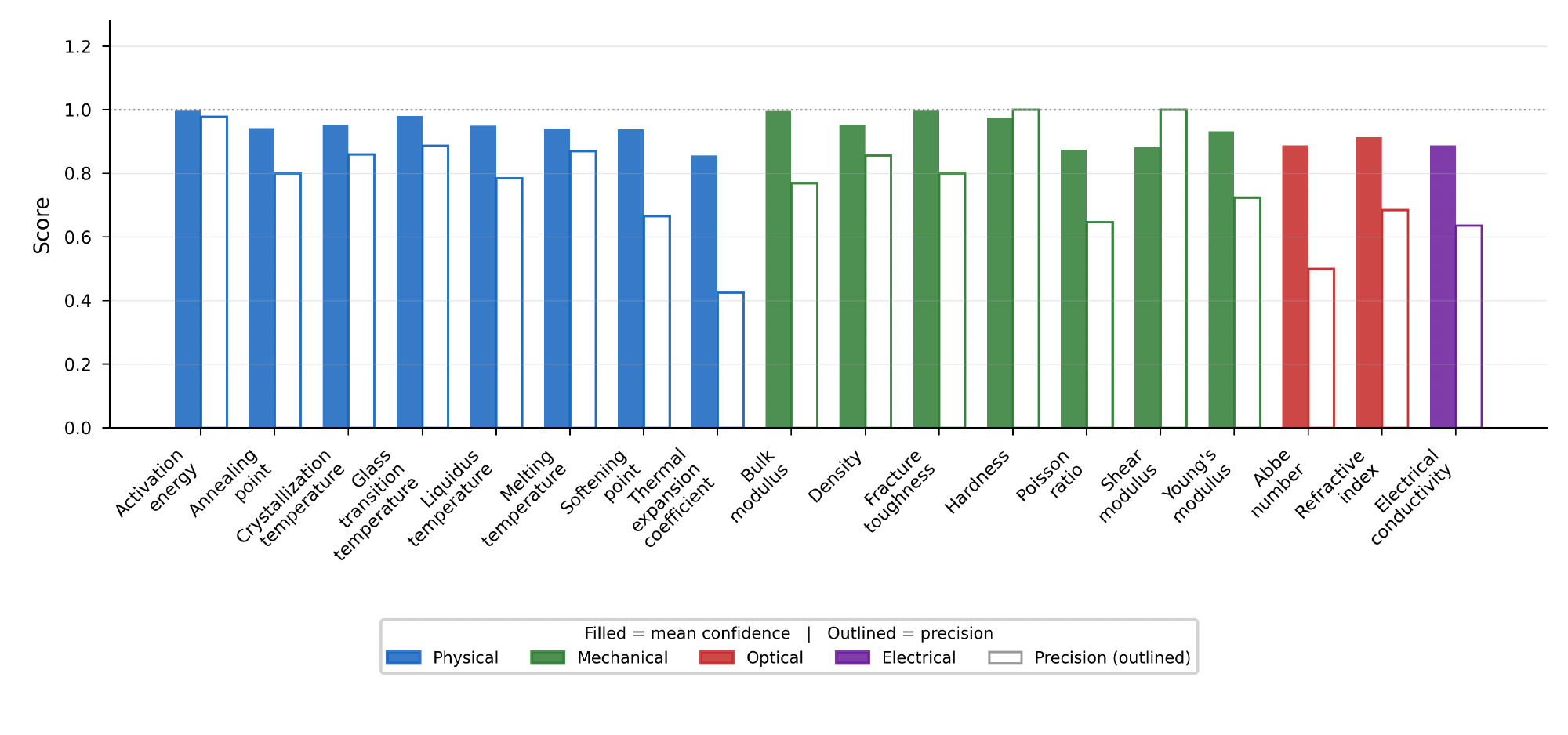}
    \caption{\textbf{Per-property calibration gap.} Mean predicted softmax confidence (filled bar) versus precision (outlined bar) for each extractable property class on the test split. Bar colour indicates property category — physical (blue), mechanical (green), optical (red), and electrical (purple). }
    \label{fig:c3}
\end{figure}

Despite the overconfidence, the scores carry a clear discriminative signal. We witness that incorrect predictions span a broad tail from approximately 0.25 to 1.0, while correct predictions are sharply concentrated near 1.0 (see Supplementary Figure~\ref{fig:c2}). This means the model’s errors are not uniformly distributed across confidence levels. A substantial fraction of false positives occupies the filterable intermediate range where a confidence threshold $\alpha$ can suppress them at acceptable recall cost. As shown in Supplementary Figure~\ref{fig:c3}, the overconfidence gap varies substantially across property classes, suggesting that a single global threshold may not be sufficient, thus motivating the property-adaptive acceptance thresholding developed in Supplementary Information~\ref{app:confidence_thresholding}.

\section{Data-Driven Confidence Thresholding}
\label{app:confidence_thresholding}

At the default decision boundary ($\alpha = 0$), every row/column header node is assigned the class with the highest softmax probability, regardless of how confident that prediction is. Evaluation on the validation split reveals that a subset of property classes exhibits a systematic precision–recall imbalance. We prioritize high precision for the final database while minimizing the loss of true positives. To address performance variations across properties, we implement per-property confidence thresholding.

Confidence thresholding offers a direct way to address this challenge. When a property node’s predicted probability falls below a minimum threshold $\alpha$, the prediction is suppressed back to the background class. The premise is that excess false positives concentrate among lower confidence predictions (already demonstrated in Supplementary Figure~\ref{fig:c2}), so raising the bar at inference time can recover precision without retraining the model. This methodology helps where the imbalance actually exists. For properties where precision already exceeds recall, applying a confidence gate would cost significant recall, removing a meaningful volume of valid extractions. Therefore, properties where precision exceeded recall (e.g., electrical conductivity, shear modulus, and activation energy) were excluded from confidence thresholding. All threshold selection decisions described below were made exclusively on the validation split and applied without modification to the held-out test set. The test set was never consulted during any stage of the selection procedure; test results are reported only to assess generalization.

On applying confidence thresholding, we obtained five properties (namely density, Abbe value, Poisson ratio, annealing point, thermal expansion coefficient) with 
a threshold of $0.90$, while one property (bulk modulus) was assigned a threshold of $0.50$. Validation results at the row-column level demonstrate that adaptive 
confidence thresholding corrects structural over-prediction without degrading baseline coverage. For instance, density precision advances from $87.6\%$ to 
$96.8\%$ ($+9.2\%$), while retaining a high recall of $95.7\%$ (compared to the $97.9\%$ baseline), yielding an aggregate F1 gain of $+3.8$ percentage points 
(pp). For the Poisson ratio, the baseline model achieves a perfect $100.0\%$ recall but a low precision of $70.6\%$. Applying the threshold isolates these 
false-positive errors, driving precision to $100.0\%$ ($+29.4\%$) while keeping recall completely intact at $100.0\%$, resulting in a $100.0$ F1-score. 
Suppressing predictions below the threshold thus retains only the high-confidence true positives while eliminating the bulk of false-positive predictions.

All six accepted thresholds were applied simultaneously in the final evaluation pass. The aggregate impact is reported in Supplementary Table~\ref{tab:threshold_impact}. At the 
row/column classification level, validation F1 improved from $83.79$ to $86.01$ ($+2.22$~pp) and test F1 improved from $83.63$ to $85.21$ ($+1.58$~pp). At the downstream tuple extraction level, test F1 improved 
from $88.68$ to $89.33$ ($+0.65$~pp), with test precision increasing from $90.35$ to $92.40$ ($+2.05$~pp) at a modest recall cost of $0.62$~pp.

\begin{table}[h]
\centering
\small
\begin{tabular}{llrrrrr}
\hline
\textbf{Split} & \textbf{Setting} & \textbf{Precision} & \textbf{Recall} & \textbf{F1} & \textbf{$\Delta$F1} \\
\hline
\multicolumn{6}{l}{\textit{Row/column classification}} \\
Val  & $\alpha = 0$ (baseline)  & 79.0 & 89.2 & 83.8 & ---     \\
Val  & Data-driven thresholding & 83.3 & 89.0 & 86.0 & $+$2.2  \\
Test & $\alpha = 0$ (baseline)  & 79.4 & 88.3 & 83.6 & ---     \\
Test & Data-driven thresholding & 82.9 & 87.6 & 85.2 & $+$1.6  \\
\hline
\multicolumn{6}{l}{\textit{Tuple extraction (end-to-end)}} \\
Val  & $\alpha = 0$ (baseline)  & 90.6 & 88.6 & 89.6 & ---     \\
Val  & Data-driven thresholding & 92.6 & 88.3 & 90.4 & $+$0.8  \\
Test & $\alpha = 0$ (baseline)  & 90.4 & 87.1 & 88.7 & ---     \\
Test & Data-driven thresholding & 92.4 & 86.5 & 89.3 & $+$0.7  \\
\hline
\end{tabular}
\vspace{0.1in}
\caption{\textbf{Impact of data-driven confidence thresholding on row/column
classification and tuple extraction metrics.} Threshold values were
determined exclusively on the validation split; test results reflect
generalisation to held-out data.}
\label{tab:threshold_impact}
\end{table}

This filtering logic directly altered the final large-scale knowledge base, reducing the total volume from 535,643 to 509,281 entries. Specifically, density entries decreased from 78,222 to approximately 63,129, Poisson ratio dropped from over 16,600 to approximately 10,500, and Abbe value was pruned from over 1,800 down to 680. To verify whether the discarded data represents true false positives or erroneously suppressed valid extractions, we randomly sampled and manually annotated 100 entries rejected by the confidence thresholding. Within this discarded sample, 53 entries were confirmed false positives and 47 were valid extractions. This $47.00\%$ precision among the removed entries stands in sharp contrast to the final overall knowledge-base precision of $79.04\%$, confirming that the thresholding criteria successfully isolated and suppressed predictions that fell substantially below the quality bar of the retained database. Retaining these marginal entries would have compromised overall precision; their removal represents an empirically justified trade-off where a minor sacrifice in coverage yields a cleaner, more reliable downstream knowledge resource.

\section{Comparative Evaluation against Large Language Models}
\label{sec: appendix_llms_comp}

Our comprehensive benchmarking of \matskraft{} against the leading large language models (LLMs)—spanning both established frontier systems in 2024-2025 (Gemini-1.5-Pro, GPT-4o, Claude-3.5-Sonnet, DeepSeek-V3, DeepSeek-R1) and the recently released contemporary LLMs in 2025-2026 (Gemini-3.5-Flash, Claude-Opus-4.7, GPT-5.4-Pro, DeepSeek-V4-Pro, DeepSeek-V4-Pro-Th)—establishes that domain-specialized graph neural network architectures systematically outperform generalist language models for scientific knowledge extraction, across both extraction accuracy and computational efficiency dimensions (Supplementary Table~\ref{tab:model_comparison}).

\textbf{1. Accuracy across extraction tasks.} \matskraft{} achieves F1 scores of 89.33 (detailed in Supplementary Table~\ref{tab:detailed_property_metrics}) for property extraction and 71.35 for composition extraction, establishing clear superiority over all LLM baselines despite the substantial capability advances of next-generation frontier models. For property extraction, the strongest contemporary LLM—Claude-Opus-4.7—attains an F1 of 85.65, narrowing the gap to 3.68 percentage points compared to the 15.4-point deficit observed against the prior generation LLMs (DeepSeek-V3 at 73.28 F1, see Supplementary Figure~\ref{fig:appendix_baseline_comparison}). This convergence reflects the significant progress in LLM-based table understanding; yet \matskraft{}'s concurrent precision leadership (92.40\% vs.\ 88.18\%) and superior recall (86.45\% vs.\ 82.67\%) demonstrate that the constraint-driven architecture with domain-based post-processing captures complementary advantages that scaling alone cannot replicate. Gemini-3.5-Flash achieves the highest raw precision among all models (91.45\%) for property extraction, however its recall collapses to 48.61\%, yielding an F1 of only 63.48—exposing systematic under-extraction that renders the model unsuitable for comprehensive knowledge base construction.

Similarly, during composition extraction, \matskraft{}'s 71.35 F1 score surpasses the best LLM baseline (Claude-Opus-4.7 at 64.51) by 6.84 points. The substantial precision advantage (82.31\% vs. 64.60\%) establishes \matskraft{} as the only reliable framework capable of achieving the standards required for automated knowledge base construction, where compositional accuracy directly impacts materials discovery and safety-critical engineering applications. At the database level, the integrated composition-property extraction pipeline demonstrates exceptional precision of 79.04\% across our comprehensive knowledge base of over 509,000 entries, confirming that \matskraft{} maintains stringent accuracy standards when scaling to real-world materials literature processing—a critical requirement for constructing scientifically reliable databases suitable for materials studies.

\begin{table}[htbp]
    \centering
    \resizebox{\linewidth}{!}{%
    \begin{tabular}{l|cccc|cccc}
        \hline
        & \multicolumn{4}{c|}{Extracting Properties} & \multicolumn{4}{c}{Extracting Composition} \\
        \cline{2-5} \cline{6-9}
        Models & Precision & Recall & F1 score & Time (s/table) & Precision & Recall & F1 score & Time (s/table) \\
        \hline
        Claude-3.5-Sonnet   & 78.80 & 67.64 & 72.79 &   4.90 & 47.12 & 55.38 & 50.92 &   6.56 \\
        Claude-Opus-4.7     & 88.18 & 82.67 & 85.65 &   3.35 & 64.69 & 64.32 & 64.51 &   1.38 \\
        \hline
        DeepSeek-V3         & 75.17 & 71.48 & 73.28 &  14.95 & 49.71 & 52.26 & 50.95 &  18.07 \\
        DeepSeek-R1         & 70.43 & 68.07 & 69.23 & 114.94 & 52.85 & 55.79 & 54.28 & 187.59 \\
        DeepSeek-V4-Pro     & 71.91 & 65.60 & 68.61 &   4.29 & 55.92 & 52.59 & 54.20 &   5.86 \\
        DeepSeek-V4-Pro-Th  & 89.02 & 79.91 & 84.22 &  23.31 & 50.51 & 61.35 & 55.41 &  37.44 \\
        \hline
        Gemini-1.5-Pro      & 66.09 & 58.13 & 61.85 &   8.25 & 39.60 & 44.02 & 41.69 &  11.89 \\
        Gemini-3.5-Flash    & 91.45 & 48.61 & 63.48 &   3.97 & 61.54 & 61.77 & 61.65 &   2.23 \\
        \hline
        GPT-4o              & 61.61 & 59.03 & 60.29 &   5.84 & 47.46 & 53.42 & 50.27 &   9.30 \\
        GPT-5.4         & 79.64 & 78.69 & 79.17 &   2.05 & 52.52 & 63.08 & 57.32 &   1.76 \\
        \hline
        \rowcolor{pink!30} \textbf{\matskraft{}} & \textbf{92.40} & \textbf{86.45} & \textbf{89.33} & \textbf{0.22} & \textbf{82.31} & \textbf{62.97} & \textbf{71.35} & \textbf{0.39} \\
        \hline
    \end{tabular}%
    }
    \vspace{0.1in}
    \caption{\textbf{Performance and computational efficiency comparison of MaTSKRAFT against two generations of frontier large language models for materials science information extraction from scientific tables.} Models are grouped by provider family; horizontal rules separate provider groups. Precision, recall, and F1 score are computed against expert-annotated test sets of 368 tables (property) and 737 tables (composition). Time measurements denote the mean processing duration per table.}
    \label{tab:model_comparison}
\end{table}

\textbf{2. Computational efficiency enabling knowledge extraction at large scale.} Beyond accuracy advantages, \matskraft{} demonstrates the computational efficiency that redefines the economics of scientific knowledge extraction. By processing tables at 0.22 seconds for property extraction and 0.39 seconds for composition extraction, \matskraft{} operates at speed of around 6--$10^3$ times greater than competing LLM approaches, which require 1.38-187.59 seconds per table. This high computational efficiency enables processing of around 66,267 tables across 11 journals in approximately 11.23 hours using a single 32GB V100 GPU, compared to the 70-5,792 hours (3 days to 8 months) that would be demanded by LLM API services. LLM APIs operate on massive cloud infrastructure, with OpenAI utilizing Microsoft Azure's supercomputing systems featuring distributed GPU clusters and high-performance computing (HPC) architecture, while also recently partnering with Google Cloud for additional computing capacity. These systems employ parallel processing across multiple servers and specialized GPU arrays designed for large-scale AI inference workloads. Despite this massive distributed infrastructure, \matskraft{}'s specialized graph neural network architecture achieves superior extraction performance while operating on a fraction of the computational resources, demonstrating the potential of domain-specific optimization over brute-force scaling. Furthermore, \matskraft{} supports fully parallel deployment across multiple GPUs on independent dataset partitions, enabling linear throughput scaling for multi-journal or cross-domain extraction campaigns without any architectural modification.

The efficiency advantage extends beyond processing speed to encompass economic and technical barriers that render LLM-based approaches impractical for comprehensive literature processing. \matskraft{} operates efficiently on a modest single-GPU system while LLM APIs impose prohibitive per-token costs that escalate dramatically with document length—processing our 66,267 tables would incur API costs ranging over tens of thousands of dollars, compared to \matskraft{}'s negligible computational expenses. Crucially, this cost is incurred per extraction run: any prompt refinement, re-evaluation, or model update necessitates re-processing, making iterative development workflows economically unsustainable for most academic research groups. At current API pricing structures, systematic processing of entire journal collections becomes economically prohibitive for most research organizations. LLM APIs impose per-token costs that escalate directly with prompt richness---detailed instructions, few-shot demonstrations, and article abstracts necessary for accurate extraction substantially inflate token consumption per table, making each incremental improvement to prompt quality an additional cost burden that compounds across tens of thousands of table calls. Our framework \matskraft{} enables resource-limited research groups to construct comprehensive knowledge repositories and process entire journal collections within reasonable time and budget constraints, transforming materials knowledge extraction from a computationally and economically prohibitive endeavor accessible only to well-resourced institutions into a broadly available tool for accelerating scientific discovery across academic, industrial, and governmental research organizations worldwide.

\begin{figure}
    \centering
    \includegraphics[width=0.95\textwidth, keepaspectratio]{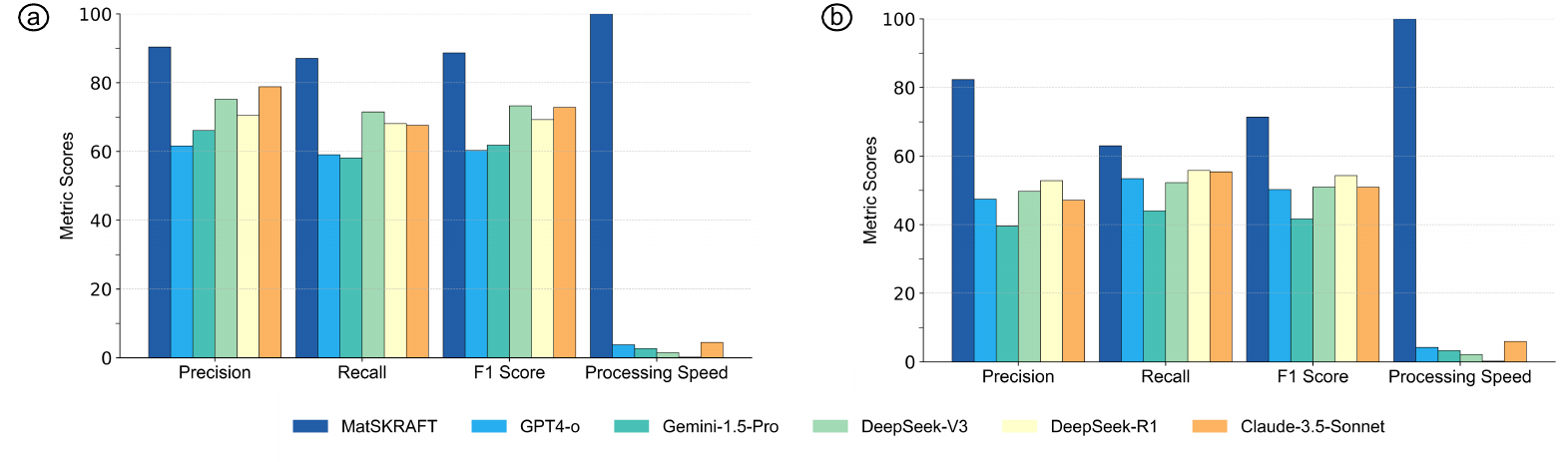}
    \vspace{-0.1in}
    \caption{\textbf{Performance comparison of MatSKRAFT against large language models (2024-2025) for materials science information extraction.} (a) Property extraction performance across precision, recall, F1 score, and processing efficiency metrics.(b) Composition extraction performance using identical evaluation criteria. \\
    \matskraft{} demonstrates superior accuracy across all metrics while achieving dramatically enhanced processing efficiency, with processing speed shown as relative throughput (number of tables processed in equivalent time, with \matskraft{} normalized to 100). \matskraft{}'s specialized graph neural network architecture consistently outperforms large language models including GPT-4o, Gemini-1.5-Pro, DeepSeek-V3, DeepSeek-R1, and Claude-3.5-Sonnet across both extraction tasks, establishing the superiority of domain-specific architectures over generalist approaches for scientific knowledge extraction.}
    \label{fig:appendix_baseline_comparison}
    \vspace{-0.1in}
\end{figure}

\textbf{3. Domain specialization advantages over generalist architectures.} The systematic performance superiority across all baselines validates our hypothesis that domain-specialized graph neural network architectures outperform generalist large language models for scientific knowledge extraction tasks. While LLMs demonstrate impressive capabilities across diverse domains, their sequence-processing paradigms struggle with the structured, relational nature of scientific tables where spatial relationships, hierarchical organization, and domain-specific constraints determine semantic meaning.

\matskraft{}'s constraint-driven learning framework, which encodes materials science principles directly into the neural architecture, enables sophisticated interpretation of tabular structures that generic text-processing models cannot achieve. The integration of domain expertise through specialized post-processing rules—handling symbol disambiguation, unit normalization, and physicochemical validation—creates extraction capabilities that approach expert-level interpretation while maintaining computational efficiency impossible with large-scale generalist models.

\textbf{4. Democratization of advanced materials knowledge extraction.} By achieving superior performance with reduced computational requirements, \matskraft{} democratizes access to advanced scientific knowledge extraction capabilities previously available only to organizations with extensive computational resources (see Supplementary Figure~\ref{fig:appendix_baseline_comparison}). The framework's efficiency enables comprehensive processing of entire journal collections within reasonable time and budget constraints, transforming materials informatics from a resource-intensive endeavor accessible to few into a broadly available tool for accelerating scientific discovery. This accessibility transformation extends to developing regions and smaller research institutions, where our open-source framework provides production-level extraction capabilities without requiring expensive cloud infrastructure or specialized hardware. The combination of superior accuracy and computational efficiency establishes \matskraft{} as a paradigm-shifting tool that can accelerate materials discovery cycles across academic, industrial, and governmental research organizations worldwide.

\begin{table*}[htbp]
\captionsetup{justification=raggedright,singlelinecheck=false}
\centering
\small
\begin{tabular*}{\textwidth}{@{\extracolsep{\fill}}lcccc@{}}
\hline
\textbf{Property} & \textbf{Precision (\%)} & \textbf{Recall (\%)} & \textbf{F1 Score (\%)} & \textbf{Support} \\
\hline
Activation energy & \cellcolor{blue!10}93.97 & 72.76 & 82.02 & 257 \\
Annealing point & 69.05 & 67.44 & 68.24 & 43 \\
Crystallization temperature & \cellcolor{blue!10}91.14 & \cellcolor{cyan!10}94.92 & \cellcolor{pink!30}92.99 & 531 \\
Glass transition temperature & \cellcolor{blue!10}94.97 & \cellcolor{cyan!10}91.12 & \cellcolor{pink!30}93.00 & 766 \\
Liquidus temperature & \cellcolor{blue!10}91.89 & \cellcolor{cyan!10}96.23 & \cellcolor{pink!30}94.01 & 106 \\
Melting temperature & \cellcolor{blue!10}92.54 & \cellcolor{cyan!10}90.51 & \cellcolor{pink!30}91.51 & 137 \\
Softening point & 81.16 & 86.15 & 83.58 & 65 \\
Thermal expansion coefficient & 71.08 & 66.29 & 68.60 & 89 \\
Bulk modulus & \cellcolor{blue!10}100.00 & 55.56 & 71.43 & 117 \\
Density & \cellcolor{blue!10}99.05 & \cellcolor{cyan!10}94.21 & \cellcolor{pink!30}96.57 & 553 \\
Fracture toughness & 70.59 & 85.71 & 77.42 & 28 \\
Hardness & \cellcolor{blue!10}100.00 & 77.52 & 87.34 & 129 \\
Poisson ratio & \cellcolor{blue!10}100.00 & 57.66 & 73.14 & 111 \\
Shear modulus & \cellcolor{blue!10}100.00 & 74.60 & 85.45 & 126 \\
Young's modulus & 86.11 & \cellcolor{cyan!10}100.00 & \cellcolor{pink!30}92.54 & 155 \\
Abbe value & \cellcolor{blue!10}100.00 & \cellcolor{cyan!10}100.00 & \cellcolor{pink!30}100.00 & 7 \\
Refractive index & 84.30 & 87.18 & 85.71 & 234 \\
Electrical conductivity & \cellcolor{blue!10}90.91 & 75.76 & 82.64 & 66 \\
\hline
\end{tabular*}
\caption{Detailed metrics of \matskraft{} in property extraction on the test dataset after thresholding. We obtain a high precision of over 90 for 12 out of 18 properties.}
\label{tab:detailed_property_metrics}
\end{table*}

\section{Comprehensive LLM Optimization and Fair Evaluation Framework}
\label{sec: appendix_llm_fairness}

To ensure rigorous and unbiased comparison with the frontier large language models (LLMs), we implemented an extensive optimization framework that provided LLMs with advantages beyond standard API usage. Our evaluation methodology was deliberately designed to maximize LLM performance through comprehensive prompt engineering, strategic few-shot learning, and robust error-handling mechanisms that align with and extend the current best practices in LLM evaluation.

\textbf{1. Systemic few-shot learning with domain-specific examples.} We developed sophisticated prompting strategies incorporating carefully curated few-shot examples that demonstrate the full complexity of materials science table structures. For property extraction, our prompts included three strategically selected examples spanning diverse property categories: physical properties (density, refractive index), mechanical properties (Young's modulus, hardness), and null cases containing no extractable properties. Each example provided complete metadata including Publisher Item Identifiers (PIIs), table indices, captions, and expected output formats, enabling LLMs to learn both positive and negative classification patterns essential for accurate scientific knowledge extraction.

For composition extraction, we implemented an even more comprehensive training paradigm with four examples representing each distinct table type classification: Single-Cell Composition tables (SCC, Table-Type 1) where entire compositions appear within individual cells, Multiple-Cell Composition tables (MCC, Table-Type 2) where constituents are distributed across separate cells, Partial-Information tables (PI, Table-Type 3) requiring contextual inference from surrounding text, and Non-Compositional tables (NC, Table-Type 0) containing no relevant compositional data. This taxonomic approach enables LLMs to understand structural variations in materials science reporting conventions, providing advantages over generic table processing capabilities.

\textbf{2. Extensive contextual information provision.} Beyond standard table content, we provided LLMs with comprehensive contextual information to maximize extraction accuracy. For composition extraction tasks, every table was supplemented with complete research article abstracts—providing semantic context that our specialized models utilize only for the challenging Partial-Information tables. Preliminary validation confirmed that abstract inclusion enhanced LLM composition extraction performance, with notable improvements in recall, as abstracts frequently contain compositional context that clarifies tabular data. This represents a substantial advantage for LLMs, as they receive complete abstract text as direct input, whereas our constraint-driven architecture processes only caption information through node embeddings for MCC and PI tables, with PI tables requiring additional sophisticated regex-based pattern matching to infer missing compositional details. Additionally, we included detailed table captions, structural metadata, and explicit instructions for handling diverse notation systems prevalent in materials science literature.

\textbf{3. Comprehensive unit normalization and domain knowledge integration.} Our prompts incorporated extensive domain expertise through detailed unit normalization instructions spanning all 18 target properties. We provided explicit conversion guidelines for density measurements (g/cm³, mg/m³, kg/m³), temperature-related properties (°C to degC, K normalization), elastic moduli (GPa, MPa, Pa standardization), and complex cases like hardness measurements requiring scale-specific handling (Vickers HV, Rockwell HRC, Knoop HK). For electrical conductivity, we included comprehensive notation conversion rules (S/cm to $\Omega^{-1}$cm$^{-1}$), while thermal expansion coefficients received specialized exponential reconstruction instructions. 

\textbf{4. Sophisticated ID construction and structural understanding.} We developed detailed identifier construction protocols that explicitly taught LLMs to understand table orientations and generate unique identifiers following our framework's conventions. Instructions differentiated between column-oriented tables (properties arranged in columns, values read horizontally) and row-oriented tables (properties arranged in rows, values read vertically), with precise indexing schemes enabling traceability to source publications. These protocols included material identifier recognition, positional encoding (row/column indices), and hierarchical ID construction (PII\_TableNumber\_Row\_Column\_MaterialID) that mirrors our specialized architecture's internal processing logic. This systematic ID generation enables both precise traceability of extracted information to source articles and facilitates the dual-pathway integration approach—linking compositions with properties through shared positional indices within tables and connecting entities across separate tables via material identifiers—essential for constructing coherent knowledge repositories from fragmented scientific literature.

\textbf{5. Robust error handling and retry mechanisms.} Our evaluation framework implemented comprehensive error-handling mechanisms with automated retry protocols designed to maximize LLM success rates. We configured 10-attempt retry loops with intelligent error classification, distinguishing between rate limiting (HTTP 429), connection timeouts, and parsing failures. For rate-limiting scenarios, we implemented exponential backoff strategies with 3-second intervals, ensuring LLMs received maximum API allocation rather than failing due to infrastructure constraints. This robust framework guarantees that performance comparisons reflect model capabilities rather than API limitations or transient connectivity issues.

\textbf{6. Iterative prompt engineering with validation-driven optimization.} We established small validation sets comprising carefully selected examples representing the full spectrum of extraction challenges. Our iterative optimization process continued until LLMs achieved perfect performance on all validation examples, ensuring that prompting strategies were fully optimized before large-scale evaluation. This validation-driven approach identified crucial improvements, such as explicit table-type classification instructions that significantly enhanced compositional extraction accuracy by teaching LLMs to recognize structural patterns before attempting content extraction, or including the abstract along with table caption for better results.

\textbf{7. Asynchronous batch processing for throughput acceleration.} To substantially reduce wall-clock evaluation time, for the extended set of frontier models evaluated in this work (2025--2026 generation: Gemini-3.5-Flash, Claude-Opus-4.7, GPT-5.4-Pro, DeepSeek-V4-Pro, and DeepSeek-V4-Pro-Th), we replaced sequential synchronous API calls with provider-native asynchronous batch processing endpoints combined with a dispatch-and-poll architecture. Requests were grouped into batches of 8 tables with up to 16 batches permitted to remain in-flight concurrently---yielding a theoretical concurrency of 128 simultaneous table-processing requests. A dedicated polling loop queried batch status at 3-second intervals; upon confirmed completion of each submitted batch, results were immediately collected and merged into the output. Failed or unparseable responses were automatically re-queued across up to 10 retry rounds, with intermediate checkpointing after each round to ensure recoverability. This architecture reduced evaluation latency by an order of magnitude relative to sequential calling, enabling processing of 368 property tables and 737 composition tables within practical time budgets.

However, this asynchronous batch processing paradigm introduces a structural reliability limitation that warrants explicit acknowledgment: individual batches can enter prolonged processing states without advancing to completion, effectively stalling the polling loop for an indeterminate duration. Since batch processing time on the commercial LLMs' infrastructure is not contractually bounded and can vary with server load, a stuck batch provides no signal distinguishable from a legitimately slow one until a timeout threshold is crossed. In practice, this resulted in non-uniform per-run completion times that could not be estimated a priori---a fundamental limitation of API-dependent evaluation pipelines that does not affect \matskraft{}'s local inference, which executes deterministically on fixed hardware with fully predictable latency.

\textbf{8. Computational resource advantages for LLMs.} Our evaluation framework maximized LLM performance through optimal configuration settings, including generous token limits (8,192 tokens) to avoid truncation, deterministic temperature settings, and comprehensive contextual information. Also, the LLMs benefited from their inherent advantages of vast pre-trained scientific knowledge, general reasoning capabilities, and distributed processing infrastructure.

This comprehensive optimization framework ensures that our performance comparisons represent our best efforts to maximize LLM capabilities when applied to scientific knowledge extraction tasks. The advantages provided to LLM baselines establish a rigorous evaluation standard that validates \matskraft{}'s superiority through fair and comprehensive comparison methodologies.

\section{Evaluation of Open-Source Large Language Models (4B-9B)}
\label{sec:appendix_qwen}

To complement the proprietary LLM benchmarking, we evaluate five open-source large language models from the Qwen family — Qwen3-4B-Instruct, Qwen3-8B~\cite{yang2025qwen3}, Qwen3-8B-VL-Instruct~\cite{bai2025qwen3}, Qwen3.5-4B-Instruct, and Qwen3.5-VL-9B-Instruct~\cite{team2026qwen3} — spanning parameter counts of 4B to 9B and covering both base and instruction-tuned variants, including vision-language models. Inference is performed locally via vLLM~\cite{kwon2023efficient} with a maximum generation length of 8,000 tokens, and a total context window of 24,000 tokens. The same three-shot prompt, system message, and output format used for proprietary LLM evaluation (Supplementary Information ~\ref{sec: appendix_llm_fairness}) are applied without modification, ensuring a controlled and directly comparable evaluation protocol. Thinking mode is disabled for all models by default. We evaluate using two matching criteria: (i) a strict match that requires the predicted tuple to reproduce the exact source cell position — identified by the table row and column index within the full identifier string — in addition to matching property name, numerical value (with tolerance), and unit; and (ii) a lenient match that relaxes the positional constraint to require only the same paper and table identifier, retaining all other matching conditions. This dual-criteria evaluation is motivated by the structural requirement of \matskraft{}: since extracted composition and property tuples are linked via shared row–column provenance indices, a model that extracts correct values but assigns incorrect cell positions produces tuples that cannot be reliably merged into the knowledge base.

\begin{table}[htbp]
    \centering
    \resizebox{\linewidth}{!}{%
    \begin{tabular}{l|cccc|cccc}
        \hline
        & \multicolumn{4}{c|}{Extracting Properties} & \multicolumn{4}{c}{Extracting Composition} \\
        \cline{2-5} \cline{6-9}
        Models & Precision & Recall & F1 score & Time (s/table) & Precision & Recall & F1 score & Time (s/table) \\
        \hline
        Qwen3-4B-Instruct       &  5.15 &  4.37 &  4.73 & 0.65 &  9.84 & 14.32 & 11.67 & 0.63 \\
        Qwen3-8B                &  2.49 &  3.79 &  2.99 & 0.72 &  4.09 &  3.01 &  3.57 & 0.61 \\
        Qwen3-8B-VL-Instruct    &  4.32 &  4.20 &  4.26 & 0.67 & 11.08 &  9.14 & 10.01 & 0.71 \\
        Qwen3.5-4B-Instruct     &  1.50 &  0.51 &  0.76 & 0.77 &  4.83 &  4.74 &  4.78 & 0.60 \\
        Qwen3.5-VL-9B-Instruct  & 12.22 &  7.59 &  9.36 & 0.85 &  9.77 &  9.40 &  9.60 & 0.99 \\
        \hline
        \rowcolor{pink!30} \textbf{\matskraft{}} & \textbf{92.40} & \textbf{86.45} & \textbf{89.33} & \textbf{0.22} & \textbf{82.31} & \textbf{62.97} & \textbf{71.35} & \textbf{0.39} \\
        \hline
    \end{tabular}%
    }
    \vspace{0.1in}
    \caption{\textbf{Performance comparison of five Qwen-family models (4B–9B parameters) and MaTSKRAFT.} Evaluated on 368 property-extraction and 737 composition-extraction tables.}
    \label{tab:qwen_strict}
\end{table}

Under strict evaluation which requires the model to reproduce the exact row and column provenance identifier of each extracted value in addition to matching property name, numerical value, and unit—all five Qwen variants exhibit near-complete failure: property extraction F1 scores range from 0.76 to 9.36, and composition F1 scores from 3.57 to 11.67, falling around 80 and 60 percentage F1 points below \matskraft{} for both the tasks (see Supplementary Table~\ref{tab:qwen_strict}). Inspection of model outputs reveals the root cause: while models often extract numerically correct property-value pairs, they systematically fail to assign correct row and column cell addresses, producing tuples that cannot be reliably used. This is a critical failure specific to our knowledge base construction pipeline: \matskraft{} links composition and property tuples by their shared row–column cell index, the same index from which both pieces of information were extracted. Therefore, a tuple carrying an incorrect cell address cannot be merged with its corresponding composition entry, rendering it unusable regardless of whether the extracted value itself is correct. This positional tracking failure proves to be beyond the capability of models in this parameter range.

\begin{table}[htbp]
    \centering
    \resizebox{\linewidth}{!}{%
    \begin{tabular}{l|cccc|cccc}
        \hline
        & \multicolumn{4}{c|}{Extracting Properties} & \multicolumn{4}{c}{Extracting Composition} \\
        \cline{2-5} \cline{6-9}
        Models & Precision & Recall & F1 score & Time (s/table) & Precision & Recall & F1 score & Time (s/table) \\
        \hline
        Qwen3-4B-Instruct       & 62.88 & 55.20 & 58.79 & 0.65 & 19.31 & 25.14 & 21.84 & 0.63 \\
        Qwen3-8B                & 33.52 & 50.37 & 40.25 & 0.72 & 29.05 & 19.55 & 23.37 & 0.61 \\
        Qwen3-8B-VL-Instruct    & 64.52 & 63.81 & 64.16 & 0.67 & 33.56 & 27.37 & 30.15 & 0.71 \\
        Qwen3.5-4B-Instruct     & 24.98 &  9.12 & 13.36 & 0.77 & 22.23 & 27.09 & 24.49 & 0.60 \\
        Qwen3.5-VL-9B-Instruct  & 72.43 & 47.53 & 57.39 & 0.85 & 31.16 & 36.03 & 33.42 & 0.99 \\
        \hline
        \rowcolor{pink!30} \textbf{\matskraft{}} & \textbf{92.40} & \textbf{86.45} & \textbf{89.33} & \textbf{0.22} & \textbf{82.31} & \textbf{62.97} & \textbf{71.35} & \textbf{0.39} \\
        \hline
    \end{tabular}%
    }
    \vspace{0.1in}
    \caption{\textbf{Lenient-match performance of open-source compact LLMs and MaTSKRAFT on property and composition extraction.} Models are evaluated on 368 property-extraction and 737 composition-extraction tables. The cell-position constraint is relaxed to paper- and table-level matching, retaining agreement on element/compound or property name, numerical value, and unit.}
    \label{tab:qwen_lenient}
\end{table}

Lenient evaluation (Supplementary Table~\ref{tab:qwen_lenient}), which relaxes the positional constraint to paper- and table-level matching, confirms this interpretation: property F1 scores increase substantially to 13.36--64.16, indicating that the extracted values and property labels are often correct in isolation.
To disentangle value extraction accuracy from positional tracking accuracy, we additionally apply a lenient evaluation that relaxes the positional constraint to only paper- and table-level matching with the extracted material composition or property tuple (see Supplementary Table~\ref{tab:qwen_lenient}).  We observe that the property F1 scores improve substantially to 13.36–64.16, confirming that small LLMs can often identify the correct property-value-unit triple in isolation. Even under this maximally permissive criterion, all Qwen variants fall significantly short of \matskraft{}'s 89.33 (property) and 71.35 (composition), with composition extraction proving particularly challenging (lenient F1: 21.84--33.42). However, we do not observe this failure of positional tracking capability in the larger commercial LLMs. \matskraft{}'s performance is identical under both evaluation criteria, reflecting its explicit graph-based provenance tracking: every extracted tuple carries a structurally guaranteed cell identifier derived from the graph node index rather than inferred from free-form generation. These results confirm that the positional tracking required for knowledge base construction---where composition and property tuples must be linked via shared cell coordinates constitutes a fundamental barrier for open-source small language models.

\section{Comparison with Open-Source Table-Specialized Models}
\label{sec:appendix_table_baselines}

To situate \matskraft{} within the broader landscape of table-understanding research, we benchmark against three tabular-based open-source architectures---TAPAS~\cite{herzig2020tapas}, TaBERT~\cite{yin2020tabert}, and TAPEX~\cite{liu2021tapex}. For each architecture, we produce two variants: a base version trained with general-domain or randomly initialized weights, and a domain-adapted version (suffix ``-MSB'') in which the language model backbone is replaced with MatSciBERT~\cite{gupta2022matscibert}. For TaBERT, we additionally evaluate an intermediate variant (suffix ``-B''), initialized from generic bert-base-uncased weights~\cite{devlin-etal-2019-bert}, isolating the effect of generic language pretraining from the vertical-attention mechanism itself. All seven models are evaluated on both tasks, property extraction and composition extraction.

\textbf{Task formulation:}
The two extraction tasks require distinct output formulations, each adapted independently to each baseline architecture. For property extraction, the task 
is formulated as a 20-class row/column classification problem: every row and column header is independently classified into one of 20 categories---representing the background class, a material identifier, or one of 18 target material properties. Row and column representations are obtained by mean-pooling over the token-level hidden states of all cells belonging to that row or column, respectively, converting the structured extraction problem into a standard sequence classification task directly comparable to \matskraft{}'s graph-based node classification objective. Property extraction baselines are evaluated on the complete validation and test split.

For composition extraction, the task is decomposed into two coupled sub-problems: (i) row/column classification, in which each header is assigned one of four labels---background, material identifier, composition, and constituent value; and (ii) edge prediction, a binary classification over cell pairs that determines which element or compound label is associated with each composition-constituent cell. Composition tuples are recovered by traversing the predicted label assignments and edges. Structurally, materials science composition tables fall into distinct types: multi-cell composition (MCC), single-cell composition (SCC), and partial-information (PI)~\cite{hira2024reconstructing}. SCC tables encode the full composition within a single cell as a chemical formula string (\emph{e.g.}, 75TeO$_2$--25ZnO), requiring regex-based decomposition, while PI tables require cross-reference with article text to reconstruct missing constituents. In MCC tables, the constituents of each compound/element is spread across different rows/columns. Since these strategies fall outside the architectural capabilities of sequence-encoding models, both PI and SCC table types are excluded from the composition baseline evaluation, scoping the comparison to MCC tables, the only composition table type tractable within a purely table-based sequence-encoding framework, together with NC (non-compositional) tables, included to test whether each baseline produces false-positive composition predictions on tables containing none. To ensure fairness, \matskraft{} is evaluated in an ablated configuration that disables its regex-based SCC parser and PI text-lookup module. 

TAPAS, TaBERT, and TAPEX base variants train all encoder weights from random initialization; TaBERT-B initializes from bert-base-uncased~\cite{devlin-etal-2019-bert}; all MSB variants initialize from MatSciBERT~\cite{gupta2022matscibert}. TAPAS and TAPEX each test a different table-structural paradigm: TAPAS through row/column index embeddings, TAPEX through explicit linearization. Both are trained from random initialization, isolating this structural contribution from any effect of language pretraining. TaBERT treats each row as a natural-language passage, and its vertical-attention mechanism is meant to test whether attending across rows helps. Starting from random representations would make it impossible to determine whether weak performance comes from poor row representations or from the attention mechanism; pretrained BERT rules out the first possibility. We therefore evaluate both variants, TaBERT and TaBERT-B. TaBERT is fully random-initialised, whereas TaBERT-B is initialised from generic bert-base-uncased weights~\cite{devlin-etal-2019-bert}. None of the three architectures receive any table-structure-specific pretraining for this structural mechanism. This mirrors \matskraft{}'s own GNN, whose graph-structural relationships are likewise learned entirely from the materials-science training data. Cell-level text embeddings are the exception in both cases: the MSB variants and \matskraft{} itself both draw on pretrained language models at this level. The classification head is trained from random initialization in all cases. All baseline models were trained on the \matskraft{} training split.

\begin{table}[htbp]
    \centering
    \small
    \begin{tabular}{l|cccc|ccc}
        \hline
        & \multicolumn{4}{c|}{Extracting Properties} & \multicolumn{3}{c}{Extracting Composition$^*$} \\
        \cline{2-8}
        Models & Precision & Recall & F1 score & Time (s/table) & Precision & Recall & F1 score \\
        \hline
        TAPAS               & 44.18 & 70.27 & 54.25 & 0.32 & 55.40 & 42.47 & 48.08 \\
        TAPAS-MSB           & 77.96 & 77.26 & 77.61 & 0.27 & 63.58 & 51.98 & 57.20 \\
        \hline
        TaBERT              & 66.86 & 79.52 & 72.64 & 0.27 & 60.93 & 37.09 & 46.11 \\
        TaBERT-B            & 73.94 & 76.49 & 75.19 & 0.28 & 60.57 & 37.62 & 46.41 \\
        TaBERT-MSB          & 80.72 & 75.79 & 78.17 & 0.27 & 67.66 & 43.88 & 53.23 \\
        \hline
        TAPEX               & 44.46 & 49.05 & 46.64 & 0.27 & 32.66 & 21.50 & 25.93 \\
        TAPEX-MSB           & 80.82 & 81.67 & 81.24 & 0.28 & 70.07 & 55.07 & 61.67 \\
        \hline
        \rowcolor{pink!30} \textbf{\matskraft{}} & \textbf{92.40} & \textbf{86.45} & \textbf{89.33} & \textbf{0.22} & \textbf{79.87} & \textbf{60.37} & \textbf{68.77} \\
        \hline
    \end{tabular}
    \vspace{0.1in}
    \caption{\textbf{Performance comparison of \textsc{MaTSKRAFT} against open table understanding model baselines.} Time measurements represent processing duration per table. ``MSB'' denotes domain-adapted variants in which the language model backbone is replaced with MatSciBERT, ``-B'' denotes the TaBERT variant initialised from generic bert-base-uncased weights. $^*$For fair comparison, only MCC and NC tables were considered for composition extraction. This differs from \matskraft{}'s full-pipeline composition F1 of 71.35\%, which is not directly comparable since none of the baselines can handle SCC or PI tables.}
    \label{tab:table_model_comparison}
\end{table}

\textbf{TAPAS and TAPAS-MSB:} The base TAPAS model adopts the native TAPAS encoding scheme~\cite{herzig2020tapas}, in which the table is tokenized with a TapasTokenizer and special token-type embeddings encode absolute row and column indices. Token-type channels encoding numerical rankings, column ranks, and previous-answer labels, specific to table question-answering and carrying no signal for property classification --- are zeroed out. All encoder weights, including the BERT backbone, are trained from random initialization, producing a model that acquires table encoding structure entirely from the MatSKRAFT training corpus. For TAPAS-MSB, the BERT backbone is replaced with MatSciBERT while the native TAPAS tokenizer is replaced by a manual linearization: the input is constructed as [CLS] caption [SEP] cell(1,1) cell(1,2) ... cell(R,C) [SEP], with cells concatenated in row-major order. Because the WordPiece tokenizer of MatSciBERT does not carry intrinsic two-dimensional positional structure, we introduce two learned embedding matrices --- one for row indices and one for column indices, whose outputs are added element-wise to the encoder hidden states of every token belonging to each cell. The total input is capped at 512 tokens.

\textbf{TaBERT, TaBERT-B and TaBERT-MSB:} In the base TaBERT variant, the encoder weights, including the underlying BERT backbone, are trained entirely from random initialisation, matching the protocol used for TAPAS and TAPEX. In TaBERT-B, the same architecture is instead initialised from generic bert-base-uncased weights~\cite{devlin-etal-2019-bert}. The TaBERT model adopts the Vertical Attention architecture~\cite{yin2020tabert}, which encodes each table row as a separate natural-language passage and applies vertical self-attention across rows to aggregate column-level context. Empty cells are represented by a dedicated [EMPTY] token appended to the vocabulary. For TaBERT-MSB, the backbone is substituted with MatSciBERT in place of bert-base-uncased, while the vertical attention architecture is retained unchanged. The same set of candidate hyperparameters was used across all seven baseline variants, with the best-performing configuration reported for each. This isolates the effect of domain-specific pretraining from architectural differences.

\textbf{TAPEX and TAPEX-MSB:}
The base TAPEX model serializes the table as \texttt{<s> caption </s> col : $h_1$ | $h_2$ | $\cdots$ row 1 : $v_{1,1}$ | $v_{1,2}$ | $\cdots$ </s>}, rows are appended greedily up to 1,024 tokens. Empty cells are rendered as the string \texttt{empty}. The BART encoder~\cite{lewis2020bart} used in the original work has been initialized entirely from scratch using the standard BART-base configuration — no pre-trained TAPEX weights are loaded — establishing the contribution of the TAPEX linearization format independent of any table-specific pre-training. For TAPEX-MSB, the BART encoder and byte-pair encoding tokenizer are replaced with MatSciBERT and its WordPiece tokenizer. The linearization format is preserved with minimal vocabulary adaptation (\texttt{[CLS]}/\texttt{[SEP]} replacing \texttt{<s>}/\texttt{</s>}), and the maximum sequence length is reduced from 1,024 to 512 tokens to match the MatSciBERT context window.

\textbf{Results:}
Aggregate results (Supplementary Table~\ref{tab:table_model_comparison}) demonstrate that MatSciBERT adaptation consistently improves over the base variants across all three architectures. Despite these gains, all MSB variants fall short of \matskraft{}, with the best-performing TAPEX-MSB reaching 81.24 versus \matskraft{}'s 89.33 — a gap of 8.1 percentage points for property extractions that widens considerably at the per-property level.
For composition extraction, \matskraft{} achieves F1$_\text{MCC-NC}$\,=\,68.77 compared to a maximum of 61.67 (TAPEX-MSB) across all baselines. The compositional extraction task — requiring simultaneous identification of multi-component stoichiometries, percentage conventions, and cross-cell element–value associations — exposes the structural limitations of linear sequence encoders.

\begin{figure}[htbp]
    \centering
    \includegraphics[width=0.9\textwidth, keepaspectratio]{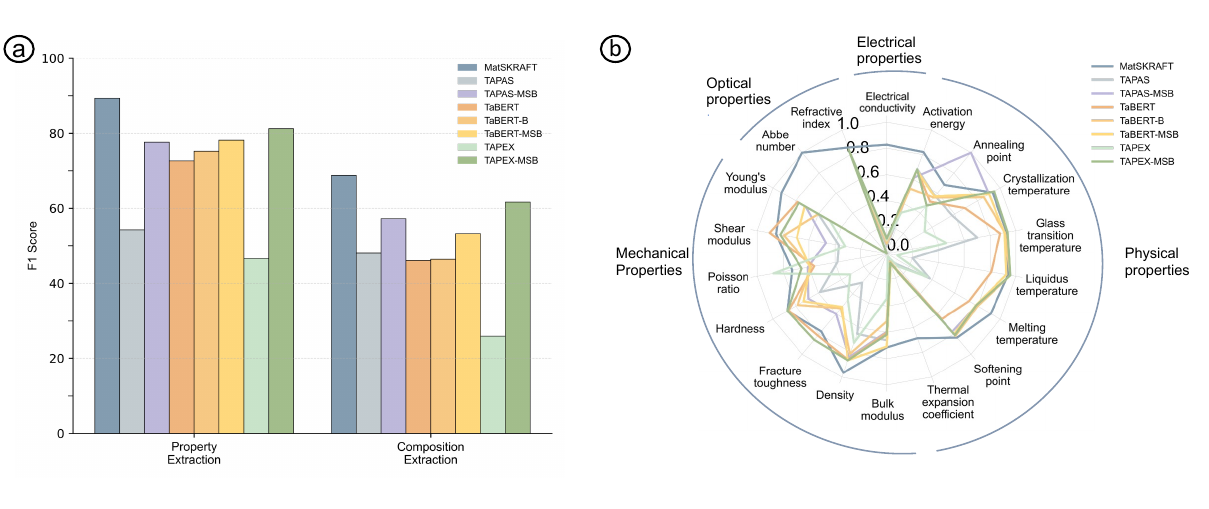}
    \caption{\textbf{Performance comparison of \textsc{MatSKRAFT} against three open-source table-understanding architectures in base and MatSciBERT-adapted configurations.} (a) Aggregate F1 scores for property and composition extraction across MatSKRAFT and seven model variants — TAPAS, TaBERT, and TAPEX, each evaluated in both base and MatSciBERT-adapted (MSB) form, and TaBERT additionally in BERT-adapted (B) form. (b) Per-property F1 scores across 18 material properties grouped by category (electrical, physical, mechanical, and optical).}
    \label{fig:d2}
\end{figure}

Per-property F1 scores reveal systematic failure modes that aggregate metrics obscure (see Supplementary Figure~\ref{fig:d2}b). Electrical conductivity and thermal expansion coefficient exhibit analogous failures: base variants achieve F1 scores of 5.7--10.4 and 2.6--6.6 respectively, with domain adaptation providing only marginal improvement (10.1--11.7 and 7.3--8.5), while \matskraft{} achieves 82.6 and 68.6. These properties share a common structural characteristic: their values appear in heterogeneous notational formats (\emph{e.g.}, S\,cm$^{-1}$, $\Omega^{-1}$\,cm$^{-1}$, $\times 10^{-6}$\,K$^{-1}$) that require unit-aware disambiguation beyond the capacity of token-level encoders. Domain adaptation is most effective for frequently reported properties with standardized notation. For crystallization temperature, TAPAS-MSB, TaBERT-MSB, and TAPEX-MSB achieve F1 scores of 90.1, 90.0, and 94.7---gains of 32.3, 21.1, and 61.3 percentage points over base variants; with the latter marginally exceeding \matskraft{}'s F1 score of 93.0.

These results establish four scientific findings. First, replacing a randomly initialized or general-domain backbone with MatSciBERT yields substantial and consistent gains across both extraction tasks. Domain adaptation improves F1 scores of property extraction by 23.4, 5.5, and 34.6; and for composition extraction by 9.1, 7.1, 35.7 percentage points for TAPAS, TaBERT, and TAPEX respectively. In TaBERT, we also observe that upgrading from generic BERT to MatSciBERT enhances property and composition extraction by around 3.0 and 6.8 percentage points respectively. These consistent improvements across all three architectures confirm that domain-specific pre-training on materials science literature is a necessary condition for table understanding in the Material Science domain. Second, the MSB variants still fall short of \matskraft{}, confirming that the performance gap reflects \matskraft{}'s constraint-driven graph architecture and domain-specialized post-processing pipeline as additional contributors. Third, the domain-adapted fine-tuned variants (TAPAS-MSB, TaBERT-MSB, TAPEX-MSB) achieve property extraction F1 scores of 77.6–81.2, surpassing all open-source small LLMs evaluated in Supplementary Information~\ref{sec:appendix_qwen} and outperforming several frontier proprietary LLMs from both the 2025 generation (GPT-4o: 60.3, Gemini-1.5-Pro: 61.9, DeepSeek-V3: 73.3) and the 2026 generation (Gemini-3.5-Flash: 63.5, DeepSeek-V4-Pro: 68.6, GPT-5.4: 79.2) — establishing domain-adapted table-understanding architectures as a viable and underexplored direction for scientific information extraction. Fourth, these fine-tuned models process tables in 0.27–0.32 seconds --- much faster than the frontier LLMs evaluated in this work, and can be deployed entirely on local hardware without API dependencies, rate-limiting constraints, or per-token costs, making them a computationally efficient and cost-effective alternative for large-scale literature mining and information extraction.

\section{Database Contents and Journal Coverage}
\label{subsec: appendix_database_coverage}

\subsection{Strategic Journal Selection for Comprehensive Materials Coverage}

 \matskraft{} creates a unique materials knowledge base from strategically selected 11 publications in the ScienceDirect database. These journals were chosen in order to maximize compositional diversity while ensuring comprehensive coverage across critical materials domains. For instance, the Journal of Nuclear Materials provides coverage of nuclear materials, whereas the Journal of Non-Crystalline Solids and Ceramics International serves as a flagship publication for ceramics and glass research, and Thin Solid Films, in conjunction with Materials Letters, broadens the focus to include nanomaterials and thin-film systems. Solid State Sciences, Solid State Communications, and the Journal of Physics and Chemistry of Solids all represent major venues of solid-state science. By considering the journals Solid State Ionics and Optical Materials, we cover the development of materials in energy and photonic applications. This strategic curation effectively captures composition-property relationships systematically underrepresented in existing commercial databases, increasing the amount of materials data that is available and thereby accelerating scientific discovery. Note that this was a conscious selection to develop a relevant dataset. The framework could be broadly applied to any journal families and publications in the materials domain.


\subsection{Massive Scale of Automated Knowledge Extraction}

The resulting \matskraft{} knowledge base comprises 509,281 individual entries extracted from 66,267 tables within 45,507 research articles—representing one of the largest automated extractions of structured materials data from scientific literature to date (Supplementary Table~\ref{tab:database_contents}). With an extraction efficiency of 7.69 entries per table, our framework demonstrates remarkable capability in capturing the rich, multi-dimensional information characteristic of materials characterization studies. This comprehensive scale enables systematic identification of rare property combinations, cross-compositional trends, and emerging research directions that are challenging to detect through traditional manual analysis approaches. The final constructed knowledge-base (KB) can be downloaded from \href{https://zenodo.org/records/20684754}{Zenodo}.

\begin{table*}[htbp]
\centering
\small
\captionsetup{justification=raggedright,singlelinecheck=false}
\begin{tabular*}{\textwidth}{@{\extracolsep{\fill}}lccccccc@{}}
\hline
\textbf{Journal Name} & \textbf{Total} & \textbf{Total} & \textbf{Comp+} & \textbf{Comp+Prop} & \textbf{Only} & \textbf{Only} & \textbf{Total} \\
& \textbf{Papers} & \textbf{Tables} & \textbf{Prop} & \textbf{Inter-Tab} & \textbf{Comp} & \textbf{Prop} & \textbf{Entries} \\
\hline
J. Solid State Chem. & 3,563 & 5,322 & 3,080 & 210 & 16,481 & 14,199 & 33,970 \\
Solid State Ionics & 2,493 & 3,517 & 4,302 & 387 & 6,112 & 17,573 & 28,374 \\
J. Phys. Chem. Solids & 2,259 & 3,528 & 5,123 & 658 & 6,346 & 21,845 & 33,972 \\
J. Non-Cryst. Solids & 6,406 & 10,199 & 29,682 & 12,104 & 23,319 & 42,186 & 107,291 \\
Ceram. Int. & 13,328 & 20,425 & 12,581 & 7,880 & 37,901 & 83,446 & 141,808 \\
Solid State Sci. & 1,675 & 2,514 & 1,873 & 698 & 6,343 & 9,351 & 18,265 \\
J. Nucl. Mater. & 4,373 & 6,237 & 1,739 & 986 & 10,417 & 29,076 & 42,218 \\
Thin Solid Films & 4,201 & 5,318 & 2,576 & 381 & 7,748 & 24,458 & 35,163 \\
Mater. Lett. & 3,246 & 3,643 & 2,197 & 214 & 5,212 & 14,051 & 21,674 \\
Solid State Commun. & 1,527 & 2,206 & 2,798 & 285 & 3,341 & 14,864 & 21,288 \\
Opt. Mater. & 2,436 & 3,358 & 2,791 & 1,658 & 5,215 & 15,594 & 25,258 \\
\hline
\textbf{Total} & \textbf{45,507} & \textbf{66,267} & \textbf{68,742} & \textbf{25,461} & \textbf{128,435} & \textbf{286,643} & \textbf{509,281} \\
\hline
\end{tabular*}
\caption{Per-journal extraction statistics, including article and table counts, and the breakdown of entries by composition and property content.}
\label{tab:database_contents}
\end{table*}

\subsection{Extracted Knowledge Integration}

\matskraft{} successfully integrates 94,203 composition-property pairs (68,742 intra-table and 25,461 inter-table), representing a breakthrough achievement in reconstructing coherent materials knowledge from fragmented scientific literature. The notable contribution of inter-table linking (27.03\% of paired entries) reveals the pervasive fragmentation of materials information across publications and validates our sophisticated cross-reference resolution methodology.

This integration success demonstrates \matskraft{}'s unique capability to transform isolated tabular data into systematic knowledge structures. The framework also captures 104,000+ material compositions entirely absent from existing databases, representing a critical expansion of known inorganic materials design space that opens remarkable opportunities for discovering novel composition-property relationships.

\subsection{Strategic Impact for Materials Discovery}

Beyond scale advantages, \matskraft{} establishes a new paradigm for materials knowledge accessibility. Our automated approach revolutionizes materials informatics from resource-intensive manual curation—previously accessible only to well-funded institutions—into a democratized capability enabling systematic exploration across diverse research environments. The comprehensive coverage spanning ceramics to specialized nuclear materials, combined with robust cross-table integration, creates an unparalleled resource for identifying underexplored compositional regions and rare property combinations critical for next-generation materials design.

While our current table-focused methodology captures the majority of systematically reported materials data, integration of text-table linkages represents a clear pathway for future enhancement. Such extensions could potentially double our composition-property integration rate by connecting synthesis details mentioned in methodology sections with characterization results presented in tables, further amplifying the remarkable potential of automated materials knowledge extraction for accelerated discovery.

\section{Algorithms used in our framework}

\subsection{Unit Extraction and Normalization for Property Tables}
\label{algo_unit_extraction}

In scientific tables, units are often embedded in ambiguous locations with various semantic structures, scattered across numeric rows, column headers, captions, or even omitted altogether. Sometimes, even the same entity is reported with different units in the same tables. In scientific domains,  accurate unit identification is essential to convert extracted property values into physically meaningful and machine-readable records. To address this, we develop a modular rule-based unit extraction pipeline that integrates symbolic matching, preventive filters, and multi-stage fallback mechanisms.

\begin{table}[htbp]
    \centering
    \small
    \begin{tabular}{ll|c||ll|c}
        \hline
        \multicolumn{3}{l||}{\textbf{Physical Properties}} & \multicolumn{3}{l}{\textbf{Mechanical and Electrical Properties}} \\
        \hline
        & \textbf{Property} & \textbf{Accuracy (\%)} & & \textbf{Property} & \textbf{Accuracy (\%)} \\
        \hline
        & Activation energy & \cellcolor{pink!30}91.67 & & Bulk modulus & \cellcolor{pink!30}100.00 \\
        & Annealing point & \cellcolor{pink!30}100.00 & & Density & \cellcolor{pink!30}100.00 \\
        & Crystallization temperature & \cellcolor{pink!30}98.00 & & Fracture toughness & 75.00 \\
        & Glass transition temperature & \cellcolor{pink!30}98.81 & & Hardness & \cellcolor{pink!30}100.00 \\
        & Liquidus temperature & \cellcolor{pink!30}90.00 & & Poisson ratio & \textemdash \\
        & Melting temperature & \cellcolor{pink!30}94.74 & & Shear modulus & \cellcolor{pink!30}100.00 \\
        & Softening point & \cellcolor{pink!30}100.00 & & Young's modulus & \cellcolor{pink!30}100.00 \\
        & Thermal expansion coefficient & 82.35 & & Electrical conductivity & \cellcolor{pink!30}100.00 \\
        \hline
    \end{tabular}
    \vspace{0.1in}
    \caption{Property-wise unit extraction accuracy of \matskraft{} on the test set. Scores $\geq$90\% are highlighted. Overall unit extraction accuracy: 97.18\%.}
    \label{tab:unit_accuracy_grouped}
\end{table}


\begin{algorithm}[H]
\caption{\textsc{SetUnits}: Robust Unit Extraction with Preventive Filters and Fallbacks}
\label{algo:set_units}
\begin{algorithmic}[1]

\STATE \textbf{Function} \textsc{SetUnits}(\texttt{table\_line}, \texttt{prop\_idx}, \texttt{pii}, \texttt{t\_idx})
\STATE \textbf{if} \texttt{prop\_idx} represents a dimensionless property:
    \STATE \hspace{1em} \textbf{return None}
\STATE Initialize $unit \gets \texttt{''}$
\STATE $unit \gets$ \textsc{InferFromHeaderOrCaption}(\texttt{table\_line}, \texttt{prop\_idx}, \texttt{pii}, \texttt{t\_idx})

\vspace{0.5em}
\STATE \textbf{if} $unit = \texttt{''}$ and \texttt{prop\_idx} is a very-high-precision property:
\STATE \hspace{1em} $unit \gets$ \textsc{DefaultValueFromMedian}(\texttt{prop\_idx}) \COMMENT{Use median-based fallback}

\STATE \textbf{if} $unit = \texttt{''}$:
\STATE \hspace{1em} $unit \gets$ \textsc{ValidateAndRefineUnit}($unit$, \texttt{prop\_idx}, \texttt{pii}, \texttt{t\_idx})

\vspace{0.5em}
\STATE \textbf{if} \textsc{CheckNonControvUnit}($unit$, $prop\_name$) is \textbf{False}:
\STATE \hspace{1em} $unit \gets$ \textsc{PostProcessUnitExtras}($unit$, $prop\_name$, $heading$, $caption$, $value$)

\vspace{0.5em}
\STATE \textbf{return} \bm{$unit$}

\end{algorithmic}
\end{algorithm}

\begin{algorithm}[H]
\caption{\textsc{InferFromHeaderOrCaption}: Unit Guessing via Contextual Header and Caption Patterns}
\label{algo:infer_unit}
\begin{algorithmic}[1]
\STATE \textbf{Function} \textsc{InferFromHeaderOrCaption}($table_line$,$prop\_idx$, $pii$, $t\_idx$)

\STATE Identify column index $c^*$ of first numeric entry in target row/column $table_line$

\STATE \textbf{if} $c^* = 0$: \COMMENT{Failsafe if numeric not found}
\STATE \hspace{1em} Set $c^* \gets \min\left(\left\lfloor \frac{c}{2} \right\rfloor, 3\right)$

\STATE Initialize $unit \gets \texttt{''}$

\STATE \textbf{for} $i = 0$ to $c^* - 1$:
    \STATE \hspace{1em} Extract token $t_i \gets \texttt{table\_line}[i]$
    \STATE \hspace{1em} \textbf{if} $t_i$ matches pattern like \texttt{(MPa)}, \texttt{$\pm$ 0.1 GPa}, or contains the known unit aliases:
        \STATE \hspace{2em} Clean it if necessary and return the candidate unit


\STATE Scan caption obtained using ($pii$, $t\_idx$) for regex-matched $unit$ and \textbf{return \bm{$unit$}} if found

\STATE \textbf{return} \texttt{''} if no match found
\end{algorithmic}
\end{algorithm}

\begin{algorithm}[H]
\caption{\textsc{ValidateAndRefineUnit}: Post-Extraction Normalization, Sanity Check, and Fallback}
\label{algo:validate_unit}
\begin{algorithmic}[1]
\STATE \textbf{Function} \textsc{ValidateAndRefineUnit}($unit$, $prop\_name$, $heading$)

\STATE $normalized \gets$ \textsc{NormUnit}($unit$, $prop\_name$)

\STATE \textbf{if} \textsc{CheckNonControvUnit}($normalized$, $prop\_name$) = \texttt{True}:
    \STATE \hspace{1em} \textbf{return} $normalized$

\STATE \textbf{else}:
    \STATE \hspace{1em} $fallback \gets$ \textsc{FindUnitFurther}($unit$, $prop\_name$, $heading$)
    \STATE \hspace{1em} $fallback\_norm \gets$ \textsc{NormUnit}($fallback$, $prop\_name$)

    \STATE \hspace{1em} \textbf{if} \textsc{CheckNonControvUnit}($fallback\_norm$, $prop\_name$) = \texttt{True}:
        \STATE \hspace{2em} \textbf{return} $fallback\_norm$

\STATE \textbf{return} \texttt{''} \COMMENT{Unit rejected after all attempts}
\end{algorithmic}
\end{algorithm}

\begin{algorithm}[H]
\caption{\textsc{NormUnit}: Canonical Normalization of Extracted Units}
\label{algo:norm_unit}
\begin{algorithmic}[1]
\STATE \textbf{Function} \textsc{NormUnit}($unit$, $prop\_name$)
\STATE \textbf{if} $unit =$ \texttt{None} or empty
    \STATE \hspace{1em} \textbf{return} \texttt{None}

\STATE Normalize format: lowercase, strip whitespace, and remove non-alphanumeric symbols (e.g., \texttt{\textbackslash cdot}, unicode dashes, or spacing artifacts)


\STATE \textbf{if} $prop\_name$ is unitless (e.g., refractive index, Abbe number)
    \STATE \hspace{1em} \textbf{return} \texttt{None}

\STATE \textbf{if} $unit$ in canonical dictionary for $prop\_name$
    \STATE \hspace{1em} \textbf{return} mapped canonical form

\STATE \textbf{return} \texttt{''} \COMMENT{Unrecognized or unsupported unit}
\end{algorithmic}
\end{algorithm}

\begin{algorithm}[H]
\caption{\textsc{CheckNonControvUnit}: Sanity Check for Non-Controversial Units}
\label{algo:check_noncontro}
\begin{algorithmic}[1]
\STATE \textbf{Function} \textsc{CheckNonControvUnit}($unit$, $prop\_name$)

\STATE Normalize input: $unit \gets$ \textsc{NormUnit}($unit$, $prop\_name$)

\STATE \textbf{if} $unit =$ \texttt{None} or empty then
    \STATE \hspace{1em} \textbf{return} \texttt{True} \COMMENT{Absence is acceptable for some properties}

\STATE \textbf{if} $unit$ matches canonical form in expert-constructed dictionary for $prop\_name$ then
    \STATE \hspace{1em} \textbf{return} \texttt{True}

\STATE \textbf{if} $unit$ matches accepted domain-specific alias (e.g., `S/cm`, `ohm$\cdot$cm`) then
    \STATE \hspace{1em} \textbf{return} \texttt{True}

\STATE \textbf{if} $prop\_name$ is a unitless property then
    \STATE \hspace{1em} \textbf{return} \texttt{True}

\STATE \textbf{return} \texttt{False} \COMMENT{Unit is unrecognized, irrelevant, or controversial}
\end{algorithmic}
\end{algorithm}

\begin{algorithm}[H]
\caption{\textsc{FindUnitFurther}: Complex Regex-Based Unit Recovery from Header or Caption}
\label{algo:find_unit_further}
\begin{algorithmic}[1]
\STATE \textbf{Function} \textsc{FindUnitFurther}($unit$, $prop\_name$, $heading$)

\STATE \textbf{if} $unit$ is valid (i.e., already recognized): 
    \STATE \hspace{1em} \textbf{return} $unit$ \COMMENT{No fallback needed}

\STATE Retrieve complex property-specific regex $\mathcal{R}_{prop}$

\STATE Apply $\mathcal{R}_{prop}$ to search for candidate units in $heading$

\STATE Extract last match (if any), clean it if required (e.g., remove parentheses or extra symbols)

\STATE \textbf{return} cleaned fallback unit if match exists; else \texttt{''}
\end{algorithmic}
\end{algorithm}

\begin{algorithm}[H]
\caption{\textsc{PostProcessUnitExtras}: Edge-Case Cleanup, Hardness-Scale Inference, and Value-Aware Filtering}
\label{algo:postprocess_unit_extras}
\begin{algorithmic}[1]
\STATE \textbf{Function} \textsc{PostProcessUnitExtras}($unit$, $prop\_name$, $heading$, $caption$, $value$)

\vspace{0.5em}
\STATE \textbf{/* Phase 1: Symbol and Formatting Cleanup */}
\STATE \textbf{if} $unit$ contains LaTeX artifacts, Greek symbols, or spacing anomalies:
\STATE \hspace{1em} $unit \gets$ \textsc{clean\_unit\_string}($unit$) 
\STATE $unit \gets$ \textsc{handle\_edge\_cases}($unit$) \COMMENT{e.g., `Pa·s', `kg·mm\textsuperscript{-2}' to canonical form, using a broader $variant\_set$ with normalization}

\vspace{0.5em}
\STATE \textbf{/* Phase 2: Property-Specific Canonicalization */}
\STATE $variant\_set \gets$ \textsc{AllUnitVariants}($unit$)

\STATE \textbf{if} $prop\_name = $ \texttt{`Hardness'}:
\STATE \hspace{1em} Remove spurious candidates from $variant\_set$ (e.g., \texttt{K}, \texttt{degC}, \texttt{g/cm$^3$})
\STATE \hspace{1em} Append scale-specific units (e.g., \texttt{HV}, \texttt{HRB}, \texttt{kgf/mm$^2$}) to $variant\_set$
\STATE \hspace{1em} $scale\_dict \gets$ predefined mapping: scale $\rightarrow$ keywords (e.g., \texttt{`vicker'} $\rightarrow$ \texttt{HV})
\STATE \hspace{1em} \textbf{for each} $(scale, keywords)$ in $scale\_dict$:
\STATE \hspace{2em} \textbf{if} any $k \in keywords$ appears in $heading.lower()$ or $caption.lower()$:
\STATE \hspace{3em} $unit \gets$ canonical unit for $scale$

\vspace{0.5em}
\STATE \textbf{/* Phase 3: Value-Aware Hardness Filtering */}
\STATE \textbf{if} $prop\_name = $ \texttt{`Hardness'}:
\STATE \hspace{1em} $range\_dict \gets$ known valid bounds for hardness units
\STATE \hspace{1em} \textbf{for each} $v \in variant\_set$:
\STATE \hspace{2em} \textbf{if} $v \in range\_dict$ and $value$ within $range\_dict[v]$:
\STATE \hspace{3em} \textbf{return} $v$ \COMMENT{Value confirms correctness of inferred unit}

\STATE \hspace{1em} \textbf{/* Phase 3.5: Caption-Based Recovery Fallback */}
\STATE \hspace{1em} \textbf{if} no unit validated:
\STATE \hspace{2em} $u_{fallback} \gets$ \textsc{AssignHardnessUnitFromValueAndText}($value$, $caption$)
\STATE \hspace{2em} \textbf{if} \textsc{ValidateHardnessValue}($value$, $u_{fallback}$): \textbf{return} $u_{fallback}$

\vspace{0.5em}
\STATE \textbf{/* Phase 4: General Canonicalization for Other Properties */}
\STATE $allowed\_units \gets$ \textsc{GetAllowedUnits}($prop\_name$)
\STATE \textbf{if} any $v \in variant\_set$ is in $allowed\_units$:
\STATE \hspace{1em} \textbf{return canonicalized} \bm{$v$}

\STATE \textbf{return} \texttt{\textbf{unit}} \COMMENT{Extracted unit which might not fit the known convention or potential error}

\end{algorithmic}
\end{algorithm}

Our pipeline begins with the \textsc{SetUnits} module. In this module, we dynamically identify header segments by scanning leftwards until the first numeric token is detected—typically marking the start of the value column. All preceding tokens are then heuristically considered as candidate unit-bearing phrases. In the absence of numeric tokens, we conservatively examine the first few cells (up to 3) as fallback, to identify tokens that resemble physical units (e.g., those enclosed in parentheses or containing ± symbols). The module filters out misleading candidates—such as composition fragments or time indicators—using a preventive rule-based classifier specific to the target properties. If no unit is found inline, we invoke \textsc{InferFromHeaderOrCaption}, a fallback strategy that scans the associated table caption and header fields using handcrafted property-specific regular expressions.

To ensure consistency across variants and abbreviations, all candidate units are passed to \textsc{NormUnit}, which maps surface forms (e.g., \texttt{MPa·m$^{1/2}$}, \texttt{MPa/sqrt(m)}, \texttt{MPa(m)$^{1/2}$} or \texttt{g/cm$^3$}, \texttt{g·cm$^{-3}$}, \texttt{g·cc$^{-1}$}, \texttt{gm/cm$^3$}, \texttt{gram per cubic centimeter}, \texttt{g/cc}) into canonical representations using a handcrafted dictionary of symbolic aliases. We validate unit correctness using \textsc{CheckNonControvUnit}, which checks whether the normalized unit is among the accepted options for the given property type. Unitless properties (e.g., refractive index, Abbe number) are gracefully handled during this step. If the unit remains ambiguous or controversial after normalization, we apply a final fallback via \textsc{FindUnitFurther}, which re-scans contextual metadata using complex property-conditioned regex templates. These steps are orchestrated through the meta-module \textsc{ValidateAndRefineUnit}, which integrates normalization, validation, and contextual search to produce a final accepted unit or defer to a default placeholder (blank string).

\textsc{PostProcessUnitExtras} serves as the final refinement module, incorporating domain-specific heuristics for complex unit cases. For all properties, the function first performs LaTeX artifact cleanup, resolves non-standard notations, and invokes value-aware inference to validate units against plausible numerical ranges. For Vickers hardness, in particular, the system detects a wide range of aliases—including \texttt{`hv'}, \texttt{`vhn'}, \texttt{`microhardness'}, and variants like \texttt{`h v'} or \texttt{`micro-hardness'}—and assigns the canonical unit \texttt{HV} if the reported value lies between 5 and 10,000 - a wide-and-safe range for accommodating extreme cases. If the value falls in a nanoindentation range (0.1–5.0), the unit \texttt{GPa} is assigned provided the surrounding text explicitly mentions ``hardness'' without a defined scale.

The system also supports multiple Rockwell subscales, mapping phrases such as \texttt{`hrb'}, \texttt{`rockwell b'}, or \texttt{`rockwell type c'} to their respective units \texttt{HRB}, \texttt{HRC}, and more, and assigning a generic \texttt{HR} label when no specific subscale is detected. Similar logic is applied for Knoop hardness (e.g., \texttt{`knoop'}, \texttt{`khn'}, \texttt{`hk'} $\rightarrow$ \texttt{HK}), Brinell hardness (\texttt{`brinell'}, \texttt{`bhn'}, \texttt{`hb'} $\rightarrow$ \texttt{HB}), and Mohs hardness (\texttt{`mohs'}, \texttt{`moh'}, \texttt{`mohs scale'} $\rightarrow$ \texttt{Mohs}). For Shore hardness, the system disambiguates \texttt{Shore A} and \texttt{Shore D} using lexical patterns such as \texttt{`shore-a'}, \texttt{`sha'}, or \texttt{`shore type d'}. These value and context-driven rules allow the framework to reliably infer the correct unit. When none of the scale-specific patterns apply, the system defers to a conservative fallback policy—returning an empty unit if no plausible match is found. If a match exists but does not conform to known conventions, the system returns the raw matched string without normalization or rejection—preserving it for downstream inspection, as it may correspond to emerging nomenclature or novel reporting styles in the literature.

Unlike joint modeling approaches that tie unit prediction directly to property classification, our framework adopts a modular, rule-based strategy where unit extraction is treated as an independent, interpretable task. This decoupled design enables detailed scientific reasoning through symbolic parsing and context-sensitive validation. For instance, thermal conductivity units—frequently expressed using diverse notations such as W/m·K, W·m$^{-1}$·K$^{-1}$, or W/(mK)—are reliably identified and mapped to a canonical format through a carefully curated regular-expression dictionary tailored per property. This normalization process also integrates contextual cues such as surrounding tokens and known value distributions, enabling the system to distinguish thermal conductivity from similar quantities like heat capacity or energy flux, even when reported with partially ambiguous units.

This design choice proves particularly critical for properties like \textbf{hardness}, which exhibit extreme variability in notation, scale, and unit conventions across the literature. The diversity of representations—ranging from scale-specific abbreviations (e.g., HV, HRB, HK) to contextual phrases (e.g., ``microhardness'', ``Rockwell type C'')—would render predictive modeling unreliable without prohibitively large, manually annotated datasets. More importantly, these variations are not mere noise: they encode physically meaningful distinctions between different hardness scales and measurement protocols. A purely data-driven model risks collapsing or erasing this information. By contrast, our modular pipeline enables precise handling of such property-specific complexities through transparent, extensible logic—ensuring that scientific insights are preserved, not discarded, while offering interpretability that is unattainable with black-box deep learning models, maintaining interoperability with downstream materials informatics workflows, and remaining open to expert inspection and verification.

The framework also enables intelligent post-processing through unit-informed cross-validation. For example, if a table column predicted as Young’s modulus contains entries with temperature-like units (e.g., °C, K), the system flags this as a misclassification and reassigns the correct property label during refinement. Likewise, if the unit suggests a reciprocal relationship—as in the case where electrical resistivity is expressed instead of conductivity—the pipeline automatically computes the inverse, validates the resulting values, and retains them only if they fall within established physical bounds. This capability arises from our strict adherence to physical semantics rather than learning correlations, allowing the model to correct itself through interpretable signals.

Our system does not hallucinate units when none are detectable. If the table, caption, or metadata lack extractable unit indicators, we assign a blank string—thus encouraging traceability to the original source rather than speculative inference. Despite this conservative evaluation policy, our framework achieves an overall 97.18\% unit extraction accuracy across 15 properties, with 9 of the 15 properties reaching 100\% accuracy and 13 out of 15 reaching over 90\% (detailed in Supplementary Table~\ref{tab:unit_accuracy_grouped}). This underscores the robustness of our domain-informed design, which not only maximizes extraction fidelity but also enhances the trustworthiness and reusability of the resulting knowledge base.


\subsection{Post-processing for Property Extraction}
\label{algo_post_processing}

\begin{algorithm}[H]
\caption{\textsc{PostProcessingBegin}: Rule-Based Correction and Contextual Label Refinement}
\label{algo:post_processing_begin}
\begin{algorithmic}[1]
\REQUIRE Table metadata $(pii, t\_idx)$, predicted labels $L_{row}, L_{col}$, table $T$, caption $C$
\ENSURE Refined property predictions $L_{row}, L_{col}$ and corrected value matrix

\STATE $L \gets L_{row} + L_{col}$ \COMMENT{Combined label vector}
\STATE Assert alignment: $\texttt{len}(L) = num\_rows + num\_cols$

\STATE Convert all label codes $2$--$20$ to $2$ to get $alt\_L$
\STATE Compute for detecting orientation: $r \gets \texttt{count}(2 \in alt\_L_{row})/num\_rows$, $c \gets \texttt{count}(2 \in alt\_L_{col})/num\_cols$

\STATE Define \textbf{PatternMap}: maps property ID $\rightarrow$ list of symbolic aliases or header patterns
\STATE Define \textbf{ContextMap}: maps property ID $\rightarrow$ conditions involving captions or units
\STATE Define \textbf{ValueFixMap}: maps property ID $\rightarrow$ logic to fix numerics (e.g., $10^{-x}$); used in properties like thermal expansion coefficient (TEC) and electrical conductivity where values are often reported in prefixed exponential forms

\STATE \textbf{if} $r \leq c$: \COMMENT{Column-major table (default)}
    \FOR{$j = 0$ to $num\_cols - 1$}
        \STATE $heading \gets T[0][j]$, $values \gets T[1:][j]$, $p \gets L_{col}[j]$

        \STATE \textbf{if} $values$ are strings:
        \STATE \hspace{1em} \textit{Evaluate each entry using a safe symbolic parser (e.g., AST), though this occurs rarely as all table cells are pre-processed through \textsc{find\_num}, our handcrafted numerical extraction function that handles scientific notation, uncertainty ranges, and embedded text}

        \STATE \textbf{if} \textsc{CheckHeading}($heading$, $C$, $p$) = \texttt{False}:
        \STATE \hspace{1em} $L_{col}[j] \gets 0$

        \STATE \textbf{if} $p = 0$: 
        \STATE \hspace{1em} $p' \gets$ \textsc{DirectMatching}($heading$)
        \STATE \hspace{1em} \textbf{if} $p' \ne 0$: $L_{col}[j] \gets p'$

        \FORALL{$(p_{id}, pattern\_list) \in$ \textbf{PatternMap}} 
            \STATE \textbf{if} $heading$ matches any pattern in $pattern\_list$ and $p \ne p_{id}$:
            \STATE \hspace{1em} $L_{col}[j] \gets p_{id}$
        \ENDFOR

        \FORALL{$(p_{id}, ctx\_condition) \in$ \textbf{ContextMap}}
            \STATE \textbf{if} $ctx\_condition$ satisfied on $(heading, C, unit)$ and $p \ne p_{id}$:
            \STATE \hspace{1em} $L_{col}[j] \gets p_{id}$
        \ENDFOR

        \FORALL{$(p_{id}, regex) \in$ \textbf{ValueFixMap}}
            \STATE \textbf{if} $heading$ matches $regex$ and $p = p_{id}$:
            \STATE \hspace{1em} $m \gets$ median of cleaned $values$
            \STATE \hspace{1em} \textbf{if} $m > 0.1$:
            \STATE \hspace{2em} multiply all $values$ by $10^{-x}$ from match
        \ENDFOR

    \ENDFOR

\STATE \textbf{else}: \COMMENT{Row-major orientation}
    \FOR{$i = 0$ to $num\_rows - 1$}
        \STATE Repeat all logic above with $T[i][:]$, $L_{row}[i]$
    \ENDFOR

\vspace{0.5em}
\STATE $f\_pred \gets$ \textsc{GenerateTuplesFromTable}($T$, $L_{row}$, $L_{col}$)
\STATE $f\_pred \gets$ \textsc{temp\_cut\_off}($f\_pred$)
\STATE $f\_pred \gets$ \textsc{remove\_tuples\_on\_units}($f\_pred$)

\RETURN Refined predictions $L_{row}, L_{col}$ and updated $f\_pred$
\end{algorithmic}
\end{algorithm}

\newpage

\begin{algorithm}[H]
\caption{\textsc{check\_heading}: Semantic Disambiguation of Predicted Property Labels}
\label{algo:check_heading}
\begin{algorithmic}[1]
\STATE \textbf{Function} \textsc{check\_heading}($pii,\ t\_idx,\ index,\ vals,\ caption,\ pred\_label$)
\STATE $heading \leftarrow vals[0]$, $num\_val \leftarrow vals[1:]$
\STATE $heading\_clean \leftarrow$ remove bracketed text from $heading$
\STATE $heading\_lower \leftarrow heading.lower()$, $caption\_lower \leftarrow caption.lower()$

\vspace{0.5em}
\STATE \textbf{if} $pred\_label = 0$: \textbf{return} \texttt{True} \COMMENT{Composition class — always valid}

\STATE \textbf{if} $heading\_lower$ contains composition units and $pred\_label \geq 2$:
\STATE \hspace{1em} Add $index$ to $comp\_v$; \textbf{return} \texttt{False}

\vspace{0.5em}
\STATE Retrieve $bad\_tokens \leftarrow \mathcal{C}_{\text{disqualify}}[pred\_label]$
\STATE \textbf{if} any token in $bad\_tokens$ appears in $heading\_lower$ or $caption\_lower$:
\STATE \hspace{1em} \textbf{return} \texttt{False}

\vspace{0.5em}
\STATE \textbf{if} $heading\_clean$ in $\mathcal{S}_{\text{ambiguous}}[pred\_label]$:
\STATE \hspace{1em} Check context from $caption$ or numerical range
\STATE \hspace{1em} \textbf{if} not resolved: \textbf{return} \texttt{False}

\vspace{0.5em}
\STATE \textbf{if} $(heading\_clean,\ pred\_label)$ is known overload (e.g., \texttt{Tm} misclassified as Melting Temp):
\STATE \hspace{1em} \textbf{if} caption does not mention canonical context:
\STATE \hspace{2em} Use \textsc{check\_paper\_for\_prop} to verify
\STATE \hspace{2em} \textbf{if} not verified: \textbf{return} \texttt{False}

\STATE \textbf{return} \texttt{True}
\end{algorithmic}
\end{algorithm}

\begin{algorithm}[H]
\caption{\textsc{direct\_matching}: High-Precision Property Identification via Canonical Phrase Matching}
\label{algo:direct_matching}
\begin{algorithmic}[1]
\STATE \textbf{Function} \textsc{direct\_matching}($element$)

\STATE $element\_lower \leftarrow$ lowercase version of $element$

\FOR{ $label,\ phrases$ in $\mathcal{D}_{\text{direct}}$ }
    \FOR{ $phrase$ in $phrases$ }
        \STATE \textbf{if} $phrase$ in $element\_lower$:
        \STATE \hspace{1em} \textbf{return} $label$
    \ENDFOR
\ENDFOR

\STATE \textbf{return} 0 \COMMENT{No direct match found — defer to fallback logic}
\end{algorithmic}
\end{algorithm}

\begin{algorithm}[H]
\caption{\textsc{check\_whether\_in\_limit}: Column-Level Median Filtering Based on Property Ranges}
\label{algo:check_whether_in_limit}
\begin{algorithmic}[1]
\STATE \textbf{Function} \textsc{check\_whether\_in\_limit}($index,\ vals,\ prop\_code$)
\STATE Retrieve $prop\_name$ from $prop\_code$ using $prop\_names$ mapping

\STATE $values \leftarrow$ numerical values extracted from $vals$ using \textsc{find\_num}
\STATE $numeric\_values \leftarrow$ convert $values$ to float
\STATE $median \leftarrow$ \texttt{np.median}($numeric\_values$)

\STATE \textbf{if} $(prop\_name \in \mathcal{M}_{\text{median}})$:
\STATE \hspace{1em} $(v_{min}, v_{max}) \leftarrow \mathcal{M}_{\text{median}}[prop\_name]$
\STATE \hspace{1em} \textbf{if} $median < v_{min}$ or $median > v_{max}$:
\STATE \hspace{2em} Add $index$ to global set tracking failures for $prop\_name$
\STATE \hspace{2em} \textbf{return} \texttt{False}

\STATE \textbf{return} \texttt{True}
\end{algorithmic}
\end{algorithm}

\begin{algorithm}[H]
\caption{\textsc{temp\_cut\_off}: Remove Physically Implausible Property Values Using Unit-Aware Bounds}
\label{algo:temp_cut_off}
\begin{algorithmic}[1]
\STATE \textbf{Function} \textsc{temp\_cut\_off}($f\_pred$)
\FORALL{ $(val,\ unit,\ prop) \in f\_pred$ }
    \STATE \textbf{if} $(prop,\ unit)$ in $\mathcal{R}_{prop}$:
    \STATE \hspace{1em} $(v_{min}, v_{max}) \leftarrow \mathcal{R}_{prop}[(prop,\ unit)]$
    \STATE \textbf{elif} $prop$ in $\mathcal{R}_{prop}$:
    \STATE \hspace{1em} $(v_{min}, v_{max}) \leftarrow \mathcal{R}_{prop}[prop]$
    \COMMENT{Properties having no units}
    \STATE \textbf{else}:
    \STATE \hspace{1em} \textbf{continue} \COMMENT{No known validation range}

    \STATE \textbf{if} $val < v_{min}$ or $val > v_{max}$:
    \STATE \hspace{1em} Remove the tuple $(val,\ unit,\ prop)$ from the list of predicted tuples $f\_pred$
\ENDFOR
\STATE \textbf{return} the updated $f\_pred$
\end{algorithmic}
\end{algorithm}

\begin{algorithm}[H]
\caption{\textsc{remove\_tuples\_on\_units}: Eliminate Tuples with Invalid Property-Unit Combinations}
\label{algo:remove_tuples_on_units}
\begin{algorithmic}[1]
\STATE \textbf{Function} \textsc{remove\_tuples\_on\_units}($f\_pred$)
\STATE indices\_to\_remove $\leftarrow$ \texttt{empty set}

\FOR{$i,\ (val,\ prop,\ valtype,\ unit)$ in enumerate($f\_pred$)}
    \STATE \textbf{if} $(prop,\ unit)$ in $\mathcal{U}_{\text{invalid}}$:
    \STATE \hspace{1em} indices\_to\_remove.add($i$)
\ENDFOR

\STATE upd\_f\_pred $\leftarrow$ all $f\_pred[i]$ where $i \notin$ indices\_to\_remove

\STATE \textbf{return} upd\_f\_pred
\end{algorithmic}
\end{algorithm}

To enhance the accuracy and interpretability of the extracted property labels, we design a modular post-processing pipeline that operates on the predictions generated by the GNN. Unlike purely predictive systems, our approach incorporates symbolic rules, statistical heuristics, and contextual reasoning to correct errors, suppress noise, and enforce physical plausibility in the extracted tuples. This layered framework is particularly critical for addressing ambiguous headers, correcting false positives, and ensuring consistency in property–unit–value combinations.

The core inference logic is orchestrated by the \textsc{PostProcessingBegin} module. It first estimates the orientation of the table (row-major or column-major) by analyzing the distribution of the property tags. Based on this, the algorithm processes either column headers or row headers as candidate property indicators. For each header, it first invokes \textsc{check\_heading}, which checks whether the predicted label is semantically valid in context. This validation involves three layers: disqualification through property-specific bad tokens (e.g."Tm as transition metal" or "Tm as maximum temperature" misclassified as `melting temperature` ), ambiguity resolution using canonical examples (e.g., distinguishing `softening point` from `annealing point`), and overload correction through cross-checking additional paper context if needed.

We apply an additional cross-checker to rectify false negatives by invoking \textsc{direct\_matching}, a lexical override mechanism that uses high-precision canonical dictionaries to map phrases like "Abbe value" or "Refractive Index" to their corresponding labels. This is followed by symbolic matching against PatternMap, a dictionary that associates each property with a set of regular expressions commonly found in real-world tables. When pattern-based recovery fails, the pipeline checks for auxiliary signals from units or caption strings through ContextMap to further disambiguate cases like distinguishing thermal conductivity from heat capacity.

The module also goes through value-aware correction through ValueFixMap, which maps specific properties to header-based regular expressions that indicate a likely need for exponent rescaling. For instance, values for thermal expansion coefficient or electrical conductivity are frequently written as "10$^{-6}$" in headers. If such a pattern is detected and the median value exceeds a safety threshold, the entire column is appropriately rescaled by the corresponding exponential factor. All values undergo a rare fallback evaluation using a symbolic parser (e.g., AST-based expression evaluation) if they remain unresolved after parsing through our handcrafted numerical extractor \textsc{find\_num}, which already handles scientific notation, uncertainty ranges, embedded citations, and text annotations. Subsequently, the \textsc{check\_whether\_in\_limit} function evaluates whether the median of the extracted numerical values lies within acceptable physical bounds for the predicted property (for instance, ensuring densities are within 0–25000, range kept higher for extracting extreme cases). If the column fails this sanity check, the label is reverted to `0' (unassigned), preemptively filtering out semantically invalid extractions that might otherwise pass downstream validation.

Once the labels are corrected, the module constructs property--value--unit triplets using \textsc{GenerateTuplesFromTable}. These tuples are passed through two critical filters to enforce physical and semantic correctness. The \textsc{temp\_cut\_off} function removes tuples whose values fall outside known physical ranges, conditioned on both property and unit (e.g., removing thermal conductivity values $>$ 1000~W/m$\cdot$K). In parallel, \textsc{remove\_tuples\_on\_units} discards tuples with invalid property--unit combinations (e.g., Abbe number in GPa) using a manually curated disallowed set.

The importance of the post-processing is clearly demonstrated by ablation studies (see Supplementary Information~\ref{sec: abl std}), where removing the post-processing module causes the sharpest performance drop among all components—reducing F1 by 9.38 points (from 88.68 to 79.30). Precision falls by 13.72 points, revealing the model’s reliance on domain-aware disambiguation to suppress false positives. The severe precision degradation signifies that these domain-specific heuristics are not superficial corrections but deeply embedded scientific reasoning that mirrors expert materials scientist decision-making.

\subsection{Algorithm for annotating multi-cell composition tables}
\label{algo_composition_annotation}

\begin{algorithm}[H]
\caption{\textsc{CompositionRelabeling}: Heuristic Re-annotation of Composition Tables with Multi-Cell Patterns}
\label{algo:mcc_relabeling}
\begin{algorithmic}[1]
\REQUIRE Original training set $T = \{t_1, t_2, ..., t_n\}$, each table $t_i$ with fields \texttt{act\_table}, \texttt{row\_label}, \texttt{col\_label}
\ENSURE Updated training set $T$ with refined composition annotations and edge lists

\FORALL{$table \in T$}
    \STATE $A \gets table[\texttt{act\_table}]$
    \STATE $(n_r, n_c) \gets table[\texttt{num\_rows}],\ table[\texttt{num\_cols}]$
    \STATE Initialize labels if missing: $table[\texttt{row\_label}] \gets [0] \times n_r,\ table[\texttt{col\_label}] \gets [0] \times n_c$
    \STATE Store backups: $table[\texttt{old\_row\_label}], table[\texttt{old\_col\_label}] \gets$ original labels

    \STATE $nums \gets$ \textsc{process\_2d\_list\_for\_numbers}($A$)
    \STATE $(comps, vars, _) \gets$ \textsc{get\_comp\_vars\_and\_nums}($A$) \COMMENT{Extract symbolic composition and placeholders}

    \vspace{0.5em}
    \STATE \textbf{if} table is column-oriented (i.e., no composition/material ID in row headers):
    \STATE \hspace{1em} $headings \gets A[0]$ 
    \FOR{$j \gets 0$ \textbf{to} $n_c - 1$}
        \STATE $text \gets headings[j]$
        \STATE \textbf{if} $text$ contains exclusion tokens (e.g., \texttt{error}, \texttt{ratio}): \textbf{continue}
        \STATE $unit \gets$ \textsc{find\_mol\_wt\_in\_text}($text$)
        \STATE \textbf{if} $unit \ne \texttt{''}$:
        \STATE \hspace{1em} $elem\_comp \gets$ \textsc{find\_elem\_comp\_var}($text$, $unit$)
        \STATE \hspace{1em} $median \gets$ median of column $j$ from $nums$
        \STATE \hspace{1em} \textbf{if} $elem\_comp$ is not empty \textbf{and} $table[\texttt{col\_label}][j] = 0$ \textbf{and} $median > 0.1$:
        \STATE \hspace{2em} $table[\texttt{col\_label}][j] \gets 2$ \COMMENT{Mark as composition column}
    \ENDFOR
    
    \vspace{0.5em}
    \STATE \textbf{if} any column is marked as composition (label 2):
    \STATE \hspace{1em} $filtered \gets$ columns with label 2 from $nums$
    \STATE \hspace{1em} $total \gets$ \textsc{sum\_rows}($filtered$, `col')
    \STATE \hspace{1em} $m \gets$ median of non-zero entries in $total$
    \STATE \hspace{1em} \textbf{if} $0.95 < m < 1.05$:
        \STATE \hspace{2em} $table[\texttt{sum\_less\_100}] \gets 0$ \COMMENT{Complete information table (tolerates doping)}
    \STATE \hspace{1em} \textbf{else}:
        \STATE \hspace{2em} $table[\texttt{sum\_less\_100}] \gets 1$ \COMMENT{Flag as partial information table}

    \STATE \hspace{1em} $medians \gets$ column-wise median for rows (per composition column)
    \STATE \hspace{1em} \textbf{for} $i = 0$ to $n_r - 1$: \textbf{if} $medians[i] \geq 0$ \textbf{and} $row\_label[i] = 0$: set $row\_label[i] \gets 1$ \COMMENT{Constituent row}

    \vspace{0.5em}
    \STATE \textbf{elif} any row is marked as composition (label 2):
        \STATE \hspace{1em} \textbf{Repeat the symmetric steps from the column-oriented branch:}
        \STATE \hspace{1em} \COMMENT{Filter composition rows $\rightarrow$ validate sum $\rightarrow$ promote constituent columns}


    \vspace{0.5em}
    \STATE $orient \gets$ `col' if $2 \in col\_label$ else `row'
    \STATE $table[\texttt{comp\_table}] \gets$ True
    \STATE $table[\texttt{edge\_list}] \gets$ \textsc{generate\_edge\_list}(\texttt{A}, row\_label, col\_label, orient)

\ENDFOR

\RETURN Updated dataset $T$ with refined composition labels
\end{algorithmic}
\end{algorithm}

In order to enhance DiSCoMaT~\cite{gupta-etal-2023-discomat}, we developed a rule-based relabeling strategy that systematically re-annotates column and row labels of the tables that are annotated as non-composition by distant supervision in the original training dataset. First, the orientation of the table is determined by the presence of existing composition or material ID labels. If none are present, we assume it to be column-oriented as 93\% of the tables reported in articles conforms to column orientation~\cite{hira2024reconstructing}. The algorithm scans header texts for chemical patterns (updated with more compounds), and molecular weight units, and assigns label~2 (composition) to a column only if the corresponding cell values have median above a small threshold, thereby avoiding erroneous annotations of metadata or purely symbolic columns. 

For column-major tables with one or more composition-labeled columns, the algorithm validates the completeness of information by summing each row across the identified composition columns. If the median of the row-wise totals falls within a tolerance window around 1.0 or 100 (i.e., \(0.95 < m < 1.05\) or \(95 < m < 105\)), the table is flagged as complete (e.g., \(\pm 5\%\) for minor misses or additional dopant concentration). Otherwise, it is marked as a partial-information table. The algorithm then computes the median of each row across the identified columns---restricted to non-zero numerical entries---to promote constituent identifiers (label~1) in previously unmarked row headers, a necessary safeguard since some rows may contain only descriptive text without any measurable composition. A symmetric logic is applied when the composition dimension lies along rows, ensuring consistent behavior across table orientations.

This re-annotation framework improves the prediction quality by marking additional source and destination nodes based on inferred table semantics. Once the refined row and column labels are finalized, an orientation-aware edge list is generated by pairing each constituent--composition tuple in the table, enabling cleaner input graphs for GNN training. This lightweight yet impactful relabeling strategy is critical for  directly enhancing composition extraction from the MCC tables by over 10.5 F1 points.

\end{document}